%% file: bhb.tex
\documentclass{aa}

\usepackage{latexsym}
\usepackage{amssymb}
\usepackage{amsmath}
\usepackage{astrobib}
\usepackage{journals}
\usepackage{graphicx}

\newcommand{\mli}[2]{#1 _{\rm {#2}}}
\newcommand{\VEC}[1]{\mbox{\boldmath${\mathrm{#1}}$\unboldmath}}
\newcommand{\mean}[1]{\langle #1 \rangle}
\def\BHBidx{\bullet\!\bullet}

\begin{document}

   \title{Binary black holes and tori in AGN}

   \subtitle{I. Ejection of stars and merging of the binary}

   \author{C. Zier
          \and
          P. L. Biermann 
          }

   \offprints{C. Zier}

   \institute{Max-Planck Institut f\"ur Radiostronomie (MPIfR), Bonn,
              Auf dem H\"ugel 69, D-53121 Bonn\\
              email: chzier@mpifr-bonn.mpg.de
             }

\date{Received, ; accepted,}

\abstract{
Observations of HST and groundbased data strongly suggest that most
galaxies harbour central supermassive black holes and that most
galaxies merge
with others. Consequently a black hole binary emerges as the two black
holes are spiraling into the center towards each other. In our work we
are investigating two basic questions of our understanding of the
central activity of galaxies and find that both can be answered with
``yes'': (1) Do the black holes actually merge?
(2) And does the effect of the torque of the black hole binary on the
surrounding stellar distribution help to explain the presence of the
ubiquitous torus of molecular material surrounding apparently all
active galactic nuclei?
The first question is the topic of the present
article, while the second question will be subject of a subsequent
paper.
Simulating the evolution of a stellar cluster in the potential of such
a binary by solving the equations of the restricted three body problem
we obtained the following results: Provided that the cluster is
about as massive as the black hole binary the two black holes coalesce
after $\sim 10^7\,{\rm yr}$ due to ejection of stars and finally via
emission of gravitational radiation. Whether a star is ejected or
not crucially depends on its angular momentum. Almost all stars
whose angular momentum is twice as large as that of a star circulating
around the binary in
a distance corresponding to that between the black holes, stay
bound to the binary.
In a sequence of models where the mass of the secondary black hole
increases while $M_1$ is
fixed, a bigger fraction of stars is ejected. For a more massive
binary also the cluster has to be more massive in order to allow the
two black holes to coalesce. The merger then proceeds on smaller time
scales.
The cluster is depleted in the central region and the final
distribution of stars assumes a torus-like structure, peaking at
three times the initial distance of the two black holes. The
relationship of the bound stars to the obscuring torus in active
galactic nuclei will be investigated in a subsequent paper.
\keywords{Galaxies: active --
                Galaxies: nuclei --
                Galaxies: interactions --
                Galaxies: kinematics and dynamics
               }
}

   \maketitle

%

\section{Introduction}

The bright radiation in the centers of galaxies is generally thought to
be produced by a powerful engine, which is located at the dynamical
center of its host. Because the region from where this high luminosity
is emitted is concentrated to a small volume, non-stellar activity is
suspected to cause such a great energy output. It is argued that these
active galactic nuclei (AGN) are powered by accretion onto a massive
black hole (BH) in the center. While matter is sinking down in the
potential of such a black hole, energy is released and this accretion
process is thought to be dominated by an accretion disk.
The maximum mass of
these central objects, of order $10^{-2.5}$ times the stellar
spheroidal mass of their host galaxies
\cite{magorrian98,richstone98,wang98,macchetto99}, is
consistent with the mass in black holes needed to produce the
observed energy density in quasar light if reasonable assumptions are
made about the efficiency of the energy production of the nucleus
\cite{novikov72,shakura73,chokshi92,blandford99}.
One of the best candidates of galaxies showing strong evidence for
harbouring a massive BH in its center is NGC 4258.
From VLBI observations of megamasers in its nucleus,
\citeN{miyoshi95} deduce the existence of a mass of
about $3.6\times 10^7 M_\odot$ which must be located inside the inner
radius of $\sim 13\,{\rm pc}$. 
\citeN{peterson99} use emission-line variability
data on NGC 5548, a Seyfert 1 galaxy, to infer the existence of a
mass of order $7\times 10^7 M_\odot$ within the inner few light days
of the nucleus. There are numerous other examples suggesting that most
galaxies keep a super-massive black hole in their center. This is
the first of our two basic assumptions.

According to hierarchical models a typical bright galaxy has merged a
few times since the quasar era
\cite{frenk88,carlberg89,baugh96,richstone98}.
For binary black holes with $M\gtrsim 10^7\, M_\odot$ the timescales
for a decay in the different regimes indicate that the binary will merge
on a scale short compared to the time when the next merger event
occurs \cite{xu94}. The rate of mergers for bright galaxies in the
observable universe may exceed $1/{\rm yr}$ \cite{richstone98}. Taking
into
account that local quasars might have been formed by major mergers
between galaxies, \citeN{taniguchi99b} proposes that all
active galactic nuclei of the local universe are products either from
minor or major mergers. This appears to be consistent
with almost all important observational properties of Seyfert galaxies.
Recent studies \cite{mulchaey97,ho97,derobertis98b} show
that Seyfert galaxies do not have always either bar-structure or companion
galaxies, what often had been considered to be the cause of effective
gas fuelling to the central engine
(see, e.g., \citeNP{adams77}, \citeNP{noguchi88}, 
\citeNP{shlosman90}, \citeNP{barnes92}).
An alternative way to supply the central engine with matter is the
merger driven fuelling, see for example
\citeN{toomre72}, \citeN{roos81}, \citeN{mihos94}, \citeN{derobertis98b}. 
In order to complete a minor merger about $10^9\,{\rm yr}$ are needed. This
is enough time to smear out the relics so that most of the advanced
minor mergers may be observed as ordinary-looking isolated galaxies
\cite{walker96} and Seyfert galaxies are observed to possess a
statistically significant excess of
faint ($M_{\rm V} \ge -18$) companion galaxies \cite{fuentes88}.
The frequent merging of galaxies is our second basic assumption and
consequently leads to massive black hole binaries in combination
with the first assumption.

Observations show that all
ellipticals have central density cusps which can be divided into
`weak' cusps ($\rho\propto r^{-(0.3-1.1)}$, bright ellipticals) and
`strong' ($\rho\propto r^{-2}$, fainter ellipticals) cusps 
\cite{kormendy94,kormendy96,faber96}.
It has already been suggested by \citeN{begelman80} that the mass
ejection due the merging of two
massive black holes should reduce the central density and that the core
expands. In several numerical simulations it has been found that
the merging of two galaxies and their central black holes can account for
the weak cusps, where the sinking of the BH is the dominant
mechanism \cite{makino96,nakano99,milos01}. 

Because the critical central regions of AGN are strongly
  nonspherical but spatially unresolved, orientation effects are
  thought to be important, see e.g. the discussion of
  quasar spectra by \citeN{sanders89b}. Thus the same object seen from
  other angles would be classified differently. In Type 1 objects
  the radiation emerging from the center is not blocked by matter in
  the line of sight of the observer. Hence the X-rays in the ${\rm
  keV}$-range, the big blue bump (BBB) and the broad emission line
  region (BLR) as
  well as the narrow emission line region (NLR) can be seen directly. On
  the other hand, objects classified as Type 2, show strongly absorbed
  X-rays and no clear evidence for the big blue bump and the BLR
  \cite{lawrence82}. But
  the NLR is indistinguishable from that detected in Type 1
  objects. This region is located at bigger radii from the center
  and thus not blocked by matter at a smaller distance in the line of
  sight. Sometimes a 
  Type 1 spectrum is seen in the polarized light of Type 2 classified
  objects and thus gives strong evidence for the existence of an obscuring
  torus, see e.g. \citeN{miller91}: this is interpreted within the
  unification model as radiation
  emerging from the center and then scattered by ionized matter in the
  opening of the torus into the line of sight
  \cite{antonucci85,miller90,tran92}. This model ascribes the
  different appearance of AGN rather to different orientations than to
  intrinsic differences. Its basic ingredients which are responsible
  for the non-spherical symmetry of the nucleus are relativistic
  beaming in a jet, the obscuration of the central engine by a
  molecular, dusty torus, and possibly the power of the central
  engine (see e.g the reviews by
  \citeNP{antonucci93,urry95,madejski98,wills99} and articles by
  \citeNP{falcke95a,falcke95b,taniguchi99b,rudge00}).
  Recent XMM-Newton observations of the radio-quiet quasar Mrk 205
  might provide the first evidence for a dust-torus in a quasar
  \cite{reeves01}. XMM-data of IRAS 13349+2438 suggest
  column-densities which are consistent with a dusty torus
  \cite{sako01}.
  Observations impose important constraints on the
  properties of such tori. From X-ray absorption a
  column density of $N_{\rm H}\approx 10^{24}\,{\rm cm^{-2}}$ is
  deduced \cite{mushotzky82,mulchaey92,madejski98},
  i.e. the tori have to be optically
  thick. Number ratios of Type 1 and 2 objects tell us that such tori
  are also geometrically thick (see e.g. \citeN{taniguchi99a} and
  references therein). The AGN spectra also show a prominent infrared
  bump between $1\,{\rm mm}$ and $2\,{\rm \mu m}$
  \cite{sanders89b}, which is thought to be generated by dust
  reprocessing the incident radiation from the center. In about
  $1\,{\rm pc}$ distance the dust in the torus is heated to its
  evaporation temperature of about $1500\,{\rm K}$ and then reemitting
  the central ratiation in the infrared
  \cite{sanders89b,haas00}. \citeN{pier92} and \citeN{pier93} find
  that a very compact and thick dust torus, supported by radiation
  pressure, is not able to provide the
  required broad temperature range of the dust in order to explain the
  infrared (IR) emission. But a clumpy torus, as
  suggested by \citeNP{krolik88}, has the advantage that the dust,
  when organized in clouds, can survive more easily the strong
  radiation field of the AGN. Also the dust in the outer parts can be
  heated to higher temperatures and a broader temperature range can be
  achieved \cite{efstathiou95}. Such a clumpy torus
  model seems to be the most promising scenario also to
  \citeN{haas00}. A torus is also used in high energy physics
  according to \citeN{protheroe97} who show for Blazars
  that the TeV $\gamma$-emission site is far from the central engine
  due to $\gamma\gamma$-interactions with the far infrared photon
  field of the torus.

Now, based on our two underlying assumptions, necessarily leading to a
black hole binary, we here investigate the influence of a stellar
cluster and a black hole binary (BHB) in its center on each other in
order to answer two  basic questions in our understanding of the
central activity of galaxies:
\begin{itemize}
\item Do the two black holes in the center of two colliding galaxies
actually merge?
\item Does the effect of the torque of the binary black holes on the
stellar distribution help to explain the presence of the ubiquituous
torus of molecular material surrounding apparently all active galactic
nuclei (AGN) in the ``unified scheme''?
\end{itemize}

Since a BHB has a non-spherical symmetry it might cause an
  initially spherical symmetric stellar distribution to possibly
  assume a torus-like structure.
Hence, to find an answer to both the questions we performed numerical
simulations of the stars in 
the potential of the BHB by solving the three-body problem for each
star of the initial distribution. Integrating over all stars of the
cluster we obtain the properties of the stellar
population and can trail its evolution under the influence of the
binary.
Following the suggestion by \citeN{kazanas89},
  \citeN{alexander_netzer94} and \citeN{alexander_netzer97} that the
  BLR is made up by stars and their winds, we propose that the gas and
  dust of the torus is bound to the stars within their winds. The
  dynamics of these stars in the potential of the binary then supports
  the torus in the vertical direction, solving the problem of how to
  keep the torus geometrically thick.

In this paper we investigate the conditions for which the black holes merge
and how long such a merger lasts.
We will first use a mass-ratio of $q=10$ for the black holes to
discuss the results in more detail. Afterwards we compare them with
those obtained for the ratios $q=1$ and 100.

In a following paper (paper 2) we will concentrate on the
configuration of the stars remaining bound to the black holes, their
ability to constitute a dusty torus and the observational consequences
in order to answer the second question.


\section{The model}

The evolution of the orbit of a comparatively low mass
secondary black hole moving around a massive BH $M_1$
in the center of a massive galaxy is investigated with
analytical and numerical means by \citeN{polnarev94}.
Dynamical friction between both
galaxies results in a loss of energy and angular momentum so that the
cores of the galaxies quickly approach each other. The
probability of merging with a given initial value of pericenter is
roughly proportional to the square of this value. Thus small initial
pericenters are unlikely and in the
most cases \citeN{polnarev94} find the eccentricity
never to exceed a value of order $0.1$. After the stars surrounding
the secondary black hole have been
stripped off by tidal effects, it moves on bound
orbits in the mean gravitational field of the core of the primary
galaxy. Dynamical friction extracts kinetic energy from
the secondary BH \cite{chandrasekhar43,binney87} which binds to $M_1$
at the radius $r_{\rm b} = (M_1/M_{\rm core})^{1/3} r_{\rm
core}$. Below this
separation $M_2$ is still surrounded by the sea of field stars of the
core. The density enhancement of surrounding stars in its wake after
gravitational interaction further decreases the velocity of $M_2$ so
that it keeps on spiralling inwards due to
dynamical friction. The kinetic energy is lost to these stars,
resulting in a heating of the core, but the stars do not gain enough
energy as to leave the system. Eventually, after some $10^7\, {\rm yr}$,
the distance between both the BHs
shrinks to the cusp-radius of $M_1$, $r_{\rm cusp} = r_{\rm core}
M_1/M_{\rm core}$, where the velocity dispersion of the stars of the core
equals the keplerian velocity in the potential of $M_1$
\cite{begelman80,mikkola92,vecchio94,quinlan96}. Inside
$r_{\rm cusp}$ the kinematics of the stars and $M_2$ is dominated by
the potential of $M_1$, and the stars are now interacting with both
BHs rather than with $M_2$ only. This is the point where we start our
calculations. Neglecting the influence of the other
stars of the core we are now dealing with the three-body problem for a
single
star in the potential of both BHs. In the interaction with the black
hole binary (BHB) some stars gain enough energy in order to be ejected
from the central region. On the other hand the more distant stars just
perturb the binary's center of mass and leave its semi-major axis
unaffected and the ejection of individual stars becomes the
dominant process for further hardening. Thus stars moving on orbits
with radii of about the semi-major axis of the binary contribute most
to the shrinking of the binary in this stage. This is also found by
\citeN{hemsendorf01} in their numerical calculations. If there are sufficient
such stars the hardening continues till eventually the black holes
coalesce due to the emission of gravitational radiation. Otherwise the
hardening stalls if the inner region is not sufficiently refilled with
matter by some other process.

Therefore, once the black holes approached each other as close as the cusp
radius, the eccentricity of their orbits is likely to be very small
\cite{milos01} so that we assume them to move on circular orbits
around each other.

\subsection{Collisionless systems}
\label{sc_collsys}

The net force acting on a star
in a galaxy is mainly determined by the large structure of the galaxy.
Consequently the gravitational force of the mass distribution in the
galaxy varies smoothly with space as it acts on a single star.

The number $\mli{n}{relax}$, how often an individual star of mass $m$
has to cross
a system of $N$ identical stars so that the stars velocity $v$ is changed
by order of itself is
\begin{equation}
\label{eq_nrelax}
\mli{n}{relax} = \frac{N}{8\ln \Lambda}\,.
\end{equation}
The parameter $\Lambda$ may be
written as $\Lambda = R/\mli{b}{min} \approx Rv^2/Gm \approx N$,
see \citeN{binney87}.
The largest possible impact parameter is limited by the
scale $R$ of the stellar distribution. If, on the other hand, the impact
parameter falls below the limit $\mli{b}{min} \equiv Gm/v^2$ the
assumption of nearly a straight-line trajectory, made to obtain expression
(\ref{eq_nrelax}), breaks down.

The time a star needs to cross the volume of the stellar
distribution is of order $\mli{t}{cross} = R/v$ so that the relaxation
time may be defined as $\mli{t}{relax} = \mli{n}{relax}\,\mli{t}{cross}$.
Now, for a total amount of $N=10^8$ solar mass stars we obtain for the
number of crossing times $\mli{n}{relax}\approx 6.8\times
10^5$. If we assume the star to be of
solar mass and the cluster to extend to radii of $\sim 5\,{\rm pc}$ we
obtain the typical velocity of a star to be $v = \sqrt{GNm/R}\approx
300\,{\rm km/s}$. This allows to compute the relaxation time,
which is $\mli{t}{relax}\approx 10^{10}\,{\rm yr}$.

Thus, compared to the trajectory a star would follow if the other matter
would be perfectly smoothly distributed, individual stellar encounters
perturb this trajectory only over of order $0.1 N/\ln N$ crossing
times. This means that it takes several crossing times to deflect a
star from
its mean trajectory and therefore it is possible to obtain some
understanding of the dynamics by investigating the orbits of the stars
in a suitable mean potential without taking into account individual
stellar encounters. Since the time range of our simulation is smaller by a
factor of $\sim 10$ than the relaxation time we can neglect
individual stellar encounters in our calculations.
Thus the evolution of a stellar distribution in the potential of
a BHB can be modelled by solving the equations of motion of the
individual stars of a cluster separately. Afterwards the
results of the single stars can be combined to model the evolution 
of the complete cluster.

\subsection{Restricted three body system}

If we assume that the potential in the central region is dominated by
the two black holes we can neglect the mean potential of the stellar
distribution when we are solving the equations of motion for the
stars. Since the cluster can be
treated as a collisionless system, we can solve the equations of motion
for each star separately, i.e. we are
dealing with the restricted three body problem.
In the following all coordinates are given relative to the center of
mass, which is identified with the origin of the coordinate
system. The axis of rotation is the $z$-axis and
we always assume $M_1 \ge M_2$ and correspondingly for the mass-ratio
$q=M_1/M_2\ge 1$.
The two black holes are moving on circular orbits around each other in a
fixed distance of $a = 1\,{\rm pc}$.

To write the equations of motion in a dimensionless form we
used the following normalization constants
\begin{eqnarray}
\label{eq_norm}
r_0 & = & a\equiv 1\,{\rm pc}\,,\nonumber\\
t_0 & = & \sqrt{\frac{r_0^3}{G (M_1 +
M_2)}}\nonumber\\
& = & 1492\,\, \sqrt{\frac{q}{1+q}}\,\, M_8^{-\frac{1}{2}}
a_{\rm pc}^{\frac{3}{2}}\,{\rm yr}\,,\\
\Phi_0 & = & \frac{G (M_1 + M_2)}{r_0}\,{\rm J/kg}\,,\nonumber
\end{eqnarray}
with the mass of the primary BH $M_1$ being fixed to $10^8\,
M_\odot$ (see appendix \ref{app_3body}). $M_8 = M_1/10^8 M_\odot$ is
the dimensionless mass of the primary BH in units of
$10^8$ solar masses and equal to one throughout this paper.
The quantities $a_{\rm pc}$ and $r_{\rm pc}$ denote the dimensionless
distance between the BHs and the radial distance to the center of mass
respectively, both scaled to one parsec. In the following 
we indicate dimensionless quantities with a tilde `$\,\,\tilde{}\,\,$'
on top.

The numerical integration of the equations is processed faster
in the comoving frame where the BH masses are stationary on the
$x$-axis and the rotation axis is pointing along the $z$-axis with the
$z=0$-plane being the equatorial plane of the BHs.
In this system the equations of motion read (see appendix
\ref{app_3body}):
\begin{eqnarray}
\label{eq_dgl6_t}
\dot{\tilde{x}} & = & \tilde{v}_x\nonumber\\
\dot{\tilde{y}} & = & \tilde{v}_y\nonumber\\
\dot{\tilde{z}} & = & \tilde{v}_z\nonumber\\
\dot{\tilde{v}}_x & = &
-\frac{1}{1+q}\left(q\frac{\tilde{x}- \tilde{x}_1}{|\tilde{\VEC{r}}-
\tilde{\VEC{r}}_1|^3}
+ \frac{\tilde{x}+q \tilde{x}_1}{|\tilde{\VEC{r}}+q
\tilde{\VEC{r}}_1|^3}\right) +2 \tilde{v}_y +\tilde{x}\\
\dot{\tilde{v}}_y & = &
-\frac{1}{1+q}\left(q\frac{\tilde{y}}{|\tilde{\VEC{r}}- \tilde{\VEC{r}}_1|^3}
+ \frac{\tilde{y}}{|\tilde{\VEC{r}}+q \tilde{\VEC{r}}_1|^3}\right) -2
\tilde{v}_x +\tilde{y}\nonumber\\ 
\dot{\tilde{v}}_z & = &
-\frac{1}{1+q}\left(q\frac{\tilde{z}}{|\tilde{\VEC{r}}- \tilde{\VEC{r}}_1|^3}
+ \frac{\tilde{z}}{|\tilde{\VEC{r}}+q \tilde{\VEC{r}}_1|^3}\right)\nonumber\,.
\end{eqnarray}
These have been solved numerically after being supplied with a set of
initial conditions for the stars which we will discuss in the next
section.

In order to check the accuracy of the numerical integration, we keep
track of the deviations of the Jacobian Integral (appendix
\ref{app_jacobi}), which is a conserved quantity in this problem.

\subsection{Initial stellar distribution in phase-space}

The surface-brightness profiles of many elliptical galaxies are well
fitted by an isothermal sphere out to a few core radii. On the other
hand rotation curves of spiral galaxies are often remarkably flat out
to great radii, and this suggests that we should construct models that
deviate from the isothermal sphere only far from their cores
\cite{mihalas81,binney87}.

The structure of
an isothermal self-gravitating sphere of gas is identical with that
of a collisionless system of stars whose density in
phase-space is given by Eq.~(\ref{eq_df_iso}), see appendix
\ref{app_isosphere}.
This correspondence between the gaseous and stellar-dynamical
isothermal sphere originates in the velocity distribution of both
systems. Integrating  Eq.~(\ref{eq_df_iso}) over the
volume yields the Maxwell-Boltzmann distribution for the
velocities. As we know from kinetic gas
theory this is the equilibrium distribution for a gas
whose particles are allowed to collide elastically with each
other. Therefore, if a system of particles is described by a
distribution function 
according to Eq.~(\ref{eq_df_iso}), it makes no difference whether
the particles collide with each other or not. This makes the
isothermal sphere a very well suited initial distribution for our
purposes.
The mean and the mean-square velocity of the stars in the isothermal
sphere, $\mean{v}  = \sqrt{8/\pi} \sigma$ and $\mean{v^2} =
3\sigma^2$, are independent of the position. Since the velocity
dispersion is isotropic everywhere its
components are the same: $\mean{v_r^2}
= \mean{v_\phi^2} = \mean{v_\theta^2} =  \sigma^2$.

Interestingly \citeN{milos01} find in their numerical calculations
that the radial density profiles of the pre-merger galaxies and of the
merged galaxy just after the formation of a hard BHB are essentially
the same for a short time. For the initial density distribution
they used a powerlaw with index $-2$. 

In our model we distribute the stars according to
the singular isothermal sphere (Eq.~(\ref{eq_sing_isosph})) in
the range $0.1\le \tilde{r}\le 50$. In order to obtain the
Maxwell-Boltzmann distribution in velocity each component has been
distributed according to a gaussian.
Since we are interested in stars which initially are
bound by the potential of the BHB ($E<0$) and do not leave the system
immediately after the calculations started we choose the free
parameter $\sigma$, the velocity dispersion, to be a third of the
escape velocity, $\sigma = \mli{v}{esc}/3 = \sqrt{2 \Phi}/3$.

The initial conditions for 8000 stars have been generated
and supplied to the Eqs.~(\ref{eq_dgl6_t}). These are solved with a
Runge-Kutta code of
fourth order based on the routine {\ttfamily rkck} by
\citeN{press95}. The stepsize is selfadapting to ensure
that the desired 
accuracy is always maintained and that the stepsize does not become
too small in order to save computing time. Due to our choice of the
initial conditions all stars are bound to the binary in the beginning,
having negative energies.
Some of them are ejected later during the run time. We
define a star as being ejected if the following three conditions are
fulfilled simultaneously:
\begin{itemize}
\item the energy of the star is positive $(\tilde{E}>0)$
\item its radial distance in orbit is bigger than $50\,\tilde{a}$
\item the radial velocity component is positive $(\tilde{v}_r >0)$.
\end{itemize}
The simulation is stopped after the time $\mli{\tilde{t}}{max}=5\times
10^5$ (in dimensionless units) is reached or if the star has been
ejected before. $\mli{\tilde{t}}{max}$ corresponds to about $80000$
revolutions of the BHB or to $\approx 7.46\times 10^8\,M_8^{-1/2}
a_{\rm pc}^{3/2} \sqrt{q/(q+1)}\,{\rm yr}$.

In the next section we present the results we obtained for the
simulation using a mass-ratio $q=10$ ($M_2 = 10^7 M_\odot$). Afterwards
we will compare them with the results we obtained for $q=1$ and
$q=100$.


\section{Results for the singular isothermal sphere}

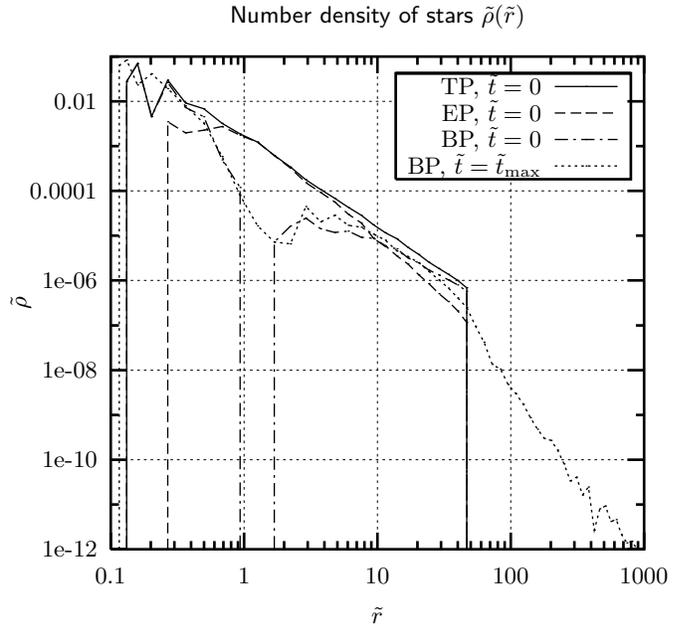
\begin{figure}
\vspace{0.7cm}
\resizebox{85mm}{!}{\input{bhb_fig01}}
\caption[]{
Initially the central region is dominated by the ejected population
(EP), the outer regions by the bound population (BP). Finally, at
$\mli{\tilde{t}}{max}$, the bound star distribution follows a powerlaw
with index $-4$ in the heated region
($\tilde{r}\gtrsim 50$), while an index $\sim -2$ is maintained in the
range $10\lesssim\tilde{r}\lesssim 50$. The inner parts are scoured
out and a maximum emerges at $\tilde{r}\approx 3$, showing the
torus-like configuration the stars assume. For $\tilde{r}\lesssim 2$
a cusp of stars bound to $M_1$ only is left. $\tilde{r}$ is
normalized to $1\,{\rm pc}$ and $\tilde{\rho}$ such that the area
underneath the solid line is $1$.} 
\label{f_dx2_rho_q10}
\end{figure}

In the following all quantities are given in and related to the
observers frame if not indicated otherwise. The origin of the
coordinate system is the center of mass of the BHB, with its rotation
axis pointing along the $z$-axis. At $\tilde{t}=0$ both black holes lie on
the $x$-axis, with $M_1$ situated on the positive and $M_2$ on the
negative seminaxis. The $y$-axis is perpendicular to both the others
so that all axes form a right handed tripod.
In spherical coordinates $\theta$ denotes the angle to the positive
$z$-axis, and $\phi$ is the azimuth in the $x-y$ plane.

A key feature is the torque exerted by the rotating black holes on the
stars, which changes the star's orbit and can eject it from the
system.

During the simulation we kept track of the single stars what enables
us at all times to assign them either to the ejected population (EP),
which is constituted
by stars being ejected before the end of the simulation, or to the
bound population (BP) when they remain tied to the binary.
Both these populations combine to
give the `total population' (TP). Quantities referring to EP, BP and TP are
denoted with the indices `ej', `bn' and `total' respectively.

\input{bhb_hlp}

\subsection{The dynamics of the stars}
\label{sec_dynst}

The density distribution of both populations EP and BP shows that
the bound stars dominate the outer regions
($\tilde{r}\gtrsim 10$, dashed-dotted line in
Fig.~\ref{f_dx2_rho_q10}). With decreasing radius the fraction of
ejected stars (long-dashed line) is increasing with both populations
being comparable in number density in a distance $\tilde{r}\approx 10$,
corresponding to $10\,{\rm pc}$. At
$\tilde{r}\approx 3$ the density of the BP has a 
maximum and after a gap in the distribution between $1$ and
$2\,{\rm pc}$, where $M_2$ is circling around $M_1$,
the density of bound stars is increasing again towards smaller
radii.
\citeN{niemeyer93} could reproduce the range of observed mm to
infrared spectra of radio-weak quasars. In their model dust is
confined to molecular clouds in a disk configuration and is heated by
the central engine.  
According to the model by \citeN{barvainis87} the dust is heated to
its evaporation temperature by the central UV luminosity in $\sim
1\,{\rm pc}$ distance. This evaporation radius is taken as the inner
radius of the torus by \citeN{lawrence91} and is in very good
agreement with the inner edge we find for the bound stellar
distribution at about $2\,{\rm pc}$ (Fig.~\ref{f_dx2_rho_q10}).
\citeN{krolik88} determine the inner radius of the
torus from the balance between cloud evaporation by central radiation
and inflow by dissipative processes in the torus and also obtain
$\sim 1\,{\rm pc}$.
At distances smaller than one parsec ($\tilde{r}\lesssim 1$)
only $62$ stars of the BP (a fraction of $1/80$) are found at
$\tilde{t}=0$. These are bound to the primary black hole only, and
till the end of the simulation the number of stars in this cusp-region
is increased by just one which has been captured by $M_1$.

The evolution of the initial density distribution (a gaussian) to its
final state is shown in Fig.~\ref{f_cont_tser}. It displays cuts
through the 3-dimensional density distribution in the comoving frame
of the binary. The right panel shows the equatorial plane ($z=0$) with
the BHs 
marked by the black points with white frames, $M_1$ to the right. The
cuts in the left panel are perpendicular to the equatorial plane and
contain the $x = 0$-plane with the $y$-axis indicated by the dashed
line. The time proceeds from top to bottom and corresponds to
$10^{4.86}$, $10^{6.57}$ and $10^{7.71}\,{\rm yr}$ respectively for
$q=10$. The first row displays the distribution very close to
the initial state. After $\approx 10^{6.57}\,{\rm yr}$
(second row) we see that mainly stars close to the secondary's
orbit ($\tilde{r}=1$) have been ejected and a gap appears at this
distance in the equatorial plane leaving a shell in the range
$1.5\lesssim \tilde{r}\lesssim 3$ behind. At about $\tilde{r}=2$ the 
density has a maximum and
we can detect a shallow torus emerging at this radius. But still the
polar regions are populated as dense as the equatorial plane, see
Fig.~\ref{f_cont_tser}, second row.

However, as is illustrated in the $3$rd row of this figure, after
about $10^{7.71}\,{\rm yr}$ the central regions have been scoured out
and the hole in the distribution has a cylindrical geometry elongated
along the binaries rotation axis. In the equatorial plane finally a
torus with a radius of about $3\,{\rm pc}$ emerges, showing that the
density distribution develops from the initial sphere over a
shell-like distribution (2nd row) to the final torus. Thus 
one important condition of the unification scheme is fulfilled, namely
that the stars do assume a torus-like distribution at a distance which
is in agreement with the origin of thermal infrared emission by dust
\cite{barvainis87,krolik88,haas00}. Whether this torus also satisfies
the other requirements, like optical thickness, will be investigated in
our paper 2.

\begin{figure}[t]
\vspace{0.7cm}
\resizebox{85mm}{!}{\input{bhb_fig08i}}%
\caption[]{
The component $\dot{\tilde{L}}_{z,1}/\tilde{r}=\tilde{r}
\ddot{\phi}\sin^2(\theta)$ of the normalized torque of the binary, 
which acts on a star moving on a fixed circular orbit with radius
$\tilde{r}$ is displayed as function of time. The angle enclosed by
the rotation-axis of the binary and the orbital spin-axis of the star
is denoted by $\theta$ and is varied in the range from $0^{\circ}$ to
$90^{\circ}$. Perturbations are stronger for orbits through the polar
regions; these stronger perturbations lead to secular changes in the
orbit, finally leading to ejection for about half of the stars in
the polar cap (see Fig.~\ref{f_cont_tser}).
}
\label{dlzdt}
\end{figure}

The basic topology of the final density distribution can be understood
in physical terms, where we consider the torque exerted by the two
black holes on an orbiting star. Fig.~\ref{dlzdt} shows of the normalized
torque the component $\dot{\tilde{L}}_{z,1}/\tilde{r}=\tilde{r}
\ddot{\phi}\sin^2(\theta)$ relative to the $z$-axis of the BHB as a
function of time for different angles 
$\theta$ between the rotation axis of the BHB and the symmetry axis of
the star's orbit. The bigger the angle $\theta$ is (i.e. for orbits
through the polar regions), the more is the star's trajectory
disturbed by the influence of the two BHs.
The curves are symmetric since for simplicity
it has been assumed that the star is moving on a keplerian circle of
radius $R$ around a point mass which corresponds to that of the binary
and is located at the origin. Of course real trajectories are
disturbed by the torque and the curves will not be symmetric any
more. The cumulative effect of these large excursions in the polarcap
regions deplete the stellar population leaving a torus behind.

Because of the conservation of the Jacobian Integral
(Eq.~(\ref{eq_djacobi})) the stars which gain energy also gain angular
momentum in the $z$-component and thus help to enhance the density in
the equatorial plane.
In paper 2 we will discuss the distribution of the BP and
investigate in the properties of the torus.

\begin{figure}[t]
\vspace{0.7cm}
\resizebox{85mm}{!}{\input{bhb_fig09i}}
\caption[]{
The angular momentum clearly separates the BP (black dots) from the EP
(grey dots), which initially have less angular momentum.
The line of transition from the EP to the BP is drawn by eye.
Its almost constant value is decreasing with increasing mass-ratio
$q$. Along this line both populations diffuse into each other, showing
that the transition is smooth.}
\label{f_dx2_ldotr_q10}
\end{figure}

At $\tilde{t}=0$ the finally ejected stars
can be clearly distinguished from the bound population by their total 
angular momentum $\tilde{L}$ or pericenter $\tilde{r}_{-}$. The mean
value of the angular momentum of the EP is smaller by a factor of
$3.22$ compared to
the BP (see Table~\ref{tb_stat} on page \pageref{tb_stat}). This
separation is clearly illustrated in Fig.~\ref{f_dx2_ldotr_q10}, where
we have plotted the initial values of angular momentum of the stars
versus their radius.
The ejected stars, marked by grey dots, have low angular momenta,
while stars staying bound to the binary till the end of the simulation
(black dots) exceed a certain value. Both populations can be separated
from each other by a transition line, as seen in
Fig.~\ref{f_dx2_ldotr_q10}.
The angular momentum has been normalized to the value of a star which
moves on circular orbits in a distance corresponding to the separation
of the BHs ($a = 1\,{\rm pc}$) around the
combined mass of the BHs $M_1 + M_1$ which is situated at the center
($L_0 = m a^2 /t_0 = m\sqrt{GM_1 a (1+q)/q}$).
Along this transition-line, which is drawn by eye and
approaches a value of about $2$ in these dimensionless units, the
different populations diffuse into each other so that the transition
from one population to the other is smooth.

This value is double the angular momentum of a star moving on a circular
orbit with the radius corresponding to the separation of the
binary. Or, the other way round, this transition-value of $\tilde{L}$
corresponds to an orbit with a radius four times as big as the
semi-major axis of the binary.

\begin{figure}[t]
\vspace{0.7cm}
\resizebox{85mm}{!}{\input{bhb_fig10i}}
\caption[]{
Ejected stars (grey dots) cover a region with $\tilde{r}_{-}\lesssim 2$
and $\tilde{r}_{+}\gtrsim 0.7$, marked by the horizontal and vertical
lines. This allows them to closely approach the orbit of $M_2$. Bound
stars (black dots) avoid this unstable region. One unit corresponds to
$1\,{\rm pc}$.}
\label{f_dx2_peridotr_q10}
\end{figure}

However, due to their much lower angular momenta the finally ejected
stars come much closer to the black holes. For the mean pericenters we
get $\mean{\tilde{r}_{\rm ej-}} = 0.86$ and 
$\mean{\tilde{r}_{\rm bn-}} = 10.42$ showing
that they differ by more than a factor of $10$.
In Fig.~\ref{f_dx2_peridotr_q10} we have plotted the initial
apocenters on the $x$-axis versus the initial pericenters on the
$y$-axis.
Without loss of information and to keep the figure distinct, we plotted
only $1/4$ of each population, chosen by random. While the
pericenters of the ejected stars stay below $\sim 2$
(i.e. $2\,{\rm pc}$, horizontal line), their
apocenters exceed $\sim 0.7$ ($0.7\,{\rm pc}$, vertical line). Hence
they cover a region which always allows them to come very
close to the orbit of the secondary BH, which rotates at a distance of
$0.9\,{\rm pc}$ around the center of mass. Here the stars undergo
violent interactions with the binary and eventually gain sufficient
energy ($\tilde{E}(\mli{\tilde{t}}{max}) > 0$) to be kicked out.
On the other hand the bound stars avoid this region and stay at
distances large enough to avoid such violent interactions
with the binary. Only very few stars, less than $1.3\%$ from the BP,
have apocenters less than $\sim 0.7\,{\rm pc}$. These are bound to the
primary black hole $M_1$ only and are not ejected because the
influence of the secondary BH is too small. They form the small
density peak around $M_1$ which is seen in Fig.~\ref{f_dx2_rho_q10}.

For the same apocenter the ejected stars have on average smaller
pericenters than the bound stars and so are moving on orbits with
higher eccentricities. As a consequence the radial velocity component
should exceed the tangential component, leading to a radially
anisotropic velocity. But due to our choice of initial conditions the
velocity anisotropy
\begin{equation}
\label{eq_beta}
\beta = 1-\frac{\mean{\tilde{v}_\phi^2} +
\mean{\tilde{v}_\theta^2}}{2\mean{\tilde{v}_r^2}}
\end{equation}
is initially zero at all radii for the total population, as can be
seen in
Fig.~\ref{f_dx2_beta_q10} (solid line).
The EP is radially anisotropic for all radii ($\beta>0$, long-dashed
line) and the velocity anisotropy is increasing with distance. While
$\beta$ is also increasing with radius for the BP, it is always
tangentially anisotropic ($\beta<0$, dashed-dotted line), the more the
smaller the radius is.
\citeN{quinlan97} also find $\beta$ of
the bound stars to grow with $\tilde{r}$ and obtain in their numerical
experiments a minimum value of $-1$, also starting with $\beta = 0$ at
all radii. They suspect the velocity anisotropy to decrease further
towards smaller radii. This is what we find in
Fig.~\ref{f_dx2_beta_q10} where $\beta$ becomes as small as $-2.5$ at
a distance $\tilde{r}\approx 4$ ($-3$ for $q=1$), the dotted
line. This is much smaller than the
values $-0.3$, predicted by the adiabatic-growth model for the formation
of massive black holes \cite{quinlan95}, and $-0.4$ obtained in
numerical simulations by \citeN{milos01}.

\begin{figure}
\vspace{0.7cm}
\resizebox{85mm}{!}{\input{bhb_fig11}}
\caption[]{The BP is shown to be strongly
tangential anisotropic while the EP is radially anisotropic. As the
eccentricity the velocity anisotropy $\beta$ increases with $\tilde{r}$
for all populations but for the TP which is initially set to zero at
all distances. At
$\mli{\tilde{t}}{max}$ the BP is radially anisotropic in the heated
region $\tilde{r}\gtrsim 50$. In the cusp region $\tilde{r}\lesssim 2$ no
clear tendency is detectable.}
\label{f_dx2_beta_q10}
\end{figure}

The eccentricity of both populations $\epsilon =
\sqrt{1+2\tilde{E}\tilde{L}^2} = (\tilde{r}_{+}-\tilde{r}_{-}) /
(\tilde{r}_{+}+\tilde{r}_{-})$ is increasing with
distance from the binary. Close to the center the bound stars have to
move on almost circular orbits in order not to come closer at the
pericenter and to avoid strong interactions with the black holes. With
increasing distance also the eccentricity can increase as long as a
big enough pericenter ($\tilde{r}_{-}\gtrsim 2\,{\rm pc}$) is maintained
i.e. not smaller than twice the separation of the BHs.
The EP must have a pericenter below this value so that
its stars can interact violently enough with the binary at the point
of the closest approch to the black holes in order to become
ejected. Thus with increasing apocenters also their eccentricity
increases, where $\epsilon \to 1$ as $\tilde{r}\to\infty$.
Consequently these stars have low angular momenta and their orbits
become the more radially anisotropic the bigger 
their distance to the center is. Hence their kinetic energy is
stored in radial rather than in tangential motion.

\subsection{Loss of $L$ and subsequent merging of the black holes}
\label{s_ang_loss}

As time proceeds there are 
almost no changes in the mean value of the total
angular momentum $\mean{\tilde{L}}$ for both populations, BP and EP. 
With the angular momentum being normalized to $L_0 = m a^2
/t_0$, where $m = M_\odot$ is the mass of a star,
the mean values of $\tilde{L}_{z}$ are $-0.045$ and $0.063$ for the BP
and EP respectively. While on average the bound stars are slightly
counterrotating, the ejected stars are corotating so that the total
population shows no net rotation in the beginning. This matches the
results obtained by \citeN{innanen79} and \citeN{innanen80}. The mean
angular momentum
$\mean{\tilde{L}_z}$ does not change for the bound stars as time
proceeds, but it increases by a factor of about $4.54$ for the ejected
population (see Table~\ref{tb_stat}, page \pageref{tb_stat}).
Such a growth is expected,
because of the conservation of the Jacobian Integral
(Eq.~(\ref{eq_djacobi})). All ejected stars gain energy and
consequently
their angular momentum in the $z$-component has to be increased, at
the expense of the BHB. Hence the ejection of stars leads to a net
loss of angular momentum and thus to hardening of the binary.

In order to get some quantitative information about the
hardening we can use the statistical values tabulated in
Table~\ref{tb_stat}. They allow to compute the mean angular momentum
that is extracted from the binary by a single star and thus also
the number of stars which is required to carry away all the
binaries angular momentum so that the black holes coalesce.
But, since we made certain assumptions and neglected some effects,
these numbers will give just an order of magnitude estimate. On the
one hand the feedback of the stars on the binary has not been taken
into account because we fixed the distance of the black holes throughout
all calculations. As a consequence the ``cross-section'' for the stars
to be ejected is not shrinking with increasing time, when ejected
stars are extracting angular momentum from the binary. Thus the
fraction of ejected stars, obtained from the simulation, is bigger
and the initial total number of stars 
needed to allow for merging will be a lower limit. For the same
reasons also the time-scale of the merger of the BHs, which we will
estimate in this section, will be shorter than in reality.
On the other hand we neglected star-star interactions, which would
support the loss-cone feeding and therefore increase the fraction of
ejected stars. Also we assumed the black holes to move on circular
orbits. If the orbits were eccentric, gravitational radiation would
become important earlier at a bigger semi-major axis resulting in a faster
merger and less stars needed in the cluster. Both the latter effects
counteract neglection of the shrinking of the black holes' distance,
but as shown previously they are of minor importance.

However, the angular momentum of the BHB is
\begin{eqnarray}
\tilde{\VEC{L}}_{\BHBidx} & = & 10^8 M_8 \left(\tilde{\VEC{r}}_1\times
\dot{\tilde{\VEC{r}}}_1 + 
\frac{1}{q}\tilde{\VEC{r}}_2\times
\dot{\tilde{\VEC{r}}}_2\right)\nonumber\\ 
& = & 10^8 M_8 (1+q)(\tilde{\VEC{r}}_1\times \dot{\tilde{\VEC{r}}}_1)
\end{eqnarray}
where we used the relation $\tilde{\VEC{r}}_2 = -q\tilde{\VEC{r}}_1$
between the positions of the two black holes. With 
the expressions $|\tilde{\VEC{r}}_1\times \dot{\tilde{\VEC{r}}}_1| =
|\tilde{\VEC{r}}_1| |\dot{\tilde{\VEC{r}}}_1|$ and
\mbox{$\tilde{a}=|\tilde{\VEC{r}}_1-\tilde{\VEC{r}}_2| = 
(1+q)|\tilde{\VEC{r}}_1|=1$} the
absolute value of the angular momentum becomes
\begin{equation}
\label{eq_lbin_stat}
\tilde{L}_{\BHBidx} = 10^8\,M_8\,a_{\rm
pc}^{1/2}\frac{1}{1+q}\,. 
\end{equation}
If $\Delta \tilde{L}_z$ is the average angular momentum extracted from
the BHB by a single ejected star, the number of ejected stars needed
to carry away all the binary's angular momentum amounts to
\begin{equation}
\label{eq_Neject_stat}
\mli{N}{ej} = \frac{\tilde{L}_{\BHBidx}}{\Delta \tilde{L}_z} =
\frac{1}{1+q}\frac{10^8 M_8}{\Delta \tilde{L}_z}\approx 4.07 \times
10^{7} M_8\,.
\end{equation}
In the last step we used the values $q=10$ and $\Delta \tilde{L}_z =
\mean{\tilde{L}_z(\mli{\tilde{t}}{max})} - 
\mean{\tilde{L}_z(\tilde{t}=0)} \approx 0.22$ taken from
Table~\ref{tb_stat}. With the ratio of bound and ejected stars
$\mli{N}{bn}/\mli{N}{ej}=1.63$ from Table~\ref{tb_stat}
the required total number of stars to let the BHs merge is:
\begin{equation}
\label{eq_Ntotal_stat}
\mli{N}{total} = \mli{N}{ej} + \mli{N}{bn} = \mli{N}{ej} \left(1 +
\frac{\mli{N}{bn}}{\mli{N}{ej}}\right) \approx 1.07 \times 10^{8}\,.
\end{equation}
If, as we assume, every star has on average the mass of the sun,
the initial star-cluster is as massive as
the binary ($M_1 + M_2 = M_1 (1+ 1/q) = 1.1\times
10^8\,M_\odot$). Hence our neglection of the mean potential, generated
by the stellar cluster compared to the potential of the BHB close
to the center, is justified.

\begin{figure}
\vspace{0.7cm}
\resizebox{85mm}{!}{\input{bhb_fig12}}
\caption[]{The loss-rate of the binary's angular momentum is largest in
the beginning. For $\tilde{t}\gtrsim 1000$ the loss-rate approximates a
powerlaw with index $\approx -3/2$, shown by the dotted line (fit
2). The dashed curve (fit 1) is used to calculate the number of
ejected stars. One time unit corresponds to $1423\,{\rm yr}$.}
\label{f_dx2_dlzdt_q10}
\end{figure}

To get an estimate of the timescales on which the black holes merge
and which serve as a lower limit as stated previously, we have
to know about the loss-rate of the angular momentum of the binary. This
is displayed in Fig.~\ref{f_dx2_dlzdt_q10} as function of
time. The data are normalized to the number of simulated
stars so that the curve represents the mean rate at which a total of
one solar mass extracts angular momentum from the binary when ejected
gradually till the end of the simulation. This mass will be referred
to in the following as a ``representative star''. A simple fit to the 
data enables us to estimate the time needed for the black holes to merge and
also to calculate the number of stars required so that the black holes can
coalesce. This is then compared with the number $\mli{N}{ej}$ obtained
above by statistical means and should be the same if the fit is
sufficiently accurate.

After a delay of $\tilde{t}\approx 26$ the loss-rate is increasing
steeply, passing through a maximum at $\tilde{t}\sim 150$
(corresponding to $2.1\times 10^5\,{\rm yr}$) and then
quickly approaches a powerlaw with index $\sim -3/2$. This initial
delay is caused by two reasons: first the stars have to interact with
the binary before they gain enough energy to be kicked out since all
stars are bound at the beginning. Afterwards the star, having now a
positive energy, has to move to a distance of $50\,{\rm pc}$ in order to
be registered as ejected by the code.
The exponential decrease of the loss-rate for $\tilde{t}\gtrsim 10^3$
($\sim 1.4\times 10^6\,{\rm yr}$) shows that the system evolves faster
in the early stages and the
angular momentum varies very slowly at later times. If the
black holes get sufficiently close due to the ejection of stars, gravitational
radiation will become strong enough to complete the merger.
For $\tilde{t}\gtrsim 10^4$ the curve is not very smooth because of the
small number statistics. As we will see later this corresponds to the
time required by the BHs in order to merge,
$\sim 1.4\times10^7 {\rm yr}$.
The lossrate can be approximated by a powerlaw with index 
$-\alpha = -3/2$, multiplied with a function which cuts it off
at $\mli{\tilde{t}}{d} = 20$:
\begin{equation}
\label{eq_dlzdt_fit}
\frac{d\tilde{L}_z}{d\tilde{t}} = k
\tilde{t}^{-\alpha}\left[1-\left(\frac{\mli{\tilde{t}}{d}}{\tilde{t}}
\right)^\eta  \right]^\kappa \,.
\end{equation}
Integration with respect to time yields
\begin{equation}
\label{eq_intdldt}
\mathcal{L}(\tilde{t}) = \int \frac{d\tilde{L}_z}{d\tilde{t}} d\tilde{t} =
-\frac{2k}{\sqrt{\tilde{t}}}\,F\left(\frac{1}{2\eta }, -\kappa ; 1 +
 \frac{1}{2\eta };
\left(\frac{\mli{\tilde{t}}{d}}{\tilde{t}}\right)^\eta\right)\, 
\end{equation}
where $F$ is the hypergeometric function. As time tends to infinity
$\mathcal{L}(\tilde{t})$ tends to zero. The angular momentum
extracted from the BHB by such a representative star in its
dependence on time is $\mathcal{L}(\tilde{t}) -
\mathcal{L}(\mli{\tilde{t}}{d})$.
The integration of Eq.~(\ref{eq_dlzdt_fit}) starts at
$\mli{\tilde{t}}{d}$ to take into account the delay till a star is
registered as ejected. The average amount of angular momentum the two
black holes can lose to the mass of the representative star is 
\begin{eqnarray}
\label{eq_lmax_fit}
\mli{\tilde{L}}{\infty} & = & \mathcal{L}(\tilde{t}\to\infty) -
\mathcal{L}(\mli{\tilde{t}}{d}) =
- \mathcal{L}(\mli{\tilde{t}}{d})\nonumber\\
& = &
\frac{2 k}{\sqrt{\mli{\tilde{t}}{d}}}\frac{\Gamma(1+\frac{1}{2\eta })
\Gamma(1+\kappa)}{\Gamma(1 + \frac{1}{2\eta } + \kappa)}\,,
\end{eqnarray}
where $\Gamma$ denotes the Gamma function.
To compute the fit-parameters we use an implementation of the
non-linear least-squares Marquardt-Levenberg algorithm and obtain
\begin{equation}
\label{eq_fit_param}
k = 1.77,\hspace{0.5cm}\eta  = 1.18 \hspace{0.5cm}\mbox{and}
\hspace{0.5cm} \kappa  = 9.84\,.
\end{equation}
These parameters provide a sufficiently good approximation to the data
as can be seen in Fig.~\ref{f_dx2_dlzdt_q10}, where the fit is
displayed by the dashed curve (fit 1). Substituting the parameter
values to Eq.~(\ref{eq_lmax_fit}) results in 
\begin{equation}
\label{eq_lmax_value}
\mli{\tilde{L}}{\infty} = 0.258\,.
\end{equation}
This is about a fourth of the angular momentum of a star
circling at $1\,{\rm pc}$ distance around a point mass corresponding
to that of the binary. 
Now the number of ejected stars required to carry away all the
binary's angular momentum is obtained in an analoguous way as in
Eq.~(\ref{eq_Neject_stat}):
\begin{equation}
\label{eq_Neject_dldt}
\mli{N}{ej} = \frac{\tilde{L}_{\BHBidx}}{\mli{\tilde{L}}{\infty}} =
\frac{1}{1+q}\frac{10^8 M_8}{\mli{\tilde{L}}{\infty}}\,.
\end{equation}
Comparison of both expressions shows that the average
angular momentum gained by a single star, $\Delta \tilde{L}_z$,
has been replaced by the angular momentum which is extracted on
average by one solar mass till the end of the simulation,
$\mli{\tilde{L}}{\infty}$. For the same ratio $\mli{N}{bn}/\mli{N}{ej}$
as above we now obtain the numbers
\begin{equation}
\label{eq_nstarsmerge}
\mli{N}{ej} \approx 3.47\times10^{7} \hspace{0.5cm}\mbox{and}
\hspace{0.5cm} \mli{N}{total} \approx 9.33\times10^{7}\,.
\end{equation}
They are in very good agreement with the numbers we got above by
statistical arguments only, confirming our finding that a comparable
amount of mass in stars as in the black holes is required to remove
all the binary's angular momentum.

Comparing both the loss-rates of angular momentum due to gravitational
radiation and ejection of stars allows us to compute the time and
distance where gravitational radiation eventually becomes the dominant
physical process for the 
merging of the black holes. As before we assume the number of stars to be
sufficient so that all angular momentum of the binary is lost as
time tends to infinity. Hence the angular momentum lost by the binary
as a function of time is
\begin{eqnarray}
\tilde{L}_{\rm lost}(\tilde{t}) & = & \mli{N}{ej} (\mathcal{L}(\tilde{t}) -
\mathcal{L}(\mli{\tilde{t}}{d})) =
\frac{\tilde{L}_{\BHBidx}}{\mli{\tilde{L}}{\infty}} (\mathcal{L}(\tilde{t}) +
\mli{\tilde{L}}{\infty})\nonumber\\
& = & \tilde{L}_{\BHBidx}(\mli{\tilde{t}}{d})
\left(\frac{\mathcal{L}(\tilde{t})}{\mli{\tilde{L}}{\infty}} + 
1\right)\,.
\end{eqnarray}
For $\tilde{t}<\mli{\tilde{t}}{d}$ we assume all quantities to be
constant. The angular momentum left to the two black holes is simply its
initial value diminished by $\tilde{L}_{\rm lost}$, and because of its
dependency on the distance $\tilde{a}(\tilde{t})$
(Eq.~(\ref{eq_lbin_stat})) we get the relation 
\begin{eqnarray}
\label{eq_lbinot}
\tilde{L}_{\BHBidx}(\tilde{t}) & = &
\tilde{L}_{\BHBidx}(\mli{\tilde{t}}{d}) \sqrt{\tilde{a}(\tilde{t})} 
\nonumber\\ 
& = & \tilde{L}_{\BHBidx}(\mli{\tilde{t}}{d}) - \tilde{L}_{\rm
lost}(\tilde{t}) = -
\frac{\tilde{L}_{\BHBidx}(\mli{\tilde{t}}{d})}{\mli{\tilde{L}}{\infty}} 
\mathcal{L}(\tilde{t})\,.
\end{eqnarray}
The separation of the black holes which we kept fixed to $1\,{\rm pc}$
($\tilde{a} = 1$) during the simulation, is now treated as
function of time 
with an initial value $\tilde{a}(\tilde{t}\le\mli{\tilde{t}}{d}) = 1$.
Solving Eq.~(\ref{eq_lbinot}) for the separation, $\tilde{a}(\tilde{t}) =
(\mathcal{L}(\tilde{t})/\mli{\tilde{L}}{\infty})^2$, the shrinking
rate of the binary due to ejection of stars can be written as
\begin{equation}
\label{eq_dadt_se}
\frac{d\tilde{a}}{d\tilde{t}} = 2\,
\frac{\mathcal{L}(\tilde{t})}{\mli{\tilde{L}}{\infty}}
\frac{d\mathcal{L}(\tilde{t})}{d\tilde{t}}\,. 
\end{equation}

This rate has now to be compared with the shrinking rate caused by the
emission of gravitational radiation. 
Using $\mli{M}{total} = M_1 (1+q)/q$ and $\mu = M_1/(1+q)$, the total
and reduced mass respectively, the energy of the binary reads
\begin{equation}
\label{eq_ebin}
\tilde{E}_{\BHBidx} = - \frac{1}{E_0} \frac{ G\mu \mli{M}{total}}{2r_0}
\frac{1}{\tilde{a}} = -\frac{1}{2\tilde{a}}\,,
\end{equation}
$\tilde{E}_{\BHBidx}$ being dimensionless and $E_0 = G\mu
\mli{M}{total}/r_0$
the normalization constant. The energy loss of the
binary due to gravitational radiation is \cite{misner73}
\begin{eqnarray}
\label{eq_debindt}
\frac{d\tilde{E}_{\BHBidx}}{d\tilde{t}} & = & -\frac{32}{5}\,\frac{G
\mu^2 r_0^4
\Omega_0^5}{E_0 c^5}\,\frac{1}{\tilde{a}^{5}}\nonumber\\
& = &-\frac{32}{5}\left(\frac{G M_1}{r_0}\right)^{\frac{5}{2}}
\left(\frac{q+1}{q^3}\right)^{\frac{1}{2}} \frac{1}{c^5 \tilde{a}^5}\,.
\end{eqnarray}
Forming the time derivative of Eq.~(\ref{eq_ebin}) and with the
help of Eq.~(\ref{eq_debindt}) the shrinking rate of the binary
results in
\begin{eqnarray}
\label{eq_dadt_gr}
\frac{d\tilde{a}}{d\tilde{t}} & = & - \frac{64}{5 c^5} \left( \frac{GM_1}{r_0}
\right)^{\frac{5}{2}}
\left(\frac{q+1}{q^3}\right)^{\frac{1}{2}} \tilde{a}^{-3}\nonumber\\
& \equiv &
-\frac{1}{\mli{\tilde{t}}{sh}} \left(\frac{q+1}{q^3}\right)^{\frac{1}{2}}
 \frac{1}{\tilde{a}^3}\,.
\end{eqnarray}
For convenience we introduced the dimensionless `shrinking time'
$\mli{\tilde{t}}{sh}$ in the last step. Multiplied with $t_0$ this reference
time scales as
\begin{equation}
\mli{\tilde{t}}{sh} t_0 = 2.33\times 10^{15}\,
M_8^{-3} r_{\rm pc}^{4}
\left(\frac{q}{1+q}\right)^{\frac{1}{2}} \,{\rm yr}
\end{equation}
Integrating Eq.~(\ref{eq_dadt_gr}) from
$\tilde{t}=0$, when the distance is $\mli{\tilde{a}}{0}$, to
$\tilde{t}$ yields
\begin{equation}
\label{eq_aotgr}
\frac{\tilde{a}(\tilde{t})}{\mli{\tilde{a}}{0}} = \left(1 -
\frac{4}{\mli{\tilde{a}}{0}^4}
\left(\frac{q+1}{q^3}\right)^{\frac{1}{2}} \frac{\tilde{t}}{
\mli{\tilde{t}}{sh}}\right)^{\frac{1}{4}}\,.
\end{equation}
We define $\mli{\tilde{a}}{tr}$ as the distance of the transition
where emission of gravitational radiation becomes more effective than
ejection of stars in extracting angular momentum from the binary.
If we now assume that the integration of the shrinking rate
(\ref{eq_dadt_gr}) starts when the BHs happen
to be just this distance apart ($\tilde{a} = \mli{\tilde{a}}{tr}$),
we get
\begin{equation}
\label{eq_amerge}
\frac{\tilde{a}(\tilde{t})}{\mli{\tilde{a}}{tr}} = \left(1 -
\frac{\tilde{t}}{\mli{\tilde{t}}{mg}} \right)^{\frac{1}{4}}\,.
\end{equation}
In this expression
\begin{equation}
\label{eq_tmerge}
\mli{\tilde{t}}{mg} = \frac{\mli{\tilde{a}}{tr}^4}{4} 
\left(\frac{q^3}{q+1}\right)^{\frac{1}{2}}\mli{\tilde{t}}{sh}
\end{equation}
is the time the black holes
still need to merge completely once gravitational radiation starts at
$\tilde{a} = \mli{\tilde{a}}{tr}$ to govern the merging process.

\begin{figure}
\vspace{0.7cm}
\resizebox{85mm}{!}{\input{bhb_fig13}}
\caption[]{The shrinking rates due to ejection of stars (solid line)
and emission of gravitational radiation (dashed line) are shown to
have very different slopes. Therefore the region of the transition is
quite narrow. The separation $\tilde{a}$ is normalized to $1\,{\rm pc}$.}
\label{f_dx2_shrink_q10}
\end{figure}

If $\mli{\tilde{a}}{0}$ in Eq.~(\ref{eq_aotgr}) is equal to $1$,
i.e. the separation where we started the simulation and where the BHs are
$1\,{\rm pc}$ apart from each other, the time needed to reduce
$\tilde{a}$ by some large fraction because of emission of gravitational
radiation only is of order of $\mli{\tilde{t}}{sh} t_0 \approx
1.56\times 10^{12}\,t_0 \approx 2\times 10^{15}\,{\rm yr}$. Thus in the
beginning, at distances as large as $1\,{\rm pc}$, the
hardening of the binary caused by emission of gravitational radiation is
completely negligible and proceeds on time-scales orders of magnitudes
higher 
than shrinking due to ejection of stars. To compare these two
rates of hardening we have to rewrite Eq.~(\ref{eq_dadt_se}) in its
dependence on distance $\tilde{a}$ instead of
$\tilde{t}$. Unfortunately, because of
its complicated form, we can not build the inverse function of
$\tilde{a}(\tilde{t})$. However, to proceed we first have to further
simplify our fit function. This is done by resetting the parameter
$\kappa$ to zero, so that the cut-off funtion in
Eq.~(\ref{eq_dlzdt_fit}) vanishes and we are left with
a pure power law only. In order not to have too large differences
between this fit and the data for small times and to take account of
the initial increase we shift the parameter $\mli{\tilde{t}}{d}$ from
$20$ to $50$. Hence the system function (\ref{eq_intdldt}) becomes
\begin{equation}
\label{eq_intdldt_pot}
\mathcal{L}(\tilde{t}) = \frac{k}{1-\alpha} \tilde{t}^{1-\alpha}\,.
\end{equation}
To maintain the number of ejected stars needed for a final
merger of the black holes we recalculate the parameter $k$ using
Eq.~(\ref{eq_lmax_fit}). Thus $\mli{\tilde{L}}{\infty} = 
k\mli{\tilde{t}}{d}^{1-\alpha}/(\alpha-1)$ and therefore we get $k =
\mli{\tilde{L}}{\infty}(\alpha -1)\,\mli{\tilde{t}}{d}^{\alpha-1} \approx
0.91$. Since we are just interested in an estimate
of the timescales of the merger, these parameters still provide a
sufficiently good fit to the data as can be 
seen in Fig.~\ref{f_dx2_dlzdt_q10}, where the power law is shown
as dotted line an denoted as `fit 2'. Now the distance as function of time, 
$\tilde{a} = (\mli{\tilde{t}}{d}/\tilde{t})^{2(\alpha -1)}$,
can be solved for $\tilde{t}$. With $\alpha=3/2$ the shrinking-rate
$da^{-1}/dt$ is constant, as is found by \citeN{milos01} in N-body
calculations. Substituting the time as function of the separation
together with Eq.~(\ref{eq_intdldt_pot}) in the the shrinking rate due
to star ejection
(\ref{eq_dadt_se}) yields
\begin{equation}
\label{eq_dadt_se_pot}
\frac{d\tilde{a}}{d\tilde{t}} = - \frac{2(\alpha
-1)}{\mli{\tilde{t}}{d}} \tilde{a}^{\frac{2\alpha -1}{2(\alpha -1)}}\,.
\end{equation}
Equating this expression with the hardening rate caused by emission of
gravitational waves (\ref{eq_dadt_gr}) we
can solve for the distance of the transition $\mli{\tilde{a}}{tr}$
where gravitational radiation takes over the process of shrinking,
\begin{equation}
\label{eq_atr_form}
\mli{\tilde{a}}{tr} = \left[\frac{1}{2(\alpha - 1)}
\left(\frac{q+1}{q^3}\right)^{\frac{1}{2}}
\frac{\mli{\tilde{t}}{d}}{\mli{\tilde{t}}{sh}}
\right]^{\frac{2(\alpha-1)}{8\alpha -7}}\,.
\end{equation}
The merging for $\tilde{a} > \mli{\tilde{a}}{tr}$ is dominated by the
ejection of
stars. Since both shrinking rates have very different slopes (see
Fig.~\ref{f_dx2_shrink_q10}) $\mli{\tilde{a}}{tr}$ does not depend
sensitively on the
parameters we chose for the fit and the transition occurs quite
suddenly over a small range in $\tilde{a}$. Substituting
$\mli{\tilde{a}}{tr}$ to $\tilde{a} =
(\mli{\tilde{t}}{d}/\tilde{t})^{2(\alpha -1)}$ we can
solve it for the time $\mli{\tilde{t}}{tr}$ needed till the distance has
shrunk from its initial value $\tilde{a} = 1$ to $\tilde{a} =
\mli{\tilde{a}}{tr}$ due to ejection of stars. For the
parameters we used, $\alpha = 3/2$ and $\mli{\tilde{t}}{d} = 50$, we
finally obtain for $q=10$
\begin{eqnarray}
\label{eq_atr_num}
\mli{\tilde{a}}{tr} &\approx & \frac{1}{197} \;\;\widehat{=}\;\; 5.1\times
10^{-3}\,{\rm pc}\,,\nonumber\\
\mli{\tilde{t}}{tr} & \approx & 9855 \;\;\widehat{=}\;\;
1.40\times10^{7}\,{\rm yr}\,.
\end{eqnarray}
Hence after about $10^{7}\,{\rm yr}$ the distance of the two
black holes has shrunk to $\sim 1/200$ of its initial value $1\,{\rm
pc}$. \citeN{merritt00} obtained for
$\mli{\tilde{a}}{tr}$ about $1/100$ of the distance where the binary
becomes hard, which translates in physical units to about $1/100\,{\rm
pc}$ according to \citeN{milos01}. This is in good agreement with the
distance of transition for the mass-ratio $q=1$, see
Table~\ref{tb_stat}.

Only the fraction $\sim 1/14$ of the initial angular momentum of the
binary is left at this distance. With $\alpha = 3/2$ for the powerlaw
index the relation between the distance and time of the transition is 
$\mli{\tilde{a}}{tr} = \mli{\tilde{t}}{d}/\mli{\tilde{t}}{tr}$, and
Eq.~(\ref{eq_atr_form}) simplifies to
\[
\mli{\tilde{a}}{tr} = \left[\left(\frac{q+1}{q^3}\right)^{\frac{1}{2}}
\frac{\mli{\tilde{t}}{d}}{\mli{\tilde{t}}{sh}}
\right]^{\frac{1}{5}}\,.
\]
Thus the ratio of the time ranges when star ejection and emission of
gravitational waves dominate the hardening is independent of the
massratio:
\begin{equation}
\label{eq_timerat}
\frac{\mli{\tilde{t}}{tr}}{\mli{\tilde{t}}{mg}}=4\,.
\end{equation}
For the time when gravitational radiation dominates we then have
\begin{equation}
\label{eq_tmerge_n}
\mli{\tilde{t}}{mg} \approx 2464 \;\;\widehat{=}\;\; 3.51\times 10^{6}
\,{\rm yr}\,,
\end{equation}
showing that the merging process is dominated most of the time by
ejection of stars.
\citeN{quinlan96} obtains for the timescales needed for
merging once the binary has become hard ($\tilde{a}\approx
\mli{\tilde{r}}{cusp}$), about $4\times 10^7\,{\rm yr}$. This is very
similar to our result, which is supposed to be a lower limit since in
the simulation the shrinking of the distance between the black holes has
not been taken into account.
This time-scale on which the BHs merge is of the same order as the
one we obtained for the torus to emerge, see
Fig.~\ref{f_cont_tser}. Thus, if a binary with a mass-ratio $10$
merges due to ejection of stars from a surrounding cluster, a
torus-shaped distribution forms on a similar time-scale.

The energy of the binary scales with its separation as
$\tilde{a}^{-1}$. If the distances of the stars in the cluster
are increased or decreased by the same factor $\chi$ as the
binary's separation, the energy of a star also scales with this
factor to the power of $-1$, i.e. $\tilde{E}_{\BHBidx}$
and $\tilde{E}_\ast \propto \chi^{-1}$. Thus the basic
results will be unchanged as long as the binary remains ``hard'', but
the time-scales of the merging due to ejection of stars scale as
$\chi^{3/2}$.

The final merger of two supermassive BHs should lead to one of
the most energetic events in the universe. As a result of such a
merger, the spin and its direction are expected to change
\cite{wilson95}. Consequently also the direction of the jet, which is
aligned with the BH's spin, will jump into the new direction of the
spin of the merged BH. This is a possible explanation for the observed
X-shaped radio galaxies. This class of objects exhibits four jets, one
pair of which shows ``young'' synchrotron emission and the other one
``old'' emission, such as seen in B2 0828+23 \cite{parma85,rottmann98}.
The spectral aging of the secondary lobes in this object suggests that
a merger has happened about $6\times 10^7\,{\rm yr}$ ago
(priv. communication with H. Rottmann). This timescale is very close
to the one we have found here for the merger of two supermassive BHs
and thus supports the idea of mergers inducing a jump of the jet.

\subsection{Final distribution and energy of ejected stars}

\begin{figure}
\resizebox{85mm}{!}{\includegraphics[80, 260][490, 600]{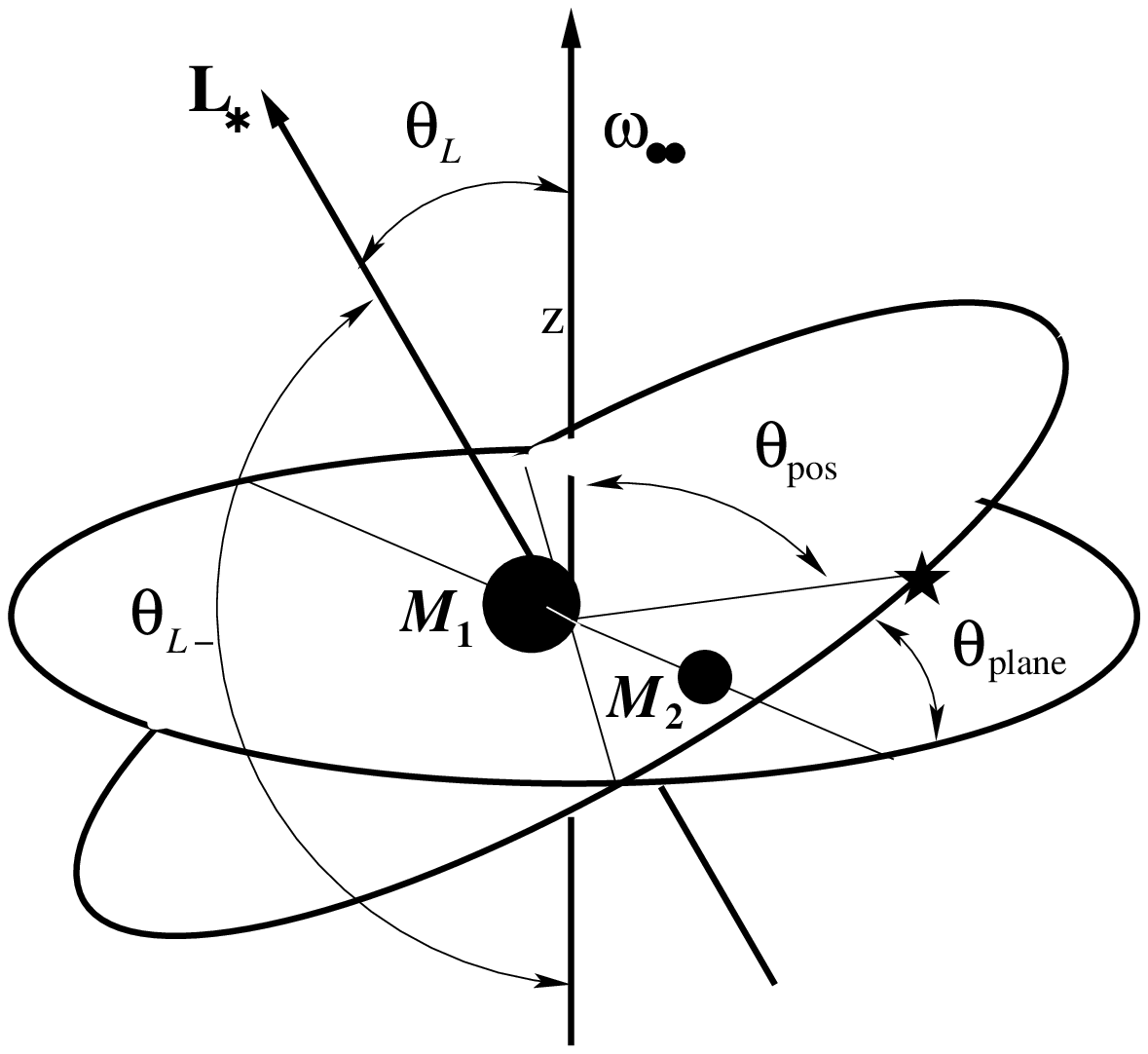}}%
\caption[]{This sketch illustrates the meaning of the different
angles. The equatorial plane is defined by the black holes rotating around
each other, with the angular frequency vector
$\tilde{\VEC{\omega}}_{\BHBidx}$
pointing along the $z$-axis. $M_1$ is displayed a little off-center
since the coordinate system is centered in the center of mass
frame. Tilted to this plane with an angle
$\mli{\theta}{plane}$ is the orbital plane of a star, whose angular
momentum $\tilde{\VEC{L}}_{\ast}$ encloses the angles $\theta_L$ with
$\tilde{\VEC{\omega}}_{\BHBidx}$ (the positive $z$-axis) and $\theta_{L_{-}}$
with the negative $z$-axis. The line from the center to the current
position of the star has the angle $\mli{\theta}{pos}$ with
$\tilde{\VEC{\omega}}_{\BHBidx}$.}
\label{f_angle_scetch}
\end{figure}

As we have shown in Sect.~\ref{sec_dynst},
both populations, the ejected and bound stars, can be clearly
distinguished by their angular momentum or their pericenter.
To learn more about their angle distribution is the task of
this section. For this purpose we introduce some angles, as is
illustrated in Fig.~\ref{f_angle_scetch}.
We define $\theta_{L}$ as the angle between
the positive $z$-axis (the rotation axis
$\tilde{\VEC{\omega}}_{\BHBidx}$) and
the angular momentum $\tilde{\VEC{L}}_\ast$ of a
star. The angle between $\tilde{\VEC{L}}_\ast$ and the negative
$z$-axis is denoted by $\theta_{L_{-}}$, so that $\theta_{L_{-}} =
180^\circ - \theta_{L}$. To the angle between the plane in which the star
rotates and the equatorial plane we refer as $\theta_{\rm plane}$.

The angle which is enclosed by the line connecting the origin with the
current position of the star and the $z$-axis is denoted by
$\theta_{\rm pos}$. The mean and the dispersion of this angle
are very similar at $\tilde{t}=0$ for all three populations (EP, BP, TP),
namely $\mean{\mli{\theta}{pos}}\approx \pi/2$ and $\Delta
\mli{\theta}{pos} \approx \sqrt{(\pi^2 -8)/4}$. These are
just the values we expect for a spherically symmetric distribution, as
the total population at $\tilde{t}=0$. Thus both populations, EP and
BP, are also spherically symmetric distributed in the beginning and
none of them shows any preferences as to certain polar angles.
However, from Table~\ref{tb_stat} we see that the mean angular momentum
$\mean{\tilde{L}}$ of the ejected stars increased by just a factor of $1.1$,
while $\mean{\tilde{L}_z}$ is by more than a factor $4.5$ bigger at
the end of the simulation. Thus we expect the geometry of
the distribution of the ejected population to be flattened and not
spherically symmetric at the end of the simulation.

\begin{figure}
\vspace{0.7cm}
\resizebox{85mm}{!}{\input{bhb_fig15}}
\caption[]{
The density of the ejected stars at $\tilde{t}=0$
(spherically symmetric distributed) and after
their ejection ($\mli{\tilde{t}}{ej}$) as function of
their polar angle is displayed.
After being ejected the cluster is more concentrated to the
equatorial plane $\mli{\theta}{pos}= 90^\circ$.
}
\label{f_dx2_upej_q10}
\end{figure}

In Fig.~\ref{f_dx2_upej_q10} we plotted the density of the stars in
dependence of
the cosine of $\theta_{\rm pos}$. As stated above we see within the
fluctuation of the data a uniform
distribution for $\tilde{t}=0$ (dashed line), according to a spherical
symmetry. But after the stars have been ejected the density has a
maximum at $\cos(\theta_{\rm pos}) = 0$ (solid line). Thus they
are concentrated towards the equatorial plane of the binary.
The same conclusion is drawn from Fig.~\ref{f_dx2_uoej_q10}.
It displays the density as a function
of the orientation of the orbital angular momentum of the stars,
$\cos(\theta_L) = \tilde{L}_z/\tilde{L}$. At 
$\tilde{t}=0$ (dashed line) the density is almost constant in
$\cos(\theta_L)$, thus there
is no special orientation prefered for the orbital plane of the
stars. The density is slightly increasing with $\cos(\theta_L)$,
showing that the ejected stars initially are weakly corotating on
average. After the ejection this curve is steeply increasing, with
only few stars having $\tilde{L}_z < 0$ (counter-rotating) and the
distribution having the maximum at $\cos(\theta_L) =
\tilde{L}_z/\tilde{L} = 1$. This means that the angular
momentum from most stars points along the rotation axis and thus their
orbital plane is aligned with the equatorial plane. Therefore the
density of the stars has a maximum in the equatorial plane and a
minimum at the poles, as seen before in Fig.~\ref{f_dx2_upej_q10}.
The angle $\theta_{\rm plane}$ between orbital and equatorial plane
(see Fig.~\ref{f_angle_scetch}) is the same for clockwise ($L_z <0$)
and counter-clockwise ($\tilde{L}_z >0$) rotation and is equal to the minimum
of $\theta_L$ and $\theta_{L_{-}}$. Gaining now angular momentum in
the $z$-component leads to a decreasing angle $\theta_L$ and
increasing $\theta_{L_{-}}$. A formerly counter-rotating star with
$\cos(\theta_L) = -1$ could now cross the poles ($\cos(\theta_L) = 0$)
and therefore
increase the density there. But Fig.~\ref{f_dx2_uoej_q10} shows a
decrease in $\tilde{\rho}$ up to $\cos(\theta_L) \approx 0.2$ and an
increase
towards smaller angles. Hence the gains in $\tilde{L}_z$ are big enough in
order to deplete the polar regions and to tilt the orbital planes
of the stars after their ejection to the equatorial plane. Moving in
its plane, $\theta_{\rm pos}$ of
a star changes in the range $90^\circ -\theta_{\rm plane} \le
\theta_{\rm pos} \le 90^\circ +\theta_{\rm plane}$, consequently
resulting in a maximum of the density at $\theta_{\rm pos} =
90^\circ$ (Fig.~\ref{f_dx2_upej_q10}).
So the EP changes its symmetry from spherical to cylindrical after the
interaction with the binary.

\begin{figure}
\vspace{0.7cm}
\resizebox{85mm}{!}{\input{bhb_fig16}}
\caption[]{The orientation of the angular momenta of the EP at
$\tilde{t}=0$ and after the ejection at $\mli{\tilde{t}}{ej}$ is
displayed. They show a strong concentration towards large positive
values, so that the orbital plane is close to the equatorial plane.}
\label{f_dx2_uoej_q10}
\end{figure}

The stars which gain the most energy also gain the most angular
momentum (see eq.~(\ref{eq_djacobi})).
In Fig.~\ref{f_dx2_eoutm_q10} we plot the energy versus $\cos
(\theta_L) = \tilde{L}_z/\tilde{L}$ after the ejection of stars. The
figure shows a
strong correlation between the energy and $\tilde{L}_z/\tilde{L}$,
confirming the
prediction of Eq.~(\ref{eq_djacobi}): The orbital planes of the
stars with the highest (kinetic) energies are more concentrated
towards the equatorial plane of the binary. A similar correlation is
found when the eccentricity $\epsilon =
\sqrt{1+2\tilde{E}\tilde{L}^2}$ is plotted
versus the orientation of the orbits, showing that the most eccentric
trajectories ($\epsilon \approx 5.5$) are most
aligned with the plane in which the black holes rotate.

The strong changes in $\tilde{L}$ and $\tilde{E}$ of the ejected stars
compared to the
almost constant values for the BP show that the BHB is
mainly influenced by the ejected stars. Provided that initially there
are enough stars, the EP can extract sufficient angular momentum
to allow the black holes to merge, as shown in Sect.~\ref{s_ang_loss}.

\begin{figure}
\vspace{0.7cm}
\resizebox{85mm}{!}{\input{bhb_fig17}}
\caption[]{The energy at $\mli{\tilde{t}}{ej}$ of the EP is plotted
versus the orientation of the angular momentum. The more
energetic the stars are, the more they are concentrated to the
equatorial plane.}
\label{f_dx2_eoutm_q10}
\end{figure}


\subsection{Does the torus survive the past-merger time?}
Finally we want to give in this section a crude estimate of the
stability of the remaining torus-like distribution of the bound stars,
once the BHs have merged. In a first approximation we can use the
formulae from Sect.~\ref{sc_collsys} for collisionless systems.
With the following values, $\mli{N}{bn} = \mli{N}{tot} / (1 +
\mli{N}{ej} / \mli{N}{bn}) = 6.4\times 10^7$, $R=50\,{\rm pc}$,
$\mli{b}{min} = 2\,{\rm pc}$ and the velocity set to the keplerian in
a distance of $5\,{\rm pc}$, $v=308\,{\rm km/s}$, we obtain for the
ralaxation time
\[
\mli{t}{relax} = 4\times 10^{10}\,{\rm yr}\,.
\]
In order to get a more precise time-scale we can
make use of the diffusion coefficients $D$, which are a measure for the
rate at which stars diffuse through phase space as a result of
encounters, see \citeN{binney87}. In the local approximation all encounters
are assumed to be local, i.e. the impact parameter $b$ is assumed to
be much less than the system size $R$. Employing now the
Fokker-Planck approximation and assuming a spherical symmetry with a
Maxwellian velocity distribution \citeN{binney87} obtain for the
relaxation time, defined as $v^2 / D(\Delta v_\parallel^2)$,
\[
\mli{t}{relax} = 0.34\frac{\sigma^3}{G^2 m \rho \ln\Lambda}\,,
\]
where $\Lambda = \mli{b}{max} \mli{v}{typ}^2/2Gm$. Here $\mli{v}{typ}$ is
the typical velocity of stars of the system and set to the keplerian
velocity in $5\,{\rm pc}$ distance, $308\,{\rm km/s}$. $\mli{b}{max}$
is the maximum impact parameter and assumed to be $50\,{\rm pc}$ since
here the index of the density powerlaw changes from $2$ to $4$ and the
density drops quickly with increasing radius
(Fig.~\ref{f_dx2_rho_q10}). For a mean density of $10^3 \, M_\odot
{\rm pc^{-3}}$ around the maximum value of $3.4 \times 10^3\,M_\odot{\rm
pc^{-3}}$ at
$\sim 3\,{\rm pc}$ and $\sigma = \mli{v}{typ}/\sqrt{3}$ we
then obtain
\[
\mli{t}{relax} = 5\times 10^{12}\,{\rm yr}
\left(\frac{\sigma}{178\,{\rm km/s}}
\right)^3 \left(\frac{10^3\,M_\odot{\rm pc^{-3}}}{\rho}\right)\,.
\]

This time scale can now be compared to the time $\mli{t}{fric}$ which a
star needs to spiral from the outer edge to the inner edge of the
torus due to dynamical friction. In fact, the diffusion coefficient is
identical to the deceleration of a test star owing to dynamical friction
by the surrounding field stars \cite{binney87},
\[
\frac{d\VEC{v}}{dt} = -\frac{8\pi \ln\Lambda G^2 m^2 n}{v^3}
\left[{\rm erf}(X) -\frac{2X}{\sqrt{\pi}} e^{-X^2}\right]\VEC{v}\,.
\]
Again the velocity is assumed to be distributed according to a
Maxwellian and $X\equiv v/\sqrt{2}\sigma$. The total number density of
the stars, with each having the mass $m$, is denoted by $n$, and ${\rm erf}$
is the error-function. The decelerating force is acting tangentially and
thus causes the test star, also of mass $m$, to lose angular momentum
per unit mass $L$ at a rate $dL/dt = r dv/dt$. Following
\citeN{binney87} the star continues to orbit at the circular speed
$v_c$ as it spirals to the center and its angular momentum per unit
mass at radius $r$ is at all times $L=r v_c$. Equating the
time derivative with the loss rate of $L$ due to dynamical friction
given above yields
$\frac{dr}{r}=\frac{1}{v_c}\frac{dv}{dt}dt$. Integrating this equation
from a radius $10\,{\rm pc}$, where the star starts to spiral inwards,
to the inner edge at $2\,{\rm pc}$ and solving for the time yields
\[
\mli{t}{fric} = \frac{v_c^3}{8\pi\ln\Lambda G^2 m^2
\left[{\rm erf}(X) - \frac{2X}{\sqrt{\pi}} e^{-X^2}\right]}
\int_2^{10} \frac{1}{\tilde{r} n}d\tilde{r}\,.
\]
The number density $n(\tilde{r})$ is approximated by the function
\[
n(\tilde{r}) = k \left(\frac{1}{1 +
\left(\frac{\tilde{r}_a}{\tilde{r}}\right)^p}
\frac{1}{1 + \big(\frac{\tilde{r}}{\tilde{r}_b}\big)^q}
\frac{1}{\tilde{r}^2} +
\frac{1}{1 + \left(\frac{\tilde{r}_b}{\tilde{r}}\right)^q}
\frac{\tilde{r}_b^2}{\tilde{r}^4} \right)\,,
\]
which actually is a powerlaw with index $-2$ in the range
$\tilde{r}_a<\tilde{r}<\tilde{r}_b$
and index $-4$ for $\tilde{r}>\tilde{r}_b$, while it vanishes for
$\tilde{r}\ll\tilde{r}_a$.

We chose for the
parameters $\tilde{r}_a = 2.75$, $\tilde{r}_b = 50$ and $k=5.3\times
10^4\,{\rm pc^{-3}}$ so
that the volume integration from $\tilde{r}=0$ to infinity gives the right
number of bound stars. Substituted in the expression for
$\mli{t}{fric}$ and using for $\Lambda$, $v_c=\mli{v}{typ}$ and
$\sigma$ the same
values as above we obtain for the time a star needs to spiral from
$10\,{\rm pc}$ distance to the inner edge of the torus at
$2\,{\rm pc}$
\[
\mli{t}{fric} = 6.2\times 10^{12}\,{\rm yr}\,,
\]
which is in good agreement with $\mli{t}{relax}$ above. Thus we can
conclude that the torus-like structure is stable.

\section{Dependency on the mass-ratio}

\begin{figure}[t]
\vspace{0.7cm}
\resizebox{85mm}{!}{\input{bhb_fig18}}
\caption[]{
At the end of the simulation the more stars are ejected the bigger
$M_2$ is. While for $q=1$ there is almost no cusp left in the center,
the region close to the orbit of $M_2$ at $\tilde{r}=1$ is not much
depleted for $q=100$. In the range $10\lesssim \tilde{r}\lesssim50$
all density distributions follow the initial powerlaw with index
$-2$ while at bigger distances $\tilde{\rho}\propto\tilde{r}^{-4}$
independent of $q$. The initial distribution is for all mass-ratios
the same and represented by the solid line.
}
\label{f_dx2_rho_qall}
\end{figure}

Finally we will discuss the influence of the mass-ratio of the
black holes on the ejected and bound star populations.
In altering the mass-ratio $q = M_1/M_2$
we always keep the mass of the primary black hole fixed to $10^8 M_\odot$.
For the mass of the secondary we chose $M_2 = 10^8$ and
$10^6$ solar masses corresponding to the ratios $q=1, 100$
respectively. Hence also the total mass of the binary $M_{\BHBidx} =
M_1 + M_2$ is changed.

\begin{figure*}[t]
\hspace{1mm}%
\resizebox{86mm}{!}{\includegraphics[bb = 10 8 490
  502]{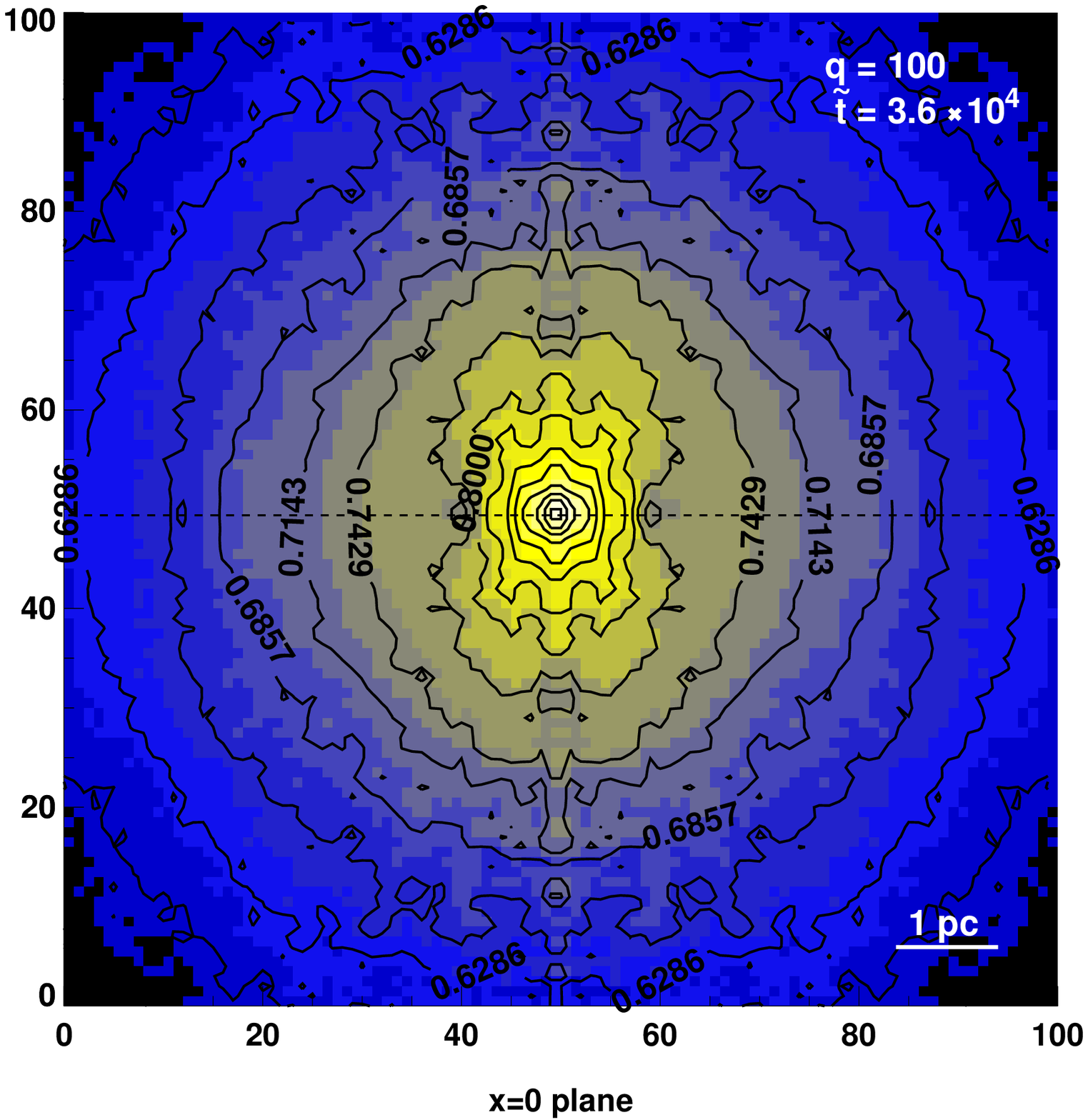}}%
\hspace{8.3mm}%
\resizebox{86mm}{!}{\includegraphics[bb = 10 8 490
  502]{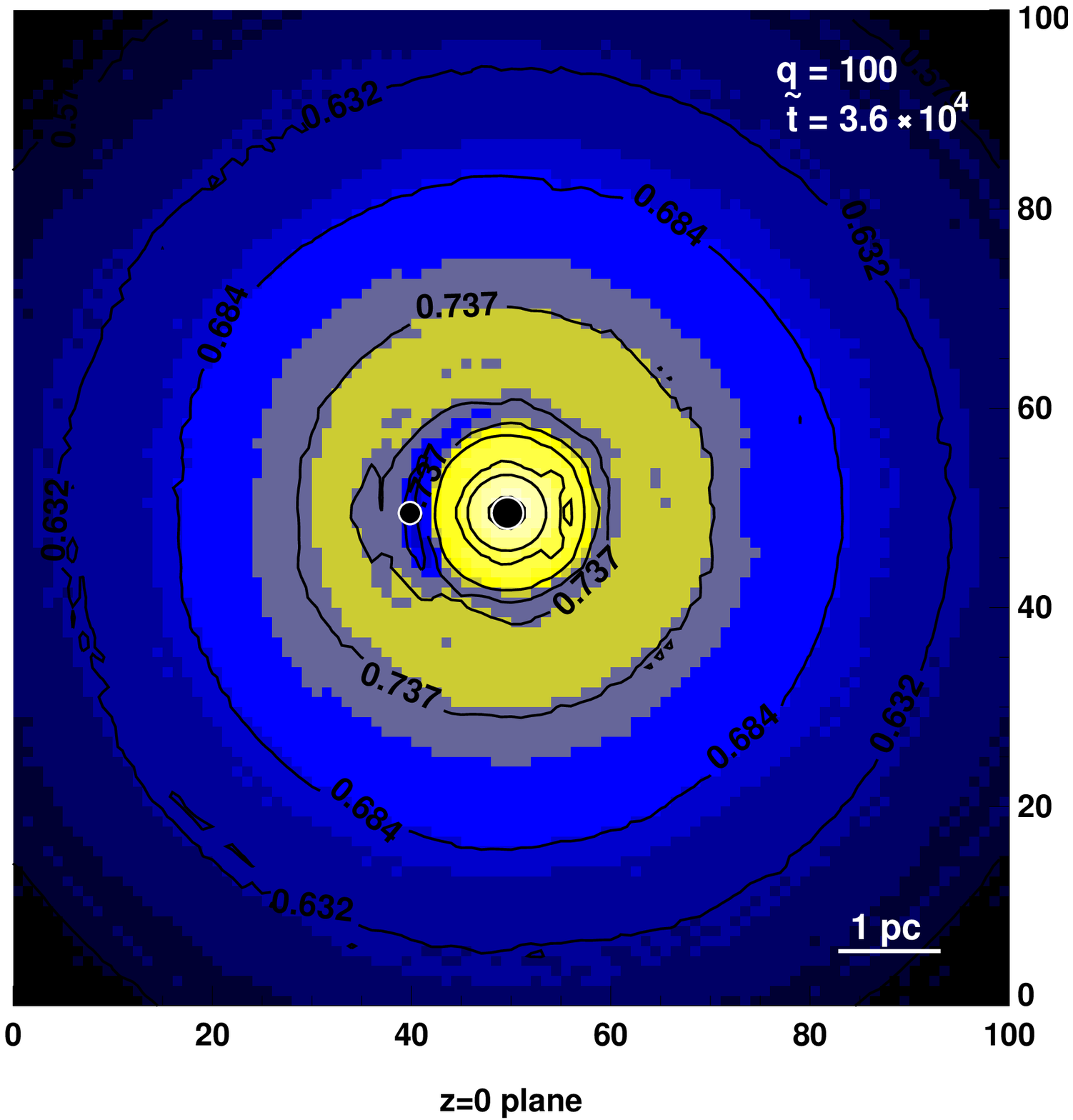}}%
\caption[]{
Same as Fig.~\ref{f_cont_tser} for the final density distributions of
the mass-ratios $q=1$ (top row) and $100$ (bottom row). For $q=1$
there is no central cusp left and a pronounced torus with
sharp defined edges emerges after $3.8\times 10^{7}\,{\rm yr}$. 
There is no torus seen in case of $q=100$ after $5.4\times
10^{7}\,{\rm yr}$, only close to the orbit of the secondary BH the
sensity is diminished.
}
\label{f_cont_qser}
\end{figure*}

As $M_2$ increases and $q$ decreases the sphere of influence of
the secondary BH expands and consequently the inner regions become
more unstable. Thus a bigger fraction of stars is ejected and the
density cusp bound to the primary BH only becomes less dense, being made
up by $0.4\%$ and $2.8\%$ for $q=1$ and $100$ respectively.
Between $\tilde{r} = 10 \,r_{\rm pc}$ and $50 \,r_{\rm pc}$ the
density is not much affected
by variations in $q$ and always follows a powerlaw with
index $-2$ (see Fig.~\ref{f_dx2_rho_qall}). In the heated region
($\tilde{r} >50 \,r_{\rm pc}$) too the
density does not depend on the mass ratio and the relation $\tilde{\rho}
\propto \tilde{r}^{-4}$ holds for all $q$.
Due to the enhanced instability of the central region for a more
massive secondary black hole the bound stars have to be more
tangentially anisotropic in order to diminish the eccentricity of their
orbits and not to come too close to the binary. This causes the inner
edge of the torus at
$\tilde{r}\approx 2 \,r_{\rm pc}$ to be much sharper defined and thus
the maximum of the density at $\tilde{r}\approx 3 \,r_{\rm pc}$ to be
much more pronounced for 
$q=1$, while it is almost smeared out for $q=100$
(Fig.~\ref{f_dx2_rho_qall}).
This is clearly visible in Fig.~\ref{f_cont_qser}, where the density
distribution is displayed at $t=10^{7.58}$ and $t=10^{7.73}\,{\rm yr}$
and corresponds within a factor of a few to the lower limits of the
merging times $t=10^{6.89}$ and $t=10^{7.77}\,{\rm yr}$ of the
BHs for $q=1$ and $100$ respectively. In fact, the bottom row for
$q=100$ shows no torus,
but the density is diminished around the orbit of $M_2$ and weakly
indicates the formation of a shell-like distribution. This is
confirmed if we let the simulation continue till $\tilde{t}_{\rm
max}$, about $10$ times longer. Then the density distribution is
comparable to that of $q=10$ after $10^{6.57}\,{\rm yr}$ (2nd row in
Fig.~\ref{f_cont_tser}).

\begin{figure*}[t]
\hspace{1mm}%
\resizebox{84.75mm}{!}{\includegraphics[bb = 10 5 482 455]{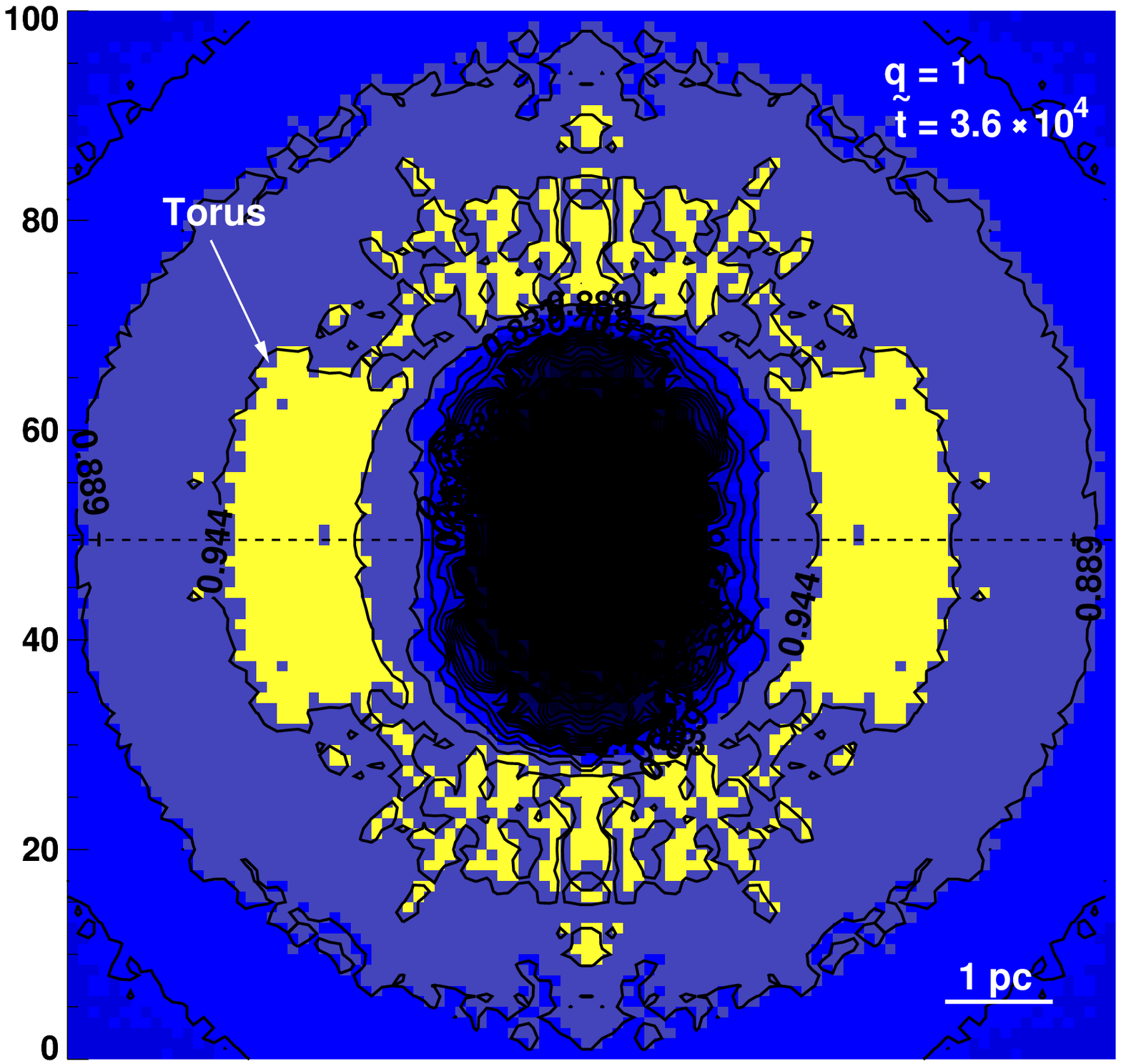}}%
\hspace{10.mm}%
\resizebox{84.75mm}{!}{\includegraphics[bb = 10 5 482 455]{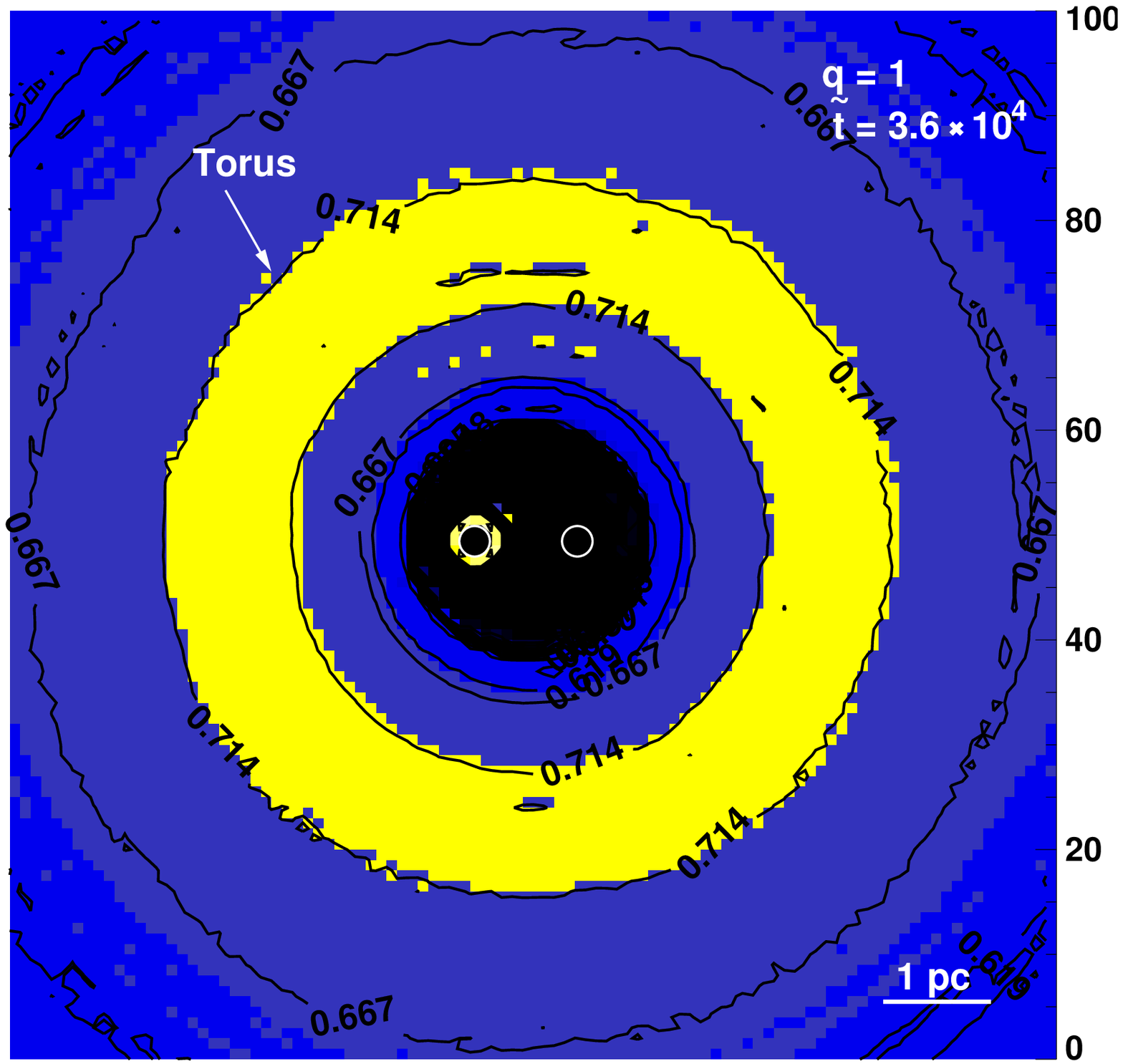}}%
\end{figure*}

Thus the central stellar cluster in the potential of two merging BHs
does not evolve into a torus-like structure for a sufficiently large
mass-ratio $q$ and rather stalls in the phase when it assumes a
shell-like structure.

On the other hand we can see that in the case of $q=1$ the final torus
($t=3.8\times 10^{7}\,{\rm yr}$) is much more pronounced than for
$q=10$, compare Fig.~\ref{f_cont_tser}  and \ref{f_cont_qser}.
This
means that for sufficiently small mass-ratios $q$ a stellar torus
forms in the late stages of a merger, while for young mergers and
mergers with a big mass-ratio a shell-like density distribution is
maintained.
For $q=1$ in Fig.~\ref{f_cont_qser} the inner edge of the torus is
more sharply defined, and there is actually no cusp left in the
center. 
Consequently the bound stars must have bigger angular momenta
as $M_2$ increases. If we compare the initial angular momentum of the
ejected stars for the mass-ratios $q=1$ and $100$ with the ratio
$q=10$, it turns out that it increases with the mass of the secondary
BH (decreasing with $q$). The relation
$\mean{\tilde{L}(q=1)}>\mean{\tilde{L}(q=10)}>\mean{\tilde{L}(q=100)}$
also holds at the end of the simulation as well as for the EP.
Plotting the initial angular momentum versus the
radius for each star for $q=1$ and $100$, just in the same way as for
$q=10$ in Fig.~\ref{f_dx2_ldotr_q10}, gives a very similar separation
between the EP and BP. But with increasing $q$ the transition
value of $\tilde{L}$, dividing both populations, decreases from a
little more than $2$ for $q=1$ to a value below $2$ for $q=100$ with
always the same initial conditions. As the value of transition of
$\tilde{L}$ in Fig.~\ref{f_dx2_ldotr_q10} is decreasing with $M_2$,
also the mean values of the angular momentum of both populations are
decreasing. Stars which have been ejected for small mass-ratios $q$
``defect'' to the BP as $q$ increases and the inner region
becomes more stable. Therefore the deviation $\Delta \tilde{L}$ of the
BP increases while it is decreasing for the ejected population (see
Table~\ref{tb_stat}).

The smaller the mass-ratio is, the more angular momentum is extracted
from the binary by a single ejected star on average
($\mean{\tilde{L}_z(\mli{\tilde{t}}{max})} -
\mean{\tilde{L}_z(\tilde{t}=0)}$, see
Table~\ref{tb_stat}). But this is not 
enough to compensate for the bigger amount of angular momentum of the
binary due to the more massive secondary BH, $\tilde{L}_{\BHBidx} =
\tilde{L}_1 + \tilde{L}_2 = 10^8 M_8/(1+q)$.
Thus the cluster has to be more massive
so that the binary can transfer its angular momentum to a larger
number of ejected stars in order to merge.
To compute the required number of stars and the time
needed to enable the merging of the black holes we proceed in the same way
as before in Sect.~\ref{s_ang_loss} for $q=10$.
With decreasing mass-ratio the maximum of the angular momentum
lossrate is more peaked and has a value $\sim 3.7$ times higher for $q=1$
compared to $q=10$ (see Fig.~\ref{f_dx2_dlzdt_all}). The maximum is
reached the sooner, the smaller $q$ is ($\tilde{t} = 94,\; 142$
corresponding to $1.34 \times 10^5$ and $2.02 \times 10^5\,{\rm yr}$
for $q = 1,\, 10$ respectively). Also the powerlaw with index
$-3/2$ is approximated earlier by the loss-rate for small mass-ratios
and therefore the system is relaxing faster.

The number of stars and the times required for the merging are
listed in Table~\ref{tb_stat}. The values we computed for the case
$q=100$ are not as accurate as for the other mass-ratios, since the
data of the angular momentum loss-rate do not allow for a fit as good
as for $q=1$ and $10$.
However, the values show that for a binary with a fixed
mass for the primary BH more stars in the cluster are required
for smaller mass-ratios in order to allow the black holes
to coalesce. On the other hand the binary with the smaller mass-ratio
is merging faster, provided that the cluster is sufficiently
massive. For instance if $q=1$, the merger is $2.4$ times
faster than the one where $q=10$, while a factor of $3.5$ times more
stars in total are required.
For a binary with smaller $q$ the distance $\tilde{a}_{\rm tr}$, where
gravitational radiation starts to dominate the merging process,
is bigger and thus reached after a shorter time due
to the larger mass of $M_2$
($a_{\rm tr} \approx 1/120,\; 1/200,\; 1/265\,{\rm pc}$ and $t_{\rm
tr} \approx 6.18 \times 10^{6},\; 1.40 \times 10^{7},\; 4.71 \times  
10^{7}\,{\rm yr}$ for $q = 1,\,10,\,100$ respectively, see
Table~\ref{tb_stat}). The ratio of the time-ranges during which the
merger is dominated by ejection of stars or the emission
gravitational radiation, $\mli{\tilde{t}}{tr}/\mli{\tilde{t}}{mg}$, is
independent of the mass-ratio and always equal to $4$,
see Eq.~(\ref{eq_timerat}).

Assuming the average stellar mass to be that of the sun, these numbers
show that the ejected mass $\mli{M}{ej}$ is of order of the
total mass of the binary $M_{\BHBidx}$. While $\mli{M}{ej}$ is
decreasing more strongly with increasing $q$ than the binary mass,
the ejected mass always exceeds that of the secondary BH.
The ratio of both masses, $\mli{M}{ej}/M_2$ is increasing with
$q$. Therefore a primary BH with mass $M_1$
ejects more mass if it merges with $N$ black holes of mass $M_1/N$ than
merging with another BH of mass $M_1$. This is the same result
which has been obtained by \citeN{quinlan96} in scattering
experiments.

\begin{figure}
\vspace{0.7cm}
\resizebox{85mm}{!}{\input{bhb_fig23}}
\caption[]{The smaller the mass-ratio $q$ is, the more peaked is the
maximum and the earlier the maximum is reached. Thus the system is
relaxing faster. For all $q$ the loss-rates finally approximate a
powerlaw with index $-3/2$. For $q=100$ the curve is not as smooth due
to the smaller number statistics.}
\label{f_dx2_dlzdt_all}
\end{figure}

\begin{table*}[t]{\mbox{\centerline{\bfseries Tabulated values of the
results for different mass-ratios}}}\\[0.1ex]
\renewcommand{\arraystretch}{1.5}
\begin{center}
\begin{tabular}{c@{\hspace{4.ex}}c@{\hspace{3.ex}}|@{\hspace{5.ex}}ccc@{\hspace{8.ex}}ccc@{\hspace{8.ex}}cccc}
\multicolumn{2}{c}{} & \multicolumn{3}{c@{\hspace{8.ex}}}{ejected stars} &
\multicolumn{3}{c@{\hspace{8.ex}}}{bound stars}& \\ [1.ex]\hline
\tabularnewline[-2.ex]
$q$ & $\tilde{t}$ & $\mean{\tilde{L}_z}$ & $\mean{\tilde{L}}$ &
$\Delta \tilde{L}$ & $\mean{\tilde{L}_z}$ & $\mean{\tilde{L}}$ &
$\Delta \tilde{L}$ &
$\frac{\mli{N}{ej}}{\mli{N}{bn}}$ &
$\mli{N}{total}$ &               
$\mli{t}{tr}\,[{\rm yr}]$ &      
$\mli{a}{tr}\,[{\rm pc}]$\\[-0.5ex]
$(1)$ & $(2)$ & $(3)$ & $(4)$ & $(5)$ & $(6)$ & $(7)$ & $(8)$ &
$(9)$ & $(10)$ & $(11)$ & $(12)$\\[1.ex]\hline

 & 0 & 0.086 & 1.165 & 0.532 & -0.064 & 3.773 & 1.432 &    \\
\raisebox{2.25ex}[-2.25ex]{1} & $\mli{\tilde{t}}{max}$ 
     & 0.422 & 1.334 & 0.481 & -0.063 & 3.780 & 1.429 &
\raisebox{2.25ex}[-2.25ex]{0.667} &
\raisebox{2.25ex}[-2.25ex]{$10^{8.58}$} &
\raisebox{2.25ex}[-2.25ex]{$10^{6.79}$} &
\raisebox{2.25ex}[-2.25ex]{$1/117$} \\ 
\hline

 & 0 & 0.063 & 1.149 & 0.507 & -0.045 & 3.701 & 1.471 & \\
\raisebox{2.25ex}[-2.25ex]{10} & $\mli{\tilde{t}}{max}$ 
     & 0.284 & 1.264 & 0.497 & -0.043 & 3.711 & 1.467 &
\raisebox{2.25ex}[-2.25ex]{0.614} &
\raisebox{2.25ex}[-2.25ex]{$10^{7.97}$} &
\raisebox{2.25ex}[-2.25ex]{$10^{7.15}$} &
\raisebox{2.25ex}[-2.25ex]{$1/197$} \\ 
\hline

 & 0 & 0.086 & 1.005 & 0.411 & -0.039 & 3.411 & 1.562 &    \\
\raisebox{2.25ex}[-2.25ex]{100} & $\mli{\tilde{t}}{max}$ 
     & 0.239 & 1.072 & 0.435 & -0.037 & 3.418 & 1.561 &
\raisebox{2.25ex}[-2.25ex]{0.395} &
\raisebox{2.25ex}[-2.25ex]{$10^{7.35}$} &
\raisebox{2.25ex}[-2.25ex]{$10^{7.67}$} &
\raisebox{2.25ex}[-2.25ex]{$1/265$} \\ 

\end{tabular}
\end{center}

\caption[]{\label{tb_stat}The lines in this table separate the results
for the different mass-ratios $q=M_1/M_2$, as indicated in the first
column. The second column subdivides each line into two lines for the
next 6 columns, with the upper and lower line referring to numbers
obtained in the beginning and the end of the simulation
respectively. The 3rd and 6th column show the mean angular momentum
of the $z$-component, the 4th and 7th column the mean total angular
momentum and the 5th and 8th column the standard deviation of the
total angular momentum for the ejected and bound stars respectively.
The next column contains the number ratio $N_{\rm ej}/N_{\rm bn}$ of
ejected to bound stars and the 10th column the total number of stars
which initially is required so that the ejected fraction extracts the
complete angular momentum from the binary. $t_{\rm tr}$ is the time
needed by the black holes to shrink from an initial distance of
$1\,{\rm pc}$ due to ejection of stars to $a_{\rm tr}$, where emission of
gravitational radiation starts to dominate further merging.}
\end{table*}



\section{Conclusions}

One basic problem in our understanding of the central activity of
galaxies is addressed in this paper: Do the black holes in the center
of two merging galaxies also merge?
Based on the two assumptions that galaxies contain supermassive black holes
in their centers and that galaxies merge we simulated a stellar
cluster in the potential of a BHB. When the distance between the
black holes has shrunk to $\sim r_{\rm cusp}$ and ejection of stars
dominates the merging, we start our calculations. Since the simulated
time is much smaller than the relaxation time of the cluster and the
potential is dominated by the mass of the two black holes we were dealing
with the restricted three body problem. The results we obtained by
solving these equations for the stars of the cluster, initially
distributed according to the singular isothermal sphere, just give an
order of magnitude estimate since in the calculations we neglected
further shrinking of the separation of the BHs due to ejection of
stars.

We find that the angular momentum divides the stellar cluster
into a bound and
ejected population. Stars staying bound in the potential of the BHs
are distributed in a torus-like shape. Its relaxation time exceeds
$10^{10-12}\,{\rm yr}$, so that this structure can be regarded as
stable. It can explain lots of the features which are postulated by
the unification model for a dusty torus in AGN and will be the topic
of paper 2.

Stars which belong to the ejected population have low angular momentum and
are moving on highly eccentric orbits, being radially
anisotropic in velocity. The distribution of their peri- and apocenter
always allows them to come very
close to the binary. Here they gain sufficient energy in violent
interactions with the black holes to become ejected.
The ejected stars gain angular momentum in their
$z$-component at the expense of the binary, which can be understood in
terms of the conservation of the Jacobian Integral ($d\tilde{E}\propto
d\tilde{L}_z$). Thus the binary continues to
shrink. Integrating its loss-rate $d\tilde{L}_z/d\tilde{t}$ allows to
calculate the number of ejected stars and the time required for the
black holes to coalesce.

Starting with $1\,{\rm pc}$ distance between the BHs,
we find that a stellar cluster with a mass of order of the binary is
required to allow the BHs to coalesce.
The merging proceeds on
timescales of order of $10^7\,{\rm yr}$ and will change the spin of
the central BH \cite{wilson95}. Consequently also the jet, emanating
from the center, will jump and flow along the new spin direction after
the merger is completed. This might be an explanation for the observed
X-shaped radio galaxies, where a spectral aging time of the the
secondary lobes suggests that a merger has happened $\sim 6\times
10^7\,{\rm yr}$ ago (priv. comm. H. Rottmann). The agreement with
our merger timescale is in favour of this idea.

Most of the time (factor 4) this merging
process is dominated by ejection of stars before emission of
gravitational radiation takes over. Thus a stellar cluster with a
total mass
comparable to that of the binary is needed to allow the black holes to
merge due to the ejection of a fraction of $\sim 0.4$ of all stars. 
After the ejection the stars are strongly concentrated to the
equatorial plane of the binary, the more the higher their energy is,
and the initial spherical geometry of the EP has been transformed
into a cylindrical geometry.

As the mass of the secondary black hole is increased, the central
region becomes more unstable and a bigger fraction of stars
is ejected. In order
to extract sufficient angular momentum from the binary, more stars are
required, so that the clusters mass still is of the order of magnitude of
the binary's mass. The system is relaxing faster and the distance
where gravitational radiation dominates the merging process is reached
earlier, so that the binary merges on smaller time scales
($t\approx 10^{6.8},\, 10^{7.2}$ and $10^{7.8}\,{\rm yr}$ for
$q=1,\,10$ and $100$ respectively). The
ratio $\mli{M}{ej}/M_2$ is increasing with $q$ and therefore a primary
black hole with mass $M_1$ ejects more mass in $N$ minor mergers with
secondary black holes of mass $M_2=M_1/N$ than in one major merger with
mass $M_2=M_1$.

In major mergers the mass distribution is likely to be completely
rearranged and the resulting galaxy will probably have a rather
elliptical than spiral shape. In their overview of the structure of
nearby AGN \cite{malkan98} find that Seyfert 1 nuclei reside in more
early-type host galaxies than Seyfert 2. This is in agreement with our
finding, that the more massive the secondary BH is, the more stars
from the cluster are ejected.
Thus the remaining torus-like distribution
becomes shallower and there is a higher probability to see the Type 1
nucleus directly.
Combinig this with the conclusions by \cite{wilson95}, that radio loud
objects are possibly the product of major mergers, it is expected that
radio loud galaxies are more likely to host Type 1 nuclei.

The key result of this paper is: Provided that a black hole binary is
surrounded by a cluster of solar mass stars with $M_{\rm cluster}
\approx M_{\BHBidx}$, i.e. the total mass of the cluster is comparable
to that of the binary, the black holes merge on time scales of about
$10^7\,{\rm yr}$. In paper 2 we will demonstrate that the stellar
torus surrounding the merged binary also meets the other requirements
of the unification scheme. We will demonstrate that the stellar winds
are pushed by radiation from the central source into long tails, and that
the ensemble of many stellar wind-tails can constitute the observed
torus.

\begin{acknowledgements}
C.Z. wishes to thank Matt Malkan, Ski Antonucci, Graeme Smith, David
Merritt and Stefan Westerhoff for their helpful discussions and their
generous and kind hospitality. We are grateful to our referee Edward
J.M. Colbert for his helpful advice and comments.
\end{acknowledgements}

\begin{appendix}
\section{Equations of the restricted three body system}
\label{app_3body}
In the restricted three body system two massive bodies (BHs)
are circulating around each other. A third particle
(a star), whose mass is negligible
($m\ll M_1, M_2$), is moving in the time dependent gravitational
potential of the massive bodies, which reads in the dimensionless form
(eq.~(\ref{eq_norm})):
\mbox{$
\tilde{\Phi} =
-\frac{1}{1+q}\left(\frac{q}{|\tilde{\VEC{r}}-\tilde{\VEC{r}}_1|} + 
\frac{1}{|\tilde{\VEC{r}}-\tilde{\VEC{r}}_2|}\right)
$}.
The star is located at $\tilde{\VEC{r}}$ while $\tilde{\VEC{r}}_1$ and
$\tilde{\VEC{r}}_2$
denote the positions of $M_1$ and $M_2$ respectively.
The origin is identified with the center of mass of the two massive
bodies and their axis of rotation is aligned with the $z$-axis and at
$t=0$ with both black holes situated on the $x$-axis.
The equations of motion for the star are
\begin{equation}
\label{eq_res3bdy}
\ddot{\tilde{\VEC{r}}} = -\VEC{\nabla} \tilde{\Phi}\,.
\end{equation}
We always assumed $M_1 \ge M_2$ (and therefore $q=
M_1/M_2\ge 1$). The separation of the black holes,
$\tilde{a} = |\tilde{\VEC{r}}_1 - \tilde{\VEC{r}}_2| = (1+q)
|\tilde{\VEC{r}}_1|$, is
constant, i.e. the black holes are moving on 
circular orbits with a constant radius around each other.
The numerical integration of the equations of motion is processed
faster in the 
comoving frame where the BHs are stationary on the
$\tilde{x}$-axis in the distances $\tilde{x}_1$ and $\tilde{x}_2$ from
the origin. The transformation into the comoving frame
introduces the centrifugal and the Coriolis force in
(\ref{eq_res3bdy}):
\begin{eqnarray}
\label{eq_dgl3}
\ddot{\tilde{\VEC{r}}}  & = & -\VEC{\nabla} \tilde{\Phi} -
\tilde{\VEC{\omega}}\times
(\tilde{\VEC{\omega}}\times\tilde{\VEC{r}}) - 2
\tilde{\VEC{\omega}}\times\dot{\tilde{\VEC{r}}}\nonumber\\
& = & -\frac{1}{1+q} \left(q\frac{\tilde{\VEC{r}} -
\tilde{\VEC{r}}_{1}}{|\tilde{\VEC{r}} - \tilde{\VEC{r}}_{1}|^3} +
\frac{\tilde{\VEC{r}} +
q \tilde{\VEC{r}}_{1}}{|\tilde{\VEC{r}} + q
\tilde{\VEC{r}}_{1}|^3}\right)\\
& & - \tilde{\VEC{\omega}}\times
(\tilde{\VEC{\omega}}\times\tilde{\VEC{r}}) - 2
\tilde{\VEC{\omega}}\times\dot{\tilde{\VEC{r}}}\,,\nonumber
\end{eqnarray}
where $\tilde{\VEC{\omega}}$ in dimensionless units is actually the
unit-vector in $z$ direction, $\hat{e}_z$.
Written in components with the vectors $\tilde{\VEC{r}} = (\tilde{x},
\tilde{y}, \tilde{z})$ and $\ddot{\tilde{\VEC{r}}} =
(\dot{\tilde{v}}_x, \dot{\tilde{v}}_y, \dot{\tilde{v}}_z)$ this
equation is transformed into a system of six coupled ordinary
differential equations of first order:
\begin{eqnarray}
\label{eq_dgl6}
\dot{\tilde{x}} & = & \tilde{v}_x\nonumber\\
\dot{\tilde{y}} & = & \tilde{v}_y\nonumber\\
\dot{\tilde{z}} & = & \tilde{v}_z\nonumber\\
\dot{\tilde{v}}_x & = &
-\frac{1}{1+q}\left(q\frac{\tilde{x}- \tilde{x}_1}{|\tilde{\VEC{r}}-
\tilde{\VEC{r}}_1|^3}
+ \frac{\tilde{x}+q \tilde{x}_1}{|\tilde{\VEC{r}}+q
\tilde{\VEC{r}}_1|^3}\right) +2 \tilde{v}_y +\tilde{x}\\
\dot{\tilde{v}}_y & = &
-\frac{1}{1+q}\left(q\frac{\tilde{y}}{|\tilde{\VEC{r}}- \tilde{\VEC{r}}_1|^3}
+ \frac{\tilde{y}}{|\tilde{\VEC{r}}+q \tilde{\VEC{r}}_1|^3}\right) -2
\tilde{v}_x +\tilde{y}\nonumber\\ 
\dot{\tilde{v}}_z & = &
-\frac{1}{1+q}\left(q\frac{\tilde{z}}{|\tilde{\VEC{r}}- \tilde{\VEC{r}}_1|^3}
+ \frac{\tilde{z}}{|\tilde{\VEC{r}}+q \tilde{\VEC{r}}_1|^3}\right)\nonumber\,.
\end{eqnarray}
After being supplied with a set of initial conditions these equations
have been integrated numerically.

\section{The Jacobian Integral}
\label{app_jacobi}

For the restricted three body problem the energy is not conserved
because the potential is time-dependent and
consequently no integral of motion. The same holds for the angular momentum, 
which is not conserved since the potential is not invariant to
rotations. But an integral of motion can still be found
if we multiply the equation of motion for a star (\ref{eq_dgl3}) with
$\dot{\tilde{\VEC{r}}}$,
\begin{eqnarray}
\dot{\tilde{\VEC{r}}}\ddot{\tilde{\VEC{r}}}  & = &
-\dot{\tilde{\VEC{r}}}\VEC{\nabla} \tilde{\Phi} - 
(\tilde{\VEC{\omega}}\times\tilde{\VEC{r}})
(\tilde{\VEC{\omega}}\times\dot{\tilde{\VEC{r}}})\nonumber\\ 
& = & \left(-\frac{d}{d\tilde{t}} + \frac{\partial}{\partial
\tilde{t}}\right)\tilde{\Phi} +
\frac{1}{2}\frac{d}{d\tilde{t}}(\tilde{\VEC{\omega}}
\times\tilde{\VEC{r}})^2\\ 
& &- (\tilde{\VEC{\omega}}
\times\tilde{\VEC{r}})(\dot{\tilde{\VEC{\omega}}}  
\times\tilde{\VEC{r}})\,.
\nonumber
\end{eqnarray}
Because $\tilde{\VEC{\omega}}$ is constant and $\tilde{\Phi}$ does not
depend explicitly on time we can rewrite this expression in a form 
showing that the quantity
\begin{equation}
\tilde{J} \equiv \tilde{\Phi} + \frac{1}{2} \dot{\tilde{\VEC{r}}}^2 -
\frac{1}{2}(\tilde{\VEC{\omega}}\times\tilde{\VEC{r}})^2\,,
\end{equation}
the Jacobian Integral, is conserved ($\frac{d\tilde{J}}{d\tilde{t}} =
0$). With the expressions for the
energy and the specific angular momentum
expressed in the comoving frame
\begin{eqnarray}
\tilde{E} & = & \frac{1}{2}\left(\dot{\tilde{\VEC{r}}}^2 +
2\dot{\tilde{\VEC{r}}}
(\tilde{\VEC{\omega}}\times\tilde{\VEC{r}}) +
(\tilde{\VEC{\omega}}\times\tilde{\VEC{r}})^2\right)
+ \tilde{\Phi}(\tilde{\VEC{r}})\,,\nonumber\\
\tilde{\VEC{L}} & = & \tilde{\VEC{r}}\times\dot{\tilde{\VEC{r}}} +
\tilde{\VEC{r}}\times(\tilde{\VEC{\omega}}\times
\tilde{\VEC{r}})\nonumber
\end{eqnarray}
and using the relation
\[
\tilde{\VEC{\omega}} \tilde{\VEC{L}} =
\dot{\tilde{\VEC{r}}}(\tilde{\VEC{\omega}}\times\tilde{\VEC{r}}) +
(\tilde{\VEC{\omega}}\times\tilde{\VEC{r}})^2
\]
we finally can write the Jacobian Integral in the simple form
\begin{equation}
\label{eq_jacobi}
\tilde{J} = \tilde{E} - \tilde{\VEC{\omega}} \tilde{\VEC{L}} = const.
\end{equation}
Thus, in a rotating non-axisymmetric potential, neither $\tilde{E}$ nor
$\tilde{\VEC{L}}$ is conserved, but the combination $\tilde{J} = E -
\tilde{\VEC{\omega}} \tilde{\VEC{L}}$.
Applying infinitesimal variations to Eq.~(\ref{eq_jacobi}), with
$\tilde{\VEC{\omega}} = \hat{e}_z$, leads to the expression
\begin{equation}
\label{eq_djacobi}
d\tilde{E} = d\tilde{L}_z\,.
\end{equation}
Consequently, stars which gain energy also gain angular momentum in the
$z$-component.
\section{The isothermal sphere}
\label{app_isosphere}

In a spherically symmetric potential there exist four isolating integrals,
the energy $E$ and the three components of the angular momentum
$\VEC{L}$. According to the \emph{Jeans Theorem} applied to spherical
systems any non-negative function of these integrals can serve as the
distribution function of a spherical stellar system. The extension of
the \emph{Strong Jeans Theorem} to spherical systems by
\citeN{lynbell62} allows to conclude that the distribution function of
any steady state spherical system can be
expressed as function a $f(E, \VEC{L})$. If the system is spherically
symmetric in all its properties, $f$ is independent on the direction
of $\VEC{L}$ so that $f = f(E, L)$.
If now the potential $\Phi$ is provided by the stellar system itself
and we understand $f$ as the mass distributionfunction (i.e. $\rho =
\int f d^3 \VEC{v}$) the Poisson equation
\begin{equation}
\Delta \Phi = 4\pi G \rho = 4\pi G \int f d^3 \VEC{v}\,,
\end{equation}
is the fundamental equation which governs spherical equilibrium
stellar systems.
In spherical symmetry this equation is
\begin{equation}
\label{eq_poisson_sph}
\frac{1}{r^2}\frac{d}{dr}\left(r^2 \frac{d\Phi}{dr}\right) = 4\pi G
\int f(\frac{1}{2} v^2 + \Phi, |\VEC{r}\times \VEC{v}|)d^3 \VEC{v}\,.
\end{equation}

Following the notation of \citeN{binney87} we
define the relative potential and the relative energy by
$\Psi \equiv -\Phi + \Phi_0$ and $\mathcal{E} \equiv -E + \Phi_0 = \Psi -
\frac{1}{2} v^2$. 
The constant $\Phi_0$ is choosen in a way that $f > 0$ for
$\mathcal{E} >0$ and $f = 0$ for $\mathcal{E} \le 0$. As
$\Phi$, the
relative potential $\Psi$ also satisfies the Poisson equation $\Delta
\Psi = -4\pi G\rho$ with the boundary condition $\Psi \to \Phi_0$ for
$|\VEC{r}| \to \infty$. A distribution function depending on the
relative energy
$\mathcal{E}$ only further simplifies the spherical model. In such
systems with $f=f(\Psi - \frac{1}{2} v^2)$ the velocity dispersion
tensor is isotropic everywhere because the velocity dispersions are
the same in all spherical components $r, \phi$ and $\theta$. Using
this distribution function and if we choose $\Phi_0$ in
a way that $f(\mathcal{E}) = 0$ for $\mathcal{E} < 0$,
Eq.~(\ref{eq_poisson_sph}) becomes
\begin{eqnarray}
\label{eq_poisson_iso}
\frac{1}{r^2}\frac{d}{dr}\left(r^2 \frac{d\Psi}{dr}\right) & = &
-16\pi^2 G\int_0^{\sqrt{2\pi}} f(\Psi -\frac{1}{2} v^2) v^2 dv
 \nonumber\\
& = & -16\pi^2 G\int_0^{\Psi} f(\mathcal{E}) \sqrt{2(\Psi -
\mathcal{E})} d\mathcal{E}\,.
\end{eqnarray}
Either we may understand this equation as a non-linear equation for
$\Psi(r)$ with given distribution function $f(\mathcal{E})$, or as a
linear equation for
$f(\mathcal{E})$ with given $\Psi(r)$.
With the choice for the distribution function for a stellar dynamical
system
\begin{equation}
\label{eq_df_iso}
f(\mathcal{E}) = \frac{\rho_1}{(2\pi\sigma^2)^{\frac{3}{2}}}
e^{\mathcal{E}/\sigma^2} = \frac{\rho_1}{(2\pi\sigma^2)^{\frac{3}{2}}}
\exp\left(\frac{\Psi -\frac{1}{2}v^2}{\sigma^2}\right)\,,
\end{equation}
Poissons equation can be rewritten as
\begin{equation}
\label{eq_poisson_mix}
\frac{d}{dr} \left(r^2 \frac{d\Psi}{dr} \right) =
4\pi G\rho\,.
\end{equation}
Integrating now the distribution function (\ref{eq_df_iso}) over all
velocities yields the
density $\rho = \rho_1 e^{\Psi/\sigma^2}$. With the help of this
expression Eq.~(\ref{eq_poisson_mix}) becomes
\begin{equation}
\label{eq_poisson_rho}
\frac{d}{dr} \left(r^2 \frac{d \ln \rho}{dr} \right) =
-\frac{4\pi G}{\sigma^2}r^2\rho\,.
\end{equation}
Comparing this with the equation of
hydrostatic equilibrium for an inviscid fluid or gas
\begin{equation}
\label{eq_hydrostatic_con}
\frac{d}{dr} \left(r^2 \frac{d\ln \rho}{dr} \right) =
-\frac{Gm}{k_B T}4\pi r^2 \rho
\end{equation}
shows that they become identical if we choose
the parameter $\sigma$ to be $\sigma = k_B T/m$.
This correspondence between the gaseous and stellar-dynamical
isothermal sphere originates in the velocity distribution of both
systems. Integrating  Eq.~(\ref{eq_df_iso}) over the
volume the distribution of velocities results in the Maxwell-Boltzmann
distribution:
\begin{equation}
\label{eq_mbdistrib}
F(v) = N e^{-\frac{1}{2} v^2/\sigma^2}\,.
\end{equation}

In order to find now a solution of Eq.~(\ref{eq_poisson_rho}) we
simply assume a powerlaw for the density ($\rho = Cr^{-b}$) and
substitute it in Eq.~(\ref{eq_poisson_rho}) and find
$b = \frac{4\pi G}{\sigma^2}C r^{2-b}$.
Thus the two parameters must have the values $b=2$ and
$C=\sigma^2/2\pi G$ to fulfil this equation and we obtain for the
density
\begin{equation}
\label{eq_sing_isosph}
\rho (r) = \frac{\sigma^2}{2\pi G r^2}\,.
\end{equation}
A system described by this function for the density is known as the
singular isothermal sphere, which is only true for big enough
radii because of its singularity at the center.

\end{appendix}

\bibliography{refs}
\bibliographystyle{aa}

\end{document}

%% file: bhb_fig01.tex
\begingroup%
  \makeatletter%
  \newcommand{\GNUPLOTspecial}{%
    \@sanitize\catcode`\%=14\relax\special}%
  \setlength{\unitlength}{0.1bp}%
{\GNUPLOTspecial{!
/gnudict 256 dict def
gnudict begin
/Color false def
/Solid false def
/gnulinewidth 5.000 def
/userlinewidth gnulinewidth def
/vshift -33 def
/dl {10 mul} def
/hpt_ 31.5 def
/vpt_ 31.5 def
/hpt hpt_ def
/vpt vpt_ def
/M {moveto} bind def
/L {lineto} bind def
/R {rmoveto} bind def
/V {rlineto} bind def
/vpt2 vpt 2 mul def
/hpt2 hpt 2 mul def
/Lshow { currentpoint stroke M
  0 vshift R show } def
/Rshow { currentpoint stroke M
  dup stringwidth pop neg vshift R show } def
/Cshow { currentpoint stroke M
  dup stringwidth pop -2 div vshift R show } def
/UP { dup vpt_ mul /vpt exch def hpt_ mul /hpt exch def
  /hpt2 hpt 2 mul def /vpt2 vpt 2 mul def } def
/DL { Color {setrgbcolor Solid {pop []} if 0 setdash }
 {pop pop pop Solid {pop []} if 0 setdash} ifelse } def
/BL { stroke gnulinewidth 2 mul setlinewidth } def
/AL { stroke gnulinewidth 2 div setlinewidth } def
/UL { gnulinewidth mul /userlinewidth exch def } def
/PL { stroke userlinewidth setlinewidth } def
/LTb { BL [] 0 0 0 DL } def
/LTa { AL [1 dl 2 dl] 0 setdash 0 0 0 setrgbcolor } def
/LT0 { PL [] 1 0 0 DL } def
/LT1 { PL [4 dl 2 dl] 0 1 0 DL } def
/LT2 { PL [2 dl 3 dl] 0 0 1 DL } def
/LT3 { PL [1 dl 1.5 dl] 1 0 1 DL } def
/LT4 { PL [5 dl 2 dl 1 dl 2 dl] 0 1 1 DL } def
/LT5 { PL [4 dl 3 dl 1 dl 3 dl] 1 1 0 DL } def
/LT6 { PL [2 dl 2 dl 2 dl 4 dl] 0 0 0 DL } def
/LT7 { PL [2 dl 2 dl 2 dl 2 dl 2 dl 4 dl] 1 0.3 0 DL } def
/LT8 { PL [2 dl 2 dl 2 dl 2 dl 2 dl 2 dl 2 dl 4 dl] 0.5 0.5 0.5 DL } def
/Pnt { stroke [] 0 setdash
   gsave 1 setlinecap M 0 0 V stroke grestore } def
/Dia { stroke [] 0 setdash 2 copy vpt add M
  hpt neg vpt neg V hpt vpt neg V
  hpt vpt V hpt neg vpt V closepath stroke
  Pnt } def
/Pls { stroke [] 0 setdash vpt sub M 0 vpt2 V
  currentpoint stroke M
  hpt neg vpt neg R hpt2 0 V stroke
  } def
/Box { stroke [] 0 setdash 2 copy exch hpt sub exch vpt add M
  0 vpt2 neg V hpt2 0 V 0 vpt2 V
  hpt2 neg 0 V closepath stroke
  Pnt } def
/Crs { stroke [] 0 setdash exch hpt sub exch vpt add M
  hpt2 vpt2 neg V currentpoint stroke M
  hpt2 neg 0 R hpt2 vpt2 V stroke } def
/TriU { stroke [] 0 setdash 2 copy vpt 1.12 mul add M
  hpt neg vpt -1.62 mul V
  hpt 2 mul 0 V
  hpt neg vpt 1.62 mul V closepath stroke
  Pnt  } def
/Star { 2 copy Pls Crs } def
/BoxF { stroke [] 0 setdash exch hpt sub exch vpt add M
  0 vpt2 neg V  hpt2 0 V  0 vpt2 V
  hpt2 neg 0 V  closepath fill } def
/TriUF { stroke [] 0 setdash vpt 1.12 mul add M
  hpt neg vpt -1.62 mul V
  hpt 2 mul 0 V
  hpt neg vpt 1.62 mul V closepath fill } def
/TriD { stroke [] 0 setdash 2 copy vpt 1.12 mul sub M
  hpt neg vpt 1.62 mul V
  hpt 2 mul 0 V
  hpt neg vpt -1.62 mul V closepath stroke
  Pnt  } def
/TriDF { stroke [] 0 setdash vpt 1.12 mul sub M
  hpt neg vpt 1.62 mul V
  hpt 2 mul 0 V
  hpt neg vpt -1.62 mul V closepath fill} def
/DiaF { stroke [] 0 setdash vpt add M
  hpt neg vpt neg V hpt vpt neg V
  hpt vpt V hpt neg vpt V closepath fill } def
/Pent { stroke [] 0 setdash 2 copy gsave
  translate 0 hpt M 4 {72 rotate 0 hpt L} repeat
  closepath stroke grestore Pnt } def
/PentF { stroke [] 0 setdash gsave
  translate 0 hpt M 4 {72 rotate 0 hpt L} repeat
  closepath fill grestore } def
/Circle { stroke [] 0 setdash 2 copy
  hpt 0 360 arc stroke Pnt } def
/CircleF { stroke [] 0 setdash hpt 0 360 arc fill } def
/C0 { BL [] 0 setdash 2 copy moveto vpt 90 450  arc } bind def
/C1 { BL [] 0 setdash 2 copy        moveto
       2 copy  vpt 0 90 arc closepath fill
               vpt 0 360 arc closepath } bind def
/C2 { BL [] 0 setdash 2 copy moveto
       2 copy  vpt 90 180 arc closepath fill
               vpt 0 360 arc closepath } bind def
/C3 { BL [] 0 setdash 2 copy moveto
       2 copy  vpt 0 180 arc closepath fill
               vpt 0 360 arc closepath } bind def
/C4 { BL [] 0 setdash 2 copy moveto
       2 copy  vpt 180 270 arc closepath fill
               vpt 0 360 arc closepath } bind def
/C5 { BL [] 0 setdash 2 copy moveto
       2 copy  vpt 0 90 arc
       2 copy moveto
       2 copy  vpt 180 270 arc closepath fill
               vpt 0 360 arc } bind def
/C6 { BL [] 0 setdash 2 copy moveto
      2 copy  vpt 90 270 arc closepath fill
              vpt 0 360 arc closepath } bind def
/C7 { BL [] 0 setdash 2 copy moveto
      2 copy  vpt 0 270 arc closepath fill
              vpt 0 360 arc closepath } bind def
/C8 { BL [] 0 setdash 2 copy moveto
      2 copy vpt 270 360 arc closepath fill
              vpt 0 360 arc closepath } bind def
/C9 { BL [] 0 setdash 2 copy moveto
      2 copy  vpt 270 450 arc closepath fill
              vpt 0 360 arc closepath } bind def
/C10 { BL [] 0 setdash 2 copy 2 copy moveto vpt 270 360 arc closepath fill
       2 copy moveto
       2 copy vpt 90 180 arc closepath fill
               vpt 0 360 arc closepath } bind def
/C11 { BL [] 0 setdash 2 copy moveto
       2 copy  vpt 0 180 arc closepath fill
       2 copy moveto
       2 copy  vpt 270 360 arc closepath fill
               vpt 0 360 arc closepath } bind def
/C12 { BL [] 0 setdash 2 copy moveto
       2 copy  vpt 180 360 arc closepath fill
               vpt 0 360 arc closepath } bind def
/C13 { BL [] 0 setdash  2 copy moveto
       2 copy  vpt 0 90 arc closepath fill
       2 copy moveto
       2 copy  vpt 180 360 arc closepath fill
               vpt 0 360 arc closepath } bind def
/C14 { BL [] 0 setdash 2 copy moveto
       2 copy  vpt 90 360 arc closepath fill
               vpt 0 360 arc } bind def
/C15 { BL [] 0 setdash 2 copy vpt 0 360 arc closepath fill
               vpt 0 360 arc closepath } bind def
/Rec   { newpath 4 2 roll moveto 1 index 0 rlineto 0 exch rlineto
       neg 0 rlineto closepath } bind def
/Square { dup Rec } bind def
/Bsquare { vpt sub exch vpt sub exch vpt2 Square } bind def
/S0 { BL [] 0 setdash 2 copy moveto 0 vpt rlineto BL Bsquare } bind def
/S1 { BL [] 0 setdash 2 copy vpt Square fill Bsquare } bind def
/S2 { BL [] 0 setdash 2 copy exch vpt sub exch vpt Square fill Bsquare } bind def
/S3 { BL [] 0 setdash 2 copy exch vpt sub exch vpt2 vpt Rec fill Bsquare } bind def
/S4 { BL [] 0 setdash 2 copy exch vpt sub exch vpt sub vpt Square fill Bsquare } bind def
/S5 { BL [] 0 setdash 2 copy 2 copy vpt Square fill
       exch vpt sub exch vpt sub vpt Square fill Bsquare } bind def
/S6 { BL [] 0 setdash 2 copy exch vpt sub exch vpt sub vpt vpt2 Rec fill Bsquare } bind def
/S7 { BL [] 0 setdash 2 copy exch vpt sub exch vpt sub vpt vpt2 Rec fill
       2 copy vpt Square fill
       Bsquare } bind def
/S8 { BL [] 0 setdash 2 copy vpt sub vpt Square fill Bsquare } bind def
/S9 { BL [] 0 setdash 2 copy vpt sub vpt vpt2 Rec fill Bsquare } bind def
/S10 { BL [] 0 setdash 2 copy vpt sub vpt Square fill 2 copy exch vpt sub exch vpt Square fill
       Bsquare } bind def
/S11 { BL [] 0 setdash 2 copy vpt sub vpt Square fill 2 copy exch vpt sub exch vpt2 vpt Rec fill
       Bsquare } bind def
/S12 { BL [] 0 setdash 2 copy exch vpt sub exch vpt sub vpt2 vpt Rec fill Bsquare } bind def
/S13 { BL [] 0 setdash 2 copy exch vpt sub exch vpt sub vpt2 vpt Rec fill
       2 copy vpt Square fill Bsquare } bind def
/S14 { BL [] 0 setdash 2 copy exch vpt sub exch vpt sub vpt2 vpt Rec fill
       2 copy exch vpt sub exch vpt Square fill Bsquare } bind def
/S15 { BL [] 0 setdash 2 copy Bsquare fill Bsquare } bind def
/D0 { gsave translate 45 rotate 0 0 S0 stroke grestore } bind def
/D1 { gsave translate 45 rotate 0 0 S1 stroke grestore } bind def
/D2 { gsave translate 45 rotate 0 0 S2 stroke grestore } bind def
/D3 { gsave translate 45 rotate 0 0 S3 stroke grestore } bind def
/D4 { gsave translate 45 rotate 0 0 S4 stroke grestore } bind def
/D5 { gsave translate 45 rotate 0 0 S5 stroke grestore } bind def
/D6 { gsave translate 45 rotate 0 0 S6 stroke grestore } bind def
/D7 { gsave translate 45 rotate 0 0 S7 stroke grestore } bind def
/D8 { gsave translate 45 rotate 0 0 S8 stroke grestore } bind def
/D9 { gsave translate 45 rotate 0 0 S9 stroke grestore } bind def
/D10 { gsave translate 45 rotate 0 0 S10 stroke grestore } bind def
/D11 { gsave translate 45 rotate 0 0 S11 stroke grestore } bind def
/D12 { gsave translate 45 rotate 0 0 S12 stroke grestore } bind def
/D13 { gsave translate 45 rotate 0 0 S13 stroke grestore } bind def
/D14 { gsave translate 45 rotate 0 0 S14 stroke grestore } bind def
/D15 { gsave translate 45 rotate 0 0 S15 stroke grestore } bind def
/DiaE { stroke [] 0 setdash vpt add M
  hpt neg vpt neg V hpt vpt neg V
  hpt vpt V hpt neg vpt V closepath stroke } def
/BoxE { stroke [] 0 setdash exch hpt sub exch vpt add M
  0 vpt2 neg V hpt2 0 V 0 vpt2 V
  hpt2 neg 0 V closepath stroke } def
/TriUE { stroke [] 0 setdash vpt 1.12 mul add M
  hpt neg vpt -1.62 mul V
  hpt 2 mul 0 V
  hpt neg vpt 1.62 mul V closepath stroke } def
/TriDE { stroke [] 0 setdash vpt 1.12 mul sub M
  hpt neg vpt 1.62 mul V
  hpt 2 mul 0 V
  hpt neg vpt -1.62 mul V closepath stroke } def
/PentE { stroke [] 0 setdash gsave
  translate 0 hpt M 4 {72 rotate 0 hpt L} repeat
  closepath stroke grestore } def
/CircE { stroke [] 0 setdash 
  hpt 0 360 arc stroke } def
/Opaque { gsave closepath 1 setgray fill grestore 0 setgray closepath } def
/DiaW { stroke [] 0 setdash vpt add M
  hpt neg vpt neg V hpt vpt neg V
  hpt vpt V hpt neg vpt V Opaque stroke } def
/BoxW { stroke [] 0 setdash exch hpt sub exch vpt add M
  0 vpt2 neg V hpt2 0 V 0 vpt2 V
  hpt2 neg 0 V Opaque stroke } def
/TriUW { stroke [] 0 setdash vpt 1.12 mul add M
  hpt neg vpt -1.62 mul V
  hpt 2 mul 0 V
  hpt neg vpt 1.62 mul V Opaque stroke } def
/TriDW { stroke [] 0 setdash vpt 1.12 mul sub M
  hpt neg vpt 1.62 mul V
  hpt 2 mul 0 V
  hpt neg vpt -1.62 mul V Opaque stroke } def
/PentW { stroke [] 0 setdash gsave
  translate 0 hpt M 4 {72 rotate 0 hpt L} repeat
  Opaque stroke grestore } def
/CircW { stroke [] 0 setdash 
  hpt 0 360 arc Opaque stroke } def
/BoxFill { gsave Rec 1 setgray fill grestore } def
end
}}%
\begin{picture}(2412,2160)(0,0)%
{\GNUPLOTspecial{"
gnudict begin
gsave
0 0 translate
0.100 0.100 scale
0 setgray
newpath
1.000 UL
LTb
0.050 UL
LT3
400 300 M
2012 0 V
1.000 UL
LTb
400 300 M
63 0 V
1949 0 R
-63 0 V
400 469 M
31 0 V
1981 0 R
-31 0 V
0.050 UL
LT3
400 638 M
2012 0 V
1.000 UL
LTb
400 638 M
63 0 V
1949 0 R
-63 0 V
400 807 M
31 0 V
1981 0 R
-31 0 V
0.050 UL
LT3
400 976 M
2012 0 V
1.000 UL
LTb
400 976 M
63 0 V
1949 0 R
-63 0 V
400 1145 M
31 0 V
1981 0 R
-31 0 V
0.050 UL
LT3
400 1315 M
2012 0 V
1.000 UL
LTb
400 1315 M
63 0 V
1949 0 R
-63 0 V
400 1484 M
31 0 V
1981 0 R
-31 0 V
0.050 UL
LT3
400 1653 M
2012 0 V
1.000 UL
LTb
400 1653 M
63 0 V
1949 0 R
-63 0 V
400 1822 M
31 0 V
1981 0 R
-31 0 V
0.050 UL
LT3
400 1991 M
1075 0 V
887 0 R
50 0 V
1.000 UL
LTb
400 1991 M
63 0 V
1949 0 R
-63 0 V
400 2160 M
31 0 V
1981 0 R
-31 0 V
0.050 UL
LT3
400 300 M
0 1860 V
1.000 UL
LTb
400 300 M
0 63 V
0 1797 R
0 -63 V
551 300 M
0 31 V
0 1829 R
0 -31 V
640 300 M
0 31 V
0 1829 R
0 -31 V
703 300 M
0 31 V
0 1829 R
0 -31 V
752 300 M
0 31 V
0 1829 R
0 -31 V
791 300 M
0 31 V
0 1829 R
0 -31 V
825 300 M
0 31 V
0 1829 R
0 -31 V
854 300 M
0 31 V
0 1829 R
0 -31 V
880 300 M
0 31 V
0 1829 R
0 -31 V
0.050 UL
LT3
903 300 M
0 1860 V
1.000 UL
LTb
903 300 M
0 63 V
0 1797 R
0 -63 V
1054 300 M
0 31 V
0 1829 R
0 -31 V
1143 300 M
0 31 V
0 1829 R
0 -31 V
1206 300 M
0 31 V
0 1829 R
0 -31 V
1255 300 M
0 31 V
0 1829 R
0 -31 V
1294 300 M
0 31 V
0 1829 R
0 -31 V
1328 300 M
0 31 V
0 1829 R
0 -31 V
1357 300 M
0 31 V
0 1829 R
0 -31 V
1383 300 M
0 31 V
0 1829 R
0 -31 V
0.050 UL
LT3
1406 300 M
0 1860 V
1.000 UL
LTb
1406 300 M
0 63 V
0 1797 R
0 -63 V
1557 300 M
0 31 V
0 1829 R
0 -31 V
1646 300 M
0 31 V
0 1829 R
0 -31 V
1709 300 M
0 31 V
0 1829 R
0 -31 V
1758 300 M
0 31 V
0 1829 R
0 -31 V
1797 300 M
0 31 V
0 1829 R
0 -31 V
1831 300 M
0 31 V
0 1829 R
0 -31 V
1860 300 M
0 31 V
0 1829 R
0 -31 V
1886 300 M
0 31 V
0 1829 R
0 -31 V
0.050 UL
LT3
1909 300 M
0 1397 V
0 400 R
0 63 V
1.000 UL
LTb
1909 300 M
0 63 V
0 1797 R
0 -63 V
2060 300 M
0 31 V
0 1829 R
0 -31 V
2149 300 M
0 31 V
0 1829 R
0 -31 V
2212 300 M
0 31 V
0 1829 R
0 -31 V
2261 300 M
0 31 V
0 1829 R
0 -31 V
2300 300 M
0 31 V
0 1829 R
0 -31 V
2334 300 M
0 31 V
0 1829 R
0 -31 V
2363 300 M
0 31 V
0 1829 R
0 -31 V
2389 300 M
0 31 V
0 1829 R
0 -31 V
0.050 UL
LT3
2412 300 M
0 1860 V
1.000 UL
LTb
2412 300 M
0 63 V
0 1797 R
0 -63 V
1.000 UL
LTb
400 300 M
2012 0 V
0 1860 V
-2012 0 V
400 300 L
1.000 UL
LTb
1475 1697 M
0 400 V
887 0 V
0 -400 V
-887 0 V
0 400 R
887 0 V
0.000 UP
1.000 UL
LT0
2075 2047 M
237 0 V
460 300 M
0 1766 V
41 67 V
53 -199 V
61 137 V
67 -86 V
69 -22 V
69 -55 V
68 -40 V
66 -31 V
63 -50 V
61 -44 V
58 -48 V
56 -37 V
53 -33 V
51 -34 V
49 -30 V
46 -36 V
45 -30 V
44 -23 V
41 -32 V
41 -27 V
38 -29 V
38 -24 V
36 -22 V
35 -23 V
34 -28 V
0 -987 V
460 2066 Circle
501 2133 Circle
554 1934 Circle
615 2071 Circle
682 1985 Circle
751 1963 Circle
820 1908 Circle
888 1868 Circle
954 1837 Circle
1017 1787 Circle
1078 1743 Circle
1136 1695 Circle
1192 1658 Circle
1245 1625 Circle
1296 1591 Circle
1345 1561 Circle
1391 1525 Circle
1436 1495 Circle
1480 1472 Circle
1521 1440 Circle
1562 1413 Circle
1600 1384 Circle
1638 1360 Circle
1674 1338 Circle
1709 1315 Circle
1743 1287 Circle
2193 2047 Circle
0.000 UP
1.000 UL
LT1
2075 1947 M
237 0 V
615 300 M
0 1614 V
67 -42 V
69 10 V
69 14 V
68 -33 V
66 -26 V
63 -51 V
61 -47 V
58 -55 V
56 -37 V
53 -36 V
51 -46 V
49 -33 V
46 -57 V
45 -29 V
44 -39 V
41 -30 V
41 -41 V
38 -33 V
38 -38 V
36 -30 V
35 -35 V
34 -44 V
0 -856 V
615 1914 TriU
682 1872 TriU
751 1882 TriU
820 1896 TriU
888 1863 TriU
954 1837 TriU
1017 1786 TriU
1078 1739 TriU
1136 1684 TriU
1192 1647 TriU
1245 1611 TriU
1296 1565 TriU
1345 1532 TriU
1391 1475 TriU
1436 1446 TriU
1480 1407 TriU
1521 1377 TriU
1562 1336 TriU
1600 1303 TriU
1638 1265 TriU
1674 1235 TriU
1709 1200 TriU
1743 1156 TriU
2193 1947 TriU
0.000 UP
1.000 UL
LT4
2075 1847 M
237 0 V
460 300 M
0 1766 V
41 67 V
53 -199 V
61 128 V
67 -94 V
69 -35 V
69 -163 V
68 -104 V
888 300 L
129 0 R
0 1160 V
61 60 V
58 30 V
56 -40 V
53 -14 V
51 5 V
49 -23 V
46 -4 V
45 -31 V
44 -9 V
41 -34 V
41 -19 V
38 -27 V
38 -18 V
36 -19 V
35 -19 V
34 -24 V
0 -974 V
460 2066 Box
501 2133 Box
554 1934 Box
615 2062 Box
682 1968 Box
751 1933 Box
820 1770 Box
888 1666 Box
1017 1460 Box
1078 1520 Box
1136 1550 Box
1192 1510 Box
1245 1496 Box
1296 1501 Box
1345 1478 Box
1391 1474 Box
1436 1443 Box
1480 1434 Box
1521 1400 Box
1562 1381 Box
1600 1354 Box
1638 1336 Box
1674 1317 Box
1709 1298 Box
1743 1274 Box
2193 1847 Box
0.000 UP
1.000 UL
LT3
2075 1747 M
237 0 V
432 300 M
0 1827 V
28 20 V
41 -94 V
53 42 V
61 -56 V
67 -65 V
69 -65 V
69 -126 V
68 -147 V
66 -115 V
63 -61 V
61 -7 V
58 141 V
56 -58 V
53 25 V
51 -38 V
49 -7 V
46 -26 V
45 -23 V
44 -41 V
41 -21 V
41 -23 V
38 -33 V
38 -32 V
36 -31 V
35 -37 V
34 -36 V
33 -71 V
32 -60 V
31 -84 V
31 -18 V
29 -58 V
29 -34 V
28 -40 V
27 -51 V
27 -44 V
26 -35 V
25 -7 V
25 -37 V
24 -52 V
24 -64 V
23 14 V
23 -67 V
22 30 V
22 -163 V
21 81 V
21 10 V
21 -58 V
20 7 V
20 -57 V
20 -31 V
19 6 V
19 -29 V
1 -1 V
432 2127 Dia
460 2147 Dia
501 2053 Dia
554 2095 Dia
615 2039 Dia
682 1974 Dia
751 1909 Dia
820 1783 Dia
888 1636 Dia
954 1521 Dia
1017 1460 Dia
1078 1453 Dia
1136 1594 Dia
1192 1536 Dia
1245 1561 Dia
1296 1523 Dia
1345 1516 Dia
1391 1490 Dia
1436 1467 Dia
1480 1426 Dia
1521 1405 Dia
1562 1382 Dia
1600 1349 Dia
1638 1317 Dia
1674 1286 Dia
1709 1249 Dia
1743 1213 Dia
1776 1142 Dia
1808 1082 Dia
1839 998 Dia
1870 980 Dia
1899 922 Dia
1928 888 Dia
1956 848 Dia
1983 797 Dia
2010 753 Dia
2036 718 Dia
2061 711 Dia
2086 674 Dia
2110 622 Dia
2134 558 Dia
2157 572 Dia
2180 505 Dia
2202 535 Dia
2224 372 Dia
2245 453 Dia
2266 463 Dia
2287 405 Dia
2307 412 Dia
2327 355 Dia
2347 324 Dia
2366 330 Dia
2385 301 Dia
2193 1747 Dia
stroke
grestore
end
showpage
}}%
\put(2025,1747){\makebox(0,0)[r]{BP, $\tilde{t}=\tilde{t}_{\rm max}$}}%
\put(2025,1847){\makebox(0,0)[r]{BP, $\tilde{t}=0$}}%
\put(2025,1947){\makebox(0,0)[r]{EP, $\tilde{t}=0$}}%
\put(2025,2047){\makebox(0,0)[r]{TP, $\tilde{t}=0$}}%
\put(1406,2310){\makebox(0,0){\textsf{Number density of stars $\tilde{\rho}(\tilde{r})$}}}%
\put(1406,50){\makebox(0,0){\sffamily $\tilde{r}$}}%
\put(100,1230){%
\special{ps: gsave currentpoint currentpoint translate
270 rotate neg exch neg exch translate}%
\makebox(0,0)[b]{\shortstack{\sffamily $\tilde{\rho}$}}%
\special{ps: currentpoint grestore moveto}%
}%
\put(2412,200){\makebox(0,0){1000}}%
\put(1909,200){\makebox(0,0){100}}%
\put(1406,200){\makebox(0,0){10}}%
\put(903,200){\makebox(0,0){1}}%
\put(400,200){\makebox(0,0){0.1}}%
\put(350,1991){\makebox(0,0)[r]{0.01}}%
\put(350,1653){\makebox(0,0)[r]{0.0001}}%
\put(350,1315){\makebox(0,0)[r]{1e-06}}%
\put(350,976){\makebox(0,0)[r]{1e-08}}%
\put(350,638){\makebox(0,0)[r]{1e-10}}%
\put(350,300){\makebox(0,0)[r]{1e-12}}%
\end{picture}%
\endgroup
 

%% file: bhb_hlp.tex
\begin{figure*}[!h]
\hspace{5mm}%
\resizebox{71mm}{!}{\includegraphics[bb = 81 165 533 635, angle =
-90]{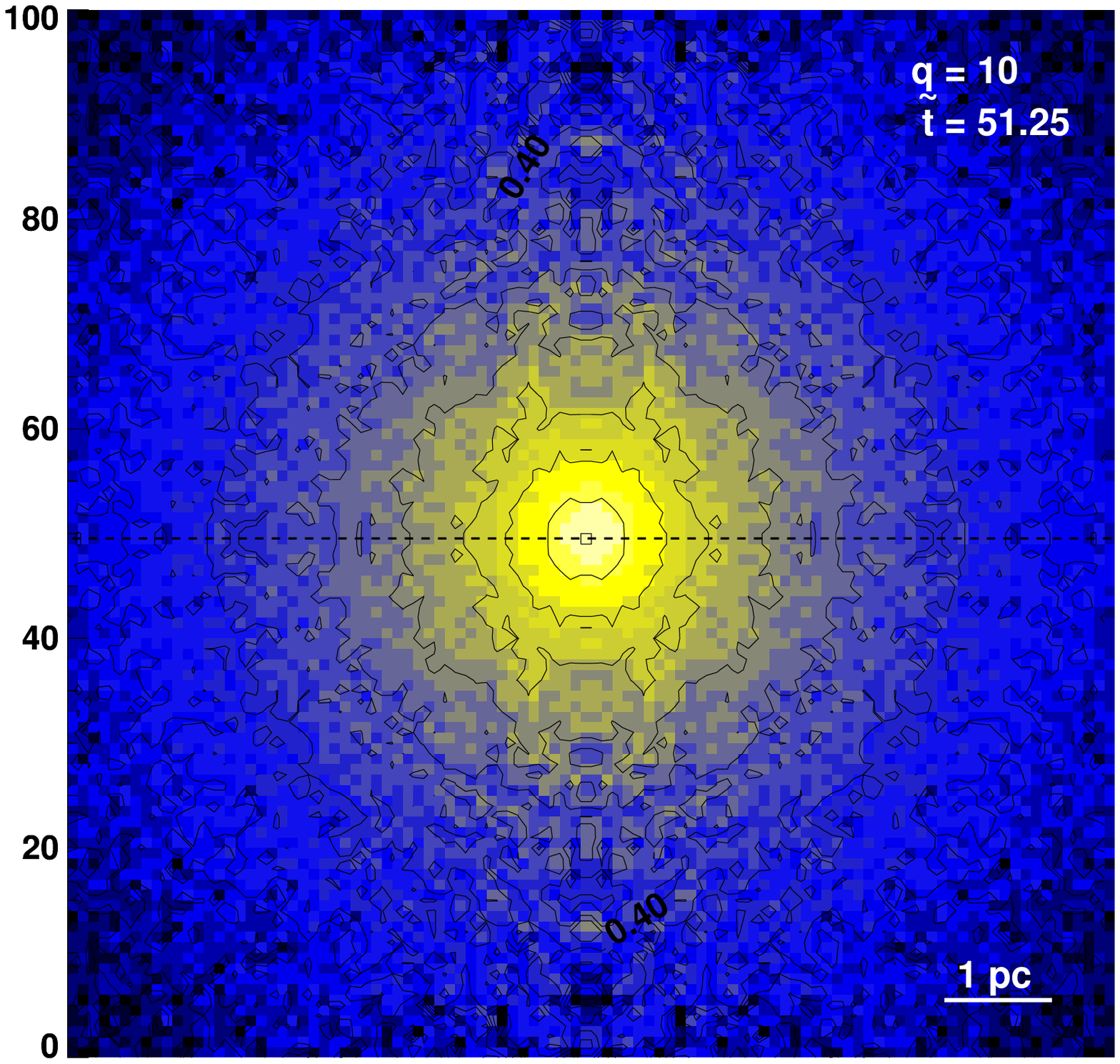}}%
\hspace{2.6cm}%
\resizebox{71mm}{!}{\includegraphics[bb = 81 165 533 635, angle =
-90]{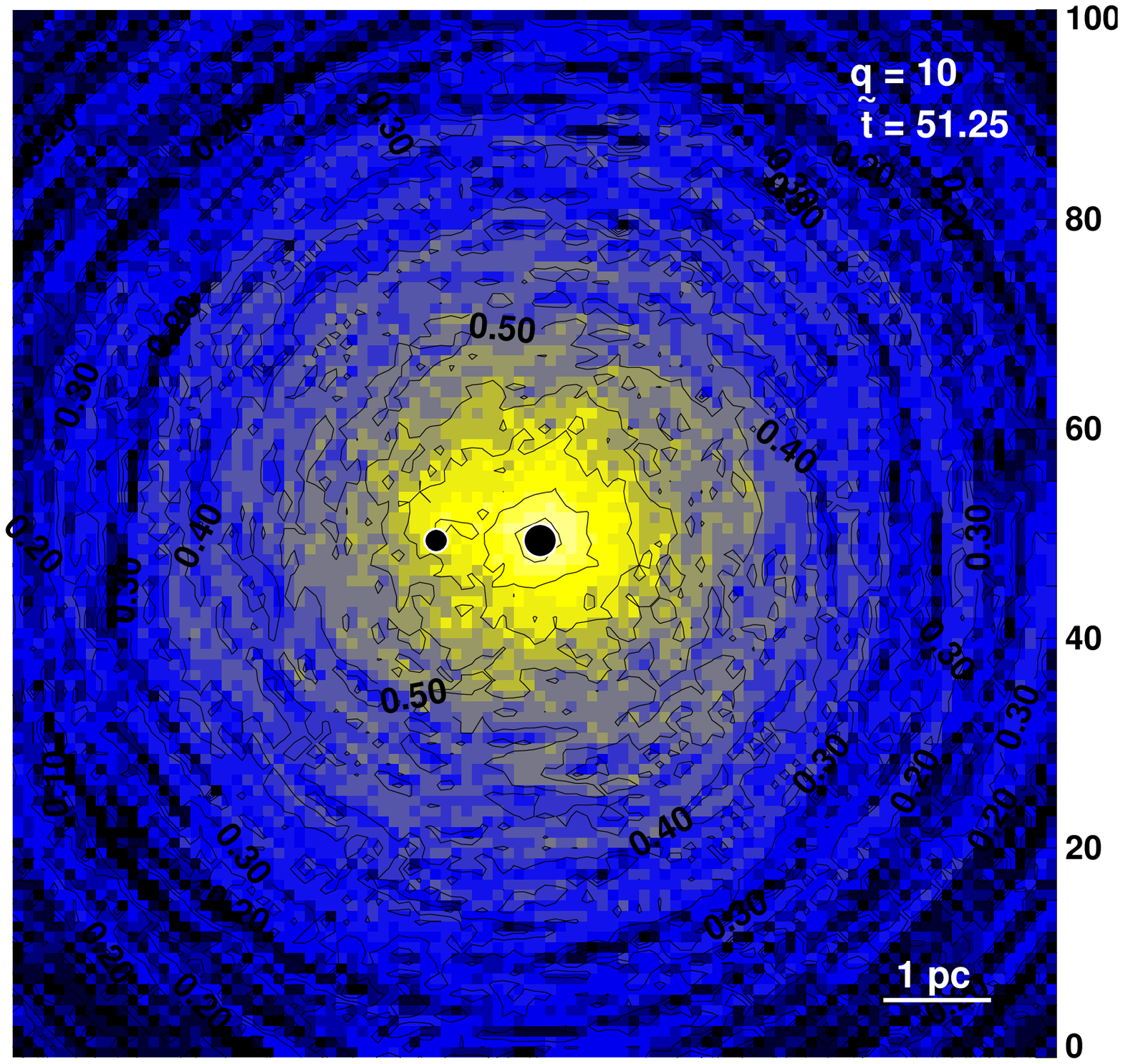}}%
\end{figure*}
\begin{figure*}[!h]
\vspace{-3.5mm}
\hspace{5mm}%
\resizebox{71mm}{!}{\includegraphics[bb = 81 165 533 635, angle =
-90]{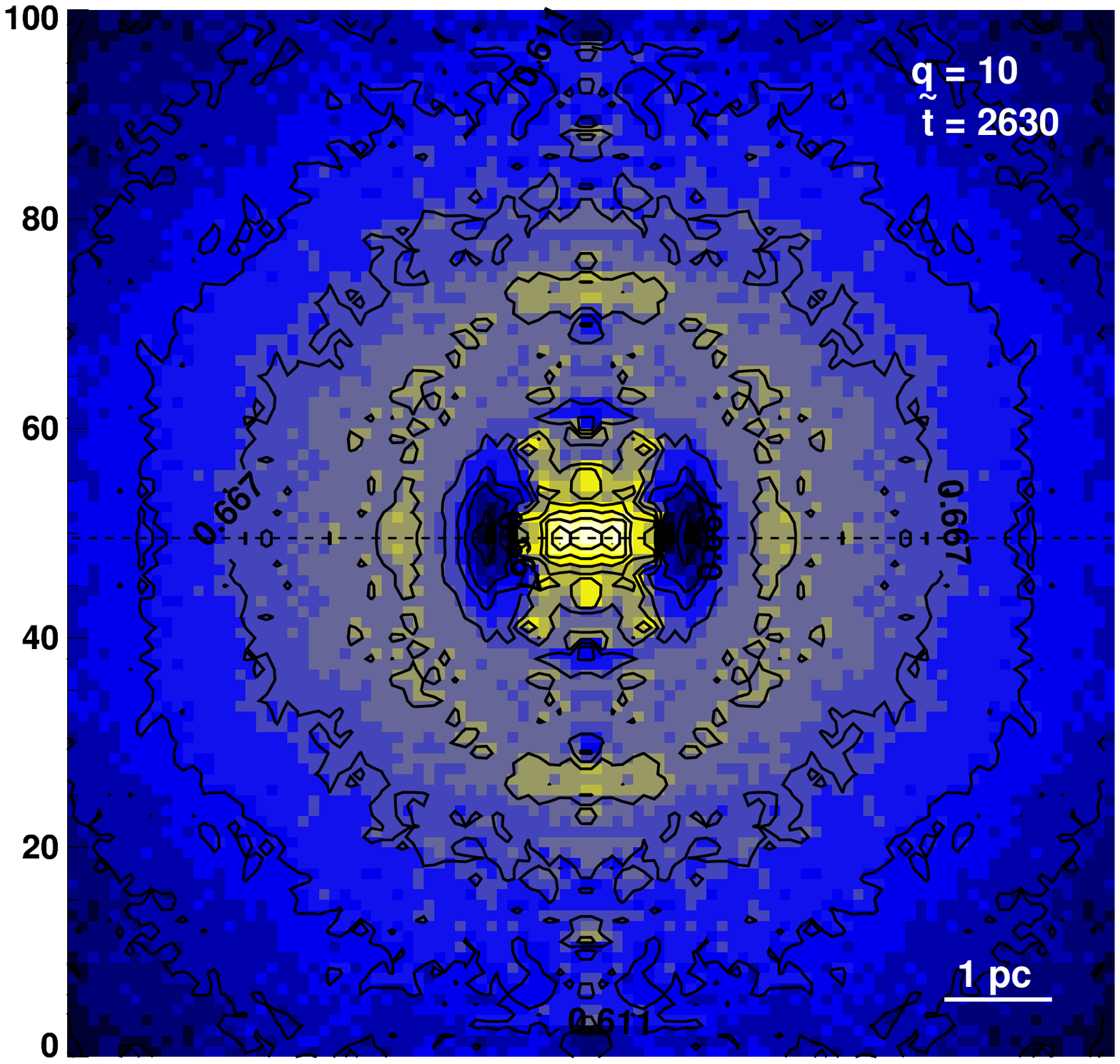}}%
\hspace{26.2mm}%
\resizebox{71mm}{!}{\includegraphics[bb = 81 165 533 635, angle =
-90]{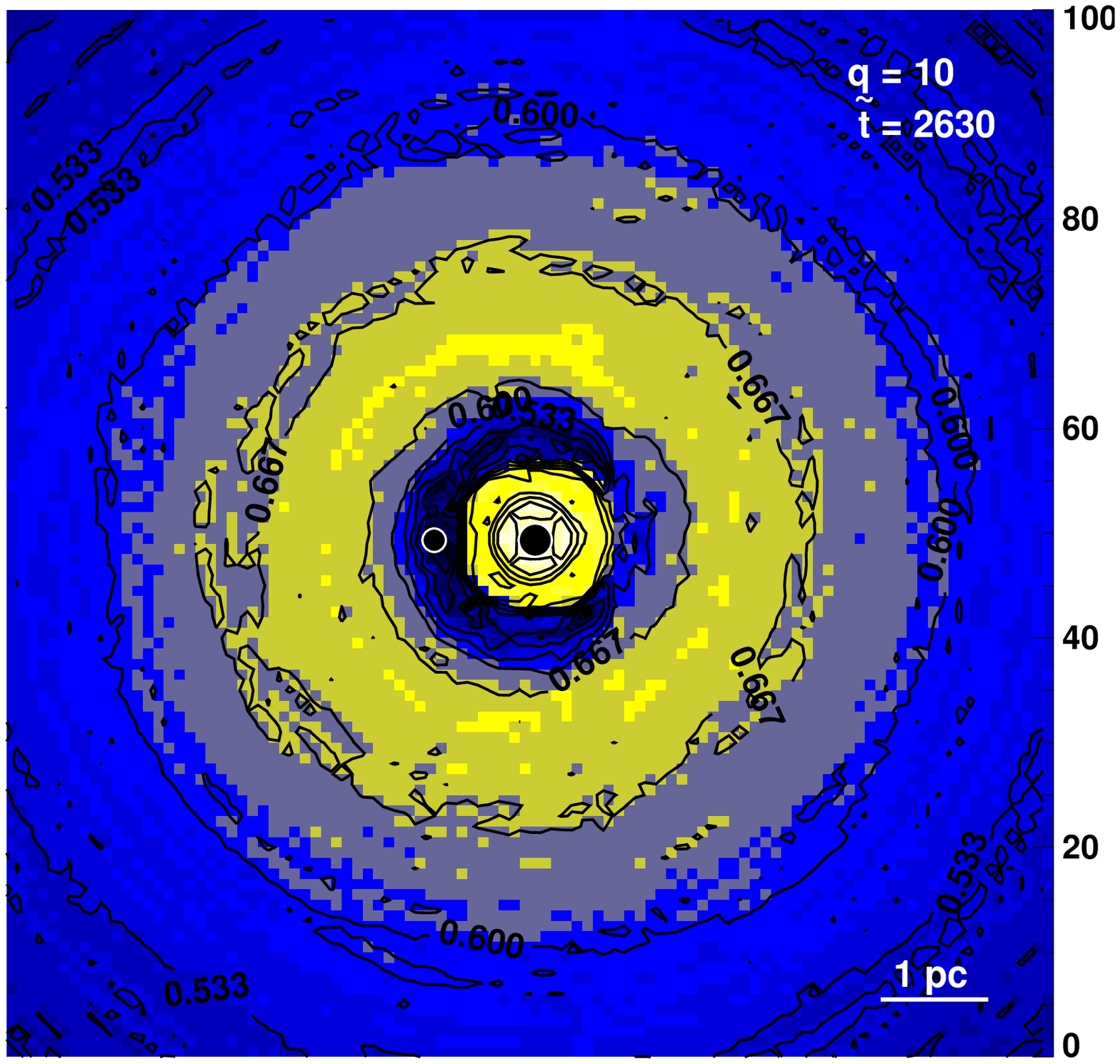}}%
\end{figure*}
\begin{figure*}[!h]
\vspace{-3.75mm}
\hspace{4.75mm}%
\resizebox{72.5mm}{!}{\includegraphics[bb = 10 7 490
  505]{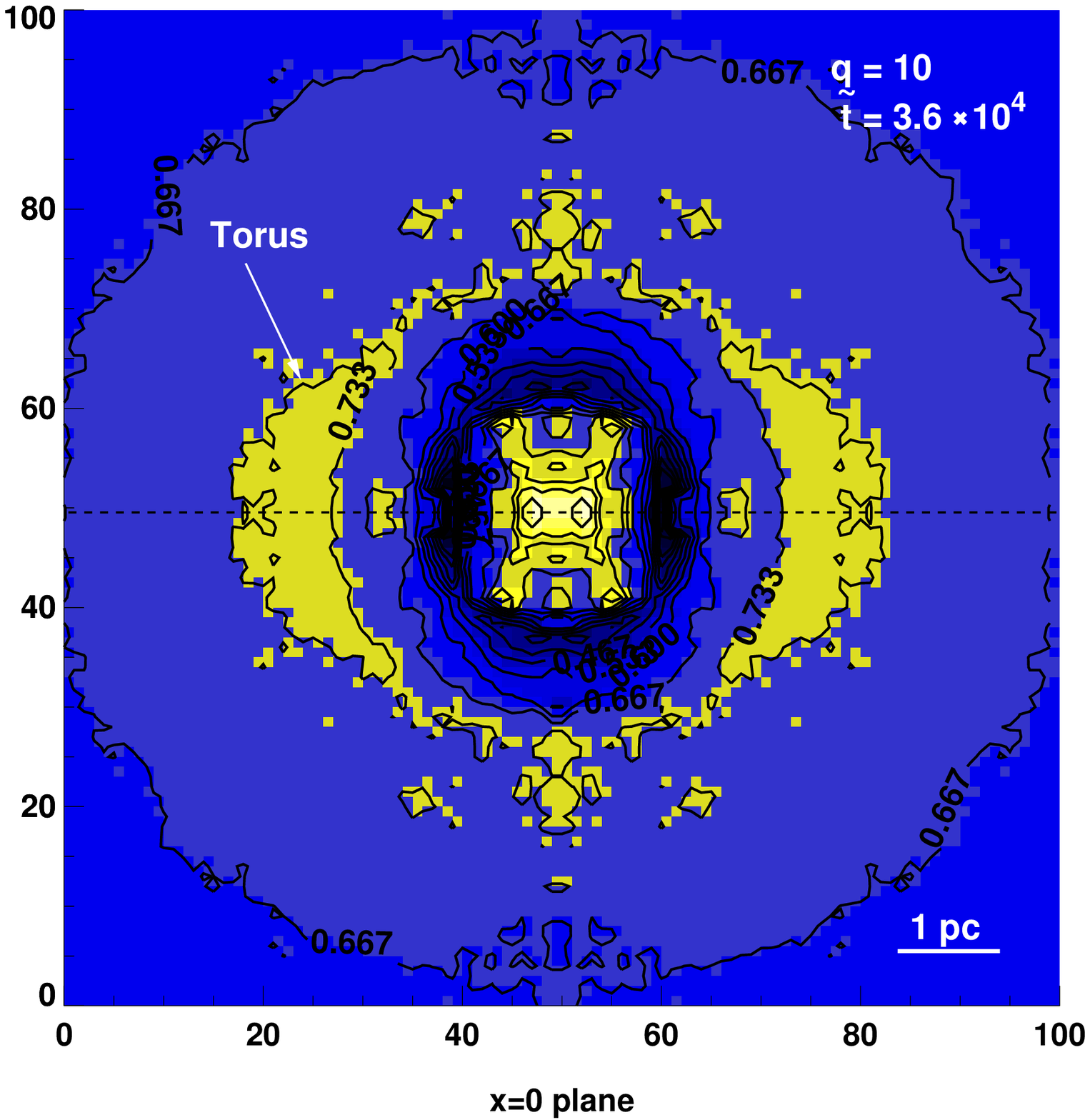}}%
\hspace{24.2mm}%
\resizebox{72.5mm}{!}{\includegraphics[bb = 10 7 490
  505]{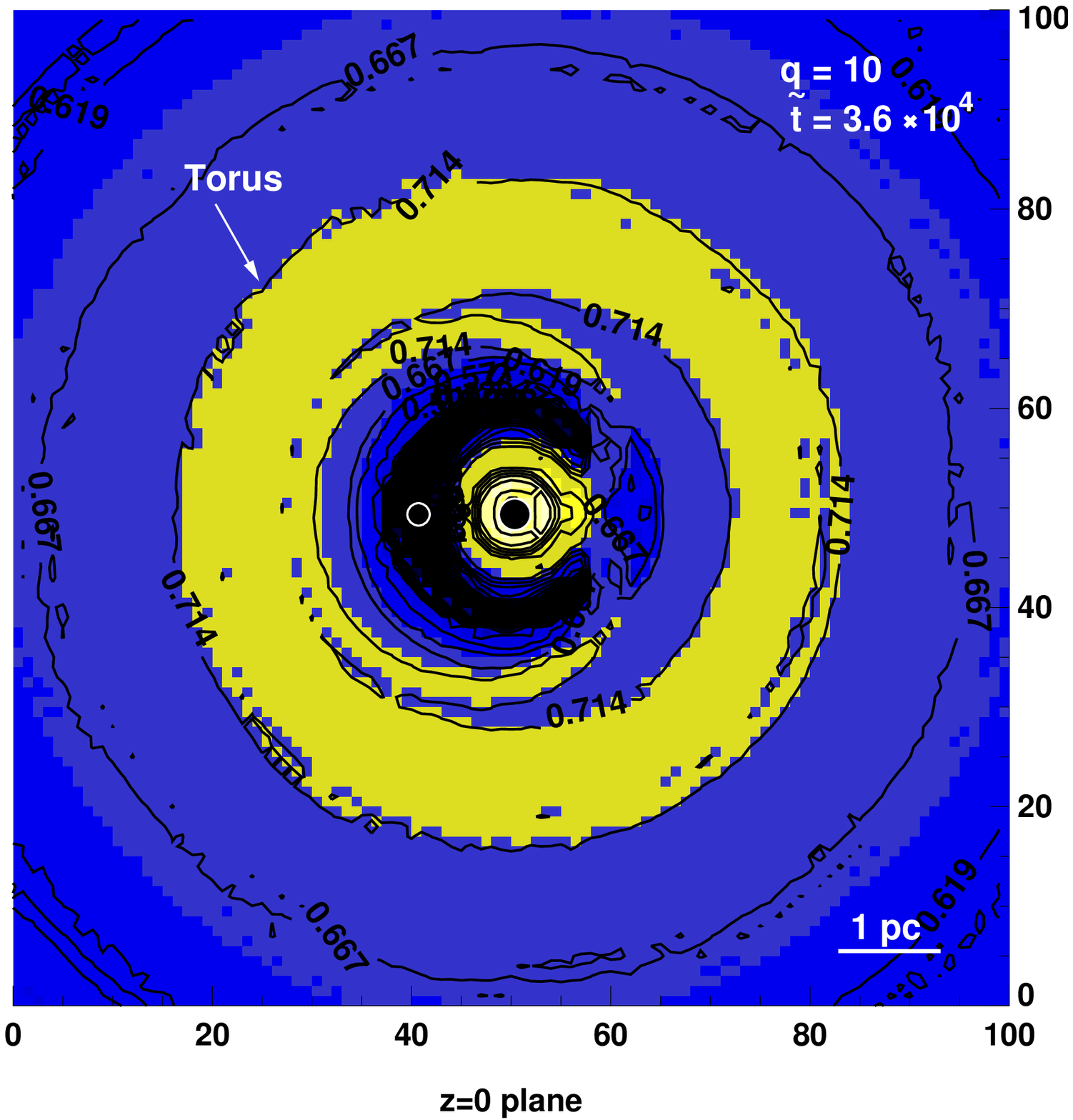}}%
\vspace{-2.0mm}
\caption[]{
Cuts through the stellar density in the comoving frame are shown with
contours logarithmically scaled. The
right panel displays the equatorial plane (BHs marked by black spots).
Perpendicular to it the $x=0$-plane is shown (left panel), with
the $y$-axis drawn as dashed line, so that the BHs are in front and
behind the paper-plane. The
initial distribution is a gaussian and the massratio is
$10$. Time increases from top to bottom (indicated by
$\tilde{t}$). After initially stars close to $M_2$'s orbit
($\approx 1\,{\rm pc}$) are ejected ($2$nd row) also the polar regions
are depleted and a torus in the equatorial plane finally emerges at
$r\sim 3\,{\rm pc}$ ($3$rd row).
}
\label{f_cont_tser}
\end{figure*}
\clearpage

%% file: bhb_fig08i.tex
\begingroup%
  \makeatletter%
  \newcommand{\GNUPLOTspecial}{%
    \@sanitize\catcode`\%=14\relax\special}%
  \setlength{\unitlength}{0.1bp}%
\begin{picture}(2412,2160)(0,0)%
\special{psfile=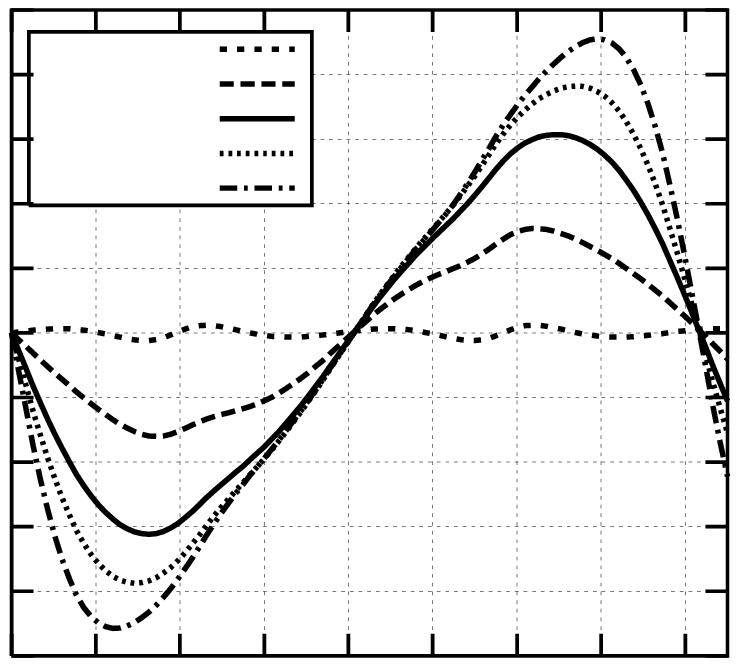 llx=0 lly=0 urx=482 ury=504 rwi=4820}
\put(900,1647){\makebox(0,0)[r]{$\theta=\pi/3$}}%
\put(900,1747){\makebox(0,0)[r]{$\theta=7\pi/24$}}%
\put(900,1847){\makebox(0,0)[r]{$\theta=\pi/4$}}%
\put(900,1947){\makebox(0,0)[r]{$\theta=\pi/6$}}%
\put(900,2047){\makebox(0,0)[r]{$\theta=0$}}%
\put(1381,2310){\makebox(0,0){\textsf{Torque divided by radius, $\tilde{r}=3$}}}%
\put(1381,50){\makebox(0,0){$\tilde{t}$}}%
\put(50,1230){%
\special{ps: gsave currentpoint currentpoint translate
270 rotate neg exch neg exch translate}%
\makebox(0,0)[b]{\shortstack{\sffamily $\dot{\tilde{L}}_{z,1}/\tilde{r}$}}%
\special{ps: currentpoint grestore moveto}%
}%
\put(2291,200){\makebox(0,0){16}}%
\put(2048,200){\makebox(0,0){14}}%
\put(1806,200){\makebox(0,0){12}}%
\put(1563,200){\makebox(0,0){10}}%
\put(1320,200){\makebox(0,0){8}}%
\put(1078,200){\makebox(0,0){6}}%
\put(835,200){\makebox(0,0){4}}%
\put(593,200){\makebox(0,0){2}}%
\put(350,200){\makebox(0,0){0}}%
\put(300,2160){\makebox(0,0)[r]{0.1}}%
\put(300,1974){\makebox(0,0)[r]{0.08}}%
\put(300,1788){\makebox(0,0)[r]{0.06}}%
\put(300,1602){\makebox(0,0)[r]{0.04}}%
\put(300,1416){\makebox(0,0)[r]{0.02}}%
\put(300,1230){\makebox(0,0)[r]{0}}%
\put(300,1044){\makebox(0,0)[r]{-0.02}}%
\put(300,858){\makebox(0,0)[r]{-0.04}}%
\put(300,672){\makebox(0,0)[r]{-0.06}}%
\put(300,486){\makebox(0,0)[r]{-0.08}}%
\put(300,300){\makebox(0,0)[r]{-0.1}}%
\end{picture}%
\endgroup
 

%% file: bhb_fig09i.tex
\begingroup%
  \makeatletter%
  \newcommand{\GNUPLOTspecial}{%
    \@sanitize\catcode`\%=14\relax\special}%
  \setlength{\unitlength}{0.1bp}%
\begin{picture}(2412,2160)(0,0)%
\special{psfile=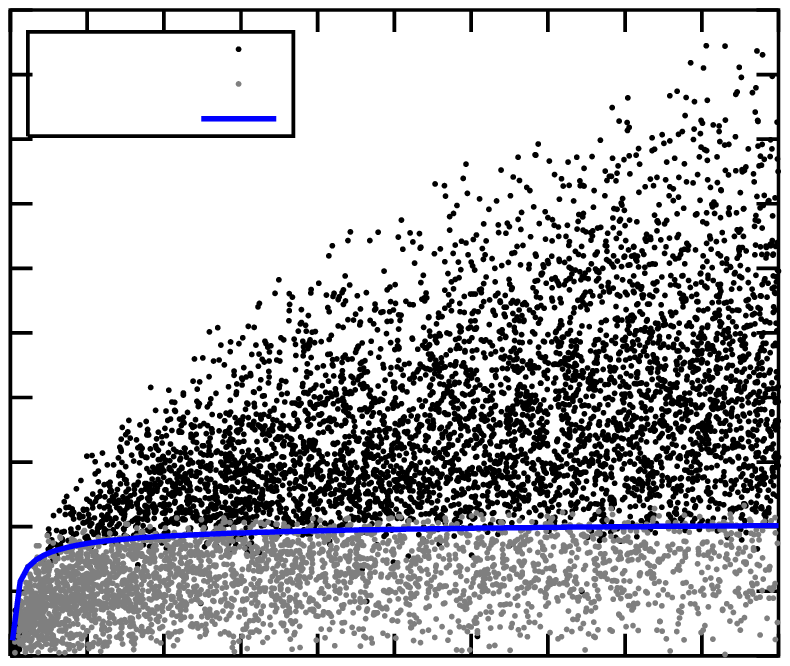 llx=0 lly=0 urx=482 ury=504 rwi=4820}
\put(700,1847){\makebox(0,0)[r]{transition}}%
\put(700,1947){\makebox(0,0)[r]{EP}}%
\put(700,2047){\makebox(0,0)[r]{BP}}%
\put(1306,2310){\makebox(0,0){\textsf{$\tilde{L}$ plotted versus $\tilde{r}$ at $\tilde{t}=0$, for $q=10$}}}%
\put(1306,50){\makebox(0,0){\sffamily $\tilde{r}$}}%
\put(50,1230){%
\special{ps: gsave currentpoint currentpoint translate
270 rotate neg exch neg exch translate}%
\makebox(0,0)[b]{\shortstack{\sffamily $\tilde{L}$}}%
\special{ps: currentpoint grestore moveto}%
}%
\put(2412,200){\makebox(0,0){50}}%
\put(2191,200){\makebox(0,0){45}}%
\put(1970,200){\makebox(0,0){40}}%
\put(1748,200){\makebox(0,0){35}}%
\put(1527,200){\makebox(0,0){30}}%
\put(1306,200){\makebox(0,0){25}}%
\put(1085,200){\makebox(0,0){20}}%
\put(864,200){\makebox(0,0){15}}%
\put(642,200){\makebox(0,0){10}}%
\put(421,200){\makebox(0,0){5}}%
\put(200,200){\makebox(0,0){0}}%
\put(150,2160){\makebox(0,0)[r]{10}}%
\put(150,1974){\makebox(0,0)[r]{9}}%
\put(150,1788){\makebox(0,0)[r]{8}}%
\put(150,1602){\makebox(0,0)[r]{7}}%
\put(150,1416){\makebox(0,0)[r]{6}}%
\put(150,1230){\makebox(0,0)[r]{5}}%
\put(150,1044){\makebox(0,0)[r]{4}}%
\put(150,858){\makebox(0,0)[r]{3}}%
\put(150,672){\makebox(0,0)[r]{2}}%
\put(150,486){\makebox(0,0)[r]{1}}%
\put(150,300){\makebox(0,0)[r]{0}}%
\end{picture}%
\endgroup
 

%% file: bhb_fig10i.tex
\begingroup%
  \makeatletter%
  \newcommand{\GNUPLOTspecial}{%
    \@sanitize\catcode`\%=14\relax\special}%
  \setlength{\unitlength}{0.1bp}%
\begin{picture}(2412,2160)(0,0)%
\special{psfile=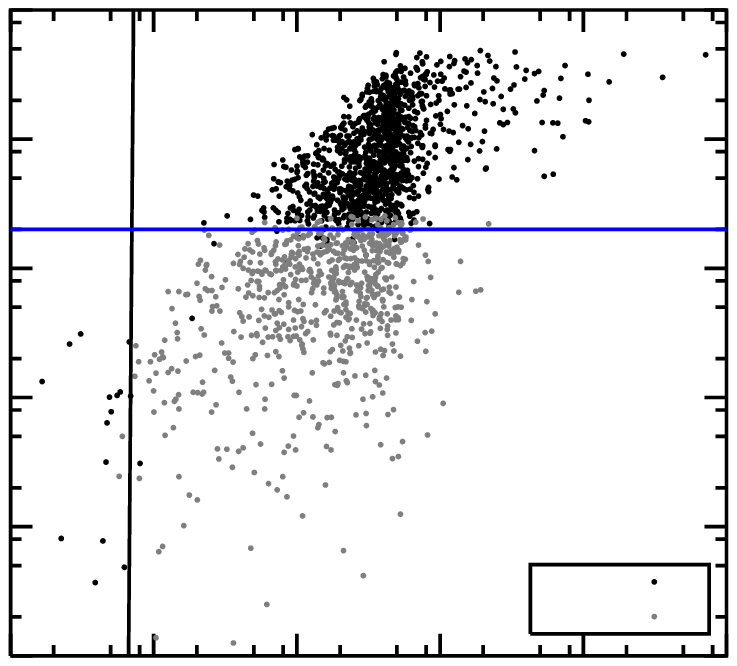 llx=0 lly=0 urx=482 ury=504 rwi=4820}
\put(2047,413){\makebox(0,0)[r]{EP}}%
\put(2047,513){\makebox(0,0)[r]{BP}}%
\put(1381,2310){\makebox(0,0){\textsf{$\tilde{r}_{-}$ plotted versus $\tilde{r}_{+}$ at $\tilde{t}=0$, for $q=10$}}}%
\put(1381,50){\makebox(0,0){\sffamily $\tilde{r}_{+}$}}%
\put(50,1230){%
\special{ps: gsave currentpoint currentpoint translate
270 rotate neg exch neg exch translate}%
\makebox(0,0)[b]{\shortstack{\sffamily $\tilde{r}_{-}$}}%
\special{ps: currentpoint grestore moveto}%
}%
\put(2412,200){\makebox(0,0){10000}}%
\put(2000,200){\makebox(0,0){1000}}%
\put(1587,200){\makebox(0,0){100}}%
\put(1175,200){\makebox(0,0){10}}%
\put(762,200){\makebox(0,0){1}}%
\put(350,200){\makebox(0,0){0.1}}%
\put(300,2160){\makebox(0,0)[r]{100}}%
\put(300,1788){\makebox(0,0)[r]{10}}%
\put(300,1416){\makebox(0,0)[r]{1}}%
\put(300,1044){\makebox(0,0)[r]{0.1}}%
\put(300,672){\makebox(0,0)[r]{0.01}}%
\put(300,300){\makebox(0,0)[r]{0.001}}%
\end{picture}%
\endgroup
 

%% file: bhb_fig11.tex
\begingroup%
  \makeatletter%
  \newcommand{\GNUPLOTspecial}{%
    \@sanitize\catcode`\%=14\relax\special}%
  \setlength{\unitlength}{0.1bp}%
{\GNUPLOTspecial{!
/gnudict 256 dict def
gnudict begin
/Color false def
/Solid false def
/gnulinewidth 5.000 def
/userlinewidth gnulinewidth def
/vshift -33 def
/dl {10 mul} def
/hpt_ 31.5 def
/vpt_ 31.5 def
/hpt hpt_ def
/vpt vpt_ def
/M {moveto} bind def
/L {lineto} bind def
/R {rmoveto} bind def
/V {rlineto} bind def
/vpt2 vpt 2 mul def
/hpt2 hpt 2 mul def
/Lshow { currentpoint stroke M
  0 vshift R show } def
/Rshow { currentpoint stroke M
  dup stringwidth pop neg vshift R show } def
/Cshow { currentpoint stroke M
  dup stringwidth pop -2 div vshift R show } def
/UP { dup vpt_ mul /vpt exch def hpt_ mul /hpt exch def
  /hpt2 hpt 2 mul def /vpt2 vpt 2 mul def } def
/DL { Color {setrgbcolor Solid {pop []} if 0 setdash }
 {pop pop pop Solid {pop []} if 0 setdash} ifelse } def
/BL { stroke gnulinewidth 2 mul setlinewidth } def
/AL { stroke gnulinewidth 2 div setlinewidth } def
/UL { gnulinewidth mul /userlinewidth exch def } def
/PL { stroke userlinewidth setlinewidth } def
/LTb { BL [] 0 0 0 DL } def
/LTa { AL [1 dl 2 dl] 0 setdash 0 0 0 setrgbcolor } def
/LT0 { PL [] 1 0 0 DL } def
/LT1 { PL [4 dl 2 dl] 0 1 0 DL } def
/LT2 { PL [2 dl 3 dl] 0 0 1 DL } def
/LT3 { PL [1 dl 1.5 dl] 1 0 1 DL } def
/LT4 { PL [5 dl 2 dl 1 dl 2 dl] 0 1 1 DL } def
/LT5 { PL [4 dl 3 dl 1 dl 3 dl] 1 1 0 DL } def
/LT6 { PL [2 dl 2 dl 2 dl 4 dl] 0 0 0 DL } def
/LT7 { PL [2 dl 2 dl 2 dl 2 dl 2 dl 4 dl] 1 0.3 0 DL } def
/LT8 { PL [2 dl 2 dl 2 dl 2 dl 2 dl 2 dl 2 dl 4 dl] 0.5 0.5 0.5 DL } def
/Pnt { stroke [] 0 setdash
   gsave 1 setlinecap M 0 0 V stroke grestore } def
/Dia { stroke [] 0 setdash 2 copy vpt add M
  hpt neg vpt neg V hpt vpt neg V
  hpt vpt V hpt neg vpt V closepath stroke
  Pnt } def
/Pls { stroke [] 0 setdash vpt sub M 0 vpt2 V
  currentpoint stroke M
  hpt neg vpt neg R hpt2 0 V stroke
  } def
/Box { stroke [] 0 setdash 2 copy exch hpt sub exch vpt add M
  0 vpt2 neg V hpt2 0 V 0 vpt2 V
  hpt2 neg 0 V closepath stroke
  Pnt } def
/Crs { stroke [] 0 setdash exch hpt sub exch vpt add M
  hpt2 vpt2 neg V currentpoint stroke M
  hpt2 neg 0 R hpt2 vpt2 V stroke } def
/TriU { stroke [] 0 setdash 2 copy vpt 1.12 mul add M
  hpt neg vpt -1.62 mul V
  hpt 2 mul 0 V
  hpt neg vpt 1.62 mul V closepath stroke
  Pnt  } def
/Star { 2 copy Pls Crs } def
/BoxF { stroke [] 0 setdash exch hpt sub exch vpt add M
  0 vpt2 neg V  hpt2 0 V  0 vpt2 V
  hpt2 neg 0 V  closepath fill } def
/TriUF { stroke [] 0 setdash vpt 1.12 mul add M
  hpt neg vpt -1.62 mul V
  hpt 2 mul 0 V
  hpt neg vpt 1.62 mul V closepath fill } def
/TriD { stroke [] 0 setdash 2 copy vpt 1.12 mul sub M
  hpt neg vpt 1.62 mul V
  hpt 2 mul 0 V
  hpt neg vpt -1.62 mul V closepath stroke
  Pnt  } def
/TriDF { stroke [] 0 setdash vpt 1.12 mul sub M
  hpt neg vpt 1.62 mul V
  hpt 2 mul 0 V
  hpt neg vpt -1.62 mul V closepath fill} def
/DiaF { stroke [] 0 setdash vpt add M
  hpt neg vpt neg V hpt vpt neg V
  hpt vpt V hpt neg vpt V closepath fill } def
/Pent { stroke [] 0 setdash 2 copy gsave
  translate 0 hpt M 4 {72 rotate 0 hpt L} repeat
  closepath stroke grestore Pnt } def
/PentF { stroke [] 0 setdash gsave
  translate 0 hpt M 4 {72 rotate 0 hpt L} repeat
  closepath fill grestore } def
/Circle { stroke [] 0 setdash 2 copy
  hpt 0 360 arc stroke Pnt } def
/CircleF { stroke [] 0 setdash hpt 0 360 arc fill } def
/C0 { BL [] 0 setdash 2 copy moveto vpt 90 450  arc } bind def
/C1 { BL [] 0 setdash 2 copy        moveto
       2 copy  vpt 0 90 arc closepath fill
               vpt 0 360 arc closepath } bind def
/C2 { BL [] 0 setdash 2 copy moveto
       2 copy  vpt 90 180 arc closepath fill
               vpt 0 360 arc closepath } bind def
/C3 { BL [] 0 setdash 2 copy moveto
       2 copy  vpt 0 180 arc closepath fill
               vpt 0 360 arc closepath } bind def
/C4 { BL [] 0 setdash 2 copy moveto
       2 copy  vpt 180 270 arc closepath fill
               vpt 0 360 arc closepath } bind def
/C5 { BL [] 0 setdash 2 copy moveto
       2 copy  vpt 0 90 arc
       2 copy moveto
       2 copy  vpt 180 270 arc closepath fill
               vpt 0 360 arc } bind def
/C6 { BL [] 0 setdash 2 copy moveto
      2 copy  vpt 90 270 arc closepath fill
              vpt 0 360 arc closepath } bind def
/C7 { BL [] 0 setdash 2 copy moveto
      2 copy  vpt 0 270 arc closepath fill
              vpt 0 360 arc closepath } bind def
/C8 { BL [] 0 setdash 2 copy moveto
      2 copy vpt 270 360 arc closepath fill
              vpt 0 360 arc closepath } bind def
/C9 { BL [] 0 setdash 2 copy moveto
      2 copy  vpt 270 450 arc closepath fill
              vpt 0 360 arc closepath } bind def
/C10 { BL [] 0 setdash 2 copy 2 copy moveto vpt 270 360 arc closepath fill
       2 copy moveto
       2 copy vpt 90 180 arc closepath fill
               vpt 0 360 arc closepath } bind def
/C11 { BL [] 0 setdash 2 copy moveto
       2 copy  vpt 0 180 arc closepath fill
       2 copy moveto
       2 copy  vpt 270 360 arc closepath fill
               vpt 0 360 arc closepath } bind def
/C12 { BL [] 0 setdash 2 copy moveto
       2 copy  vpt 180 360 arc closepath fill
               vpt 0 360 arc closepath } bind def
/C13 { BL [] 0 setdash  2 copy moveto
       2 copy  vpt 0 90 arc closepath fill
       2 copy moveto
       2 copy  vpt 180 360 arc closepath fill
               vpt 0 360 arc closepath } bind def
/C14 { BL [] 0 setdash 2 copy moveto
       2 copy  vpt 90 360 arc closepath fill
               vpt 0 360 arc } bind def
/C15 { BL [] 0 setdash 2 copy vpt 0 360 arc closepath fill
               vpt 0 360 arc closepath } bind def
/Rec   { newpath 4 2 roll moveto 1 index 0 rlineto 0 exch rlineto
       neg 0 rlineto closepath } bind def
/Square { dup Rec } bind def
/Bsquare { vpt sub exch vpt sub exch vpt2 Square } bind def
/S0 { BL [] 0 setdash 2 copy moveto 0 vpt rlineto BL Bsquare } bind def
/S1 { BL [] 0 setdash 2 copy vpt Square fill Bsquare } bind def
/S2 { BL [] 0 setdash 2 copy exch vpt sub exch vpt Square fill Bsquare } bind def
/S3 { BL [] 0 setdash 2 copy exch vpt sub exch vpt2 vpt Rec fill Bsquare } bind def
/S4 { BL [] 0 setdash 2 copy exch vpt sub exch vpt sub vpt Square fill Bsquare } bind def
/S5 { BL [] 0 setdash 2 copy 2 copy vpt Square fill
       exch vpt sub exch vpt sub vpt Square fill Bsquare } bind def
/S6 { BL [] 0 setdash 2 copy exch vpt sub exch vpt sub vpt vpt2 Rec fill Bsquare } bind def
/S7 { BL [] 0 setdash 2 copy exch vpt sub exch vpt sub vpt vpt2 Rec fill
       2 copy vpt Square fill
       Bsquare } bind def
/S8 { BL [] 0 setdash 2 copy vpt sub vpt Square fill Bsquare } bind def
/S9 { BL [] 0 setdash 2 copy vpt sub vpt vpt2 Rec fill Bsquare } bind def
/S10 { BL [] 0 setdash 2 copy vpt sub vpt Square fill 2 copy exch vpt sub exch vpt Square fill
       Bsquare } bind def
/S11 { BL [] 0 setdash 2 copy vpt sub vpt Square fill 2 copy exch vpt sub exch vpt2 vpt Rec fill
       Bsquare } bind def
/S12 { BL [] 0 setdash 2 copy exch vpt sub exch vpt sub vpt2 vpt Rec fill Bsquare } bind def
/S13 { BL [] 0 setdash 2 copy exch vpt sub exch vpt sub vpt2 vpt Rec fill
       2 copy vpt Square fill Bsquare } bind def
/S14 { BL [] 0 setdash 2 copy exch vpt sub exch vpt sub vpt2 vpt Rec fill
       2 copy exch vpt sub exch vpt Square fill Bsquare } bind def
/S15 { BL [] 0 setdash 2 copy Bsquare fill Bsquare } bind def
/D0 { gsave translate 45 rotate 0 0 S0 stroke grestore } bind def
/D1 { gsave translate 45 rotate 0 0 S1 stroke grestore } bind def
/D2 { gsave translate 45 rotate 0 0 S2 stroke grestore } bind def
/D3 { gsave translate 45 rotate 0 0 S3 stroke grestore } bind def
/D4 { gsave translate 45 rotate 0 0 S4 stroke grestore } bind def
/D5 { gsave translate 45 rotate 0 0 S5 stroke grestore } bind def
/D6 { gsave translate 45 rotate 0 0 S6 stroke grestore } bind def
/D7 { gsave translate 45 rotate 0 0 S7 stroke grestore } bind def
/D8 { gsave translate 45 rotate 0 0 S8 stroke grestore } bind def
/D9 { gsave translate 45 rotate 0 0 S9 stroke grestore } bind def
/D10 { gsave translate 45 rotate 0 0 S10 stroke grestore } bind def
/D11 { gsave translate 45 rotate 0 0 S11 stroke grestore } bind def
/D12 { gsave translate 45 rotate 0 0 S12 stroke grestore } bind def
/D13 { gsave translate 45 rotate 0 0 S13 stroke grestore } bind def
/D14 { gsave translate 45 rotate 0 0 S14 stroke grestore } bind def
/D15 { gsave translate 45 rotate 0 0 S15 stroke grestore } bind def
/DiaE { stroke [] 0 setdash vpt add M
  hpt neg vpt neg V hpt vpt neg V
  hpt vpt V hpt neg vpt V closepath stroke } def
/BoxE { stroke [] 0 setdash exch hpt sub exch vpt add M
  0 vpt2 neg V hpt2 0 V 0 vpt2 V
  hpt2 neg 0 V closepath stroke } def
/TriUE { stroke [] 0 setdash vpt 1.12 mul add M
  hpt neg vpt -1.62 mul V
  hpt 2 mul 0 V
  hpt neg vpt 1.62 mul V closepath stroke } def
/TriDE { stroke [] 0 setdash vpt 1.12 mul sub M
  hpt neg vpt 1.62 mul V
  hpt 2 mul 0 V
  hpt neg vpt -1.62 mul V closepath stroke } def
/PentE { stroke [] 0 setdash gsave
  translate 0 hpt M 4 {72 rotate 0 hpt L} repeat
  closepath stroke grestore } def
/CircE { stroke [] 0 setdash 
  hpt 0 360 arc stroke } def
/Opaque { gsave closepath 1 setgray fill grestore 0 setgray closepath } def
/DiaW { stroke [] 0 setdash vpt add M
  hpt neg vpt neg V hpt vpt neg V
  hpt vpt V hpt neg vpt V Opaque stroke } def
/BoxW { stroke [] 0 setdash exch hpt sub exch vpt add M
  0 vpt2 neg V hpt2 0 V 0 vpt2 V
  hpt2 neg 0 V Opaque stroke } def
/TriUW { stroke [] 0 setdash vpt 1.12 mul add M
  hpt neg vpt -1.62 mul V
  hpt 2 mul 0 V
  hpt neg vpt 1.62 mul V Opaque stroke } def
/TriDW { stroke [] 0 setdash vpt 1.12 mul sub M
  hpt neg vpt 1.62 mul V
  hpt 2 mul 0 V
  hpt neg vpt -1.62 mul V Opaque stroke } def
/PentW { stroke [] 0 setdash gsave
  translate 0 hpt M 4 {72 rotate 0 hpt L} repeat
  Opaque stroke grestore } def
/CircW { stroke [] 0 setdash 
  hpt 0 360 arc Opaque stroke } def
/BoxFill { gsave Rec 1 setgray fill grestore } def
end
}}%
\begin{picture}(2412,2160)(0,0)%
{\GNUPLOTspecial{"
gnudict begin
gsave
0 0 translate
0.100 0.100 scale
0 setgray
newpath
1.000 UL
LTb
0.050 UL
LT3
200 300 M
2212 0 V
1.000 UL
LTb
200 300 M
63 0 V
2149 0 R
-63 0 V
0.050 UL
LT3
200 610 M
1275 0 V
887 0 R
50 0 V
1.000 UL
LTb
200 610 M
63 0 V
2149 0 R
-63 0 V
0.050 UL
LT3
200 920 M
2212 0 V
1.000 UL
LTb
200 920 M
63 0 V
2149 0 R
-63 0 V
0.050 UL
LT3
200 1230 M
2212 0 V
1.000 UL
LTb
200 1230 M
63 0 V
2149 0 R
-63 0 V
0.050 UL
LT3
200 1540 M
2212 0 V
1.000 UL
LTb
200 1540 M
63 0 V
2149 0 R
-63 0 V
0.050 UL
LT3
200 1850 M
2212 0 V
1.000 UL
LTb
200 1850 M
63 0 V
2149 0 R
-63 0 V
0.050 UL
LT3
200 2160 M
2212 0 V
1.000 UL
LTb
200 2160 M
63 0 V
2149 0 R
-63 0 V
0.050 UL
LT3
200 300 M
0 1860 V
1.000 UL
LTb
200 300 M
0 63 V
0 1797 R
0 -63 V
366 300 M
0 31 V
0 1829 R
0 -31 V
464 300 M
0 31 V
0 1829 R
0 -31 V
533 300 M
0 31 V
0 1829 R
0 -31 V
587 300 M
0 31 V
0 1829 R
0 -31 V
630 300 M
0 31 V
0 1829 R
0 -31 V
667 300 M
0 31 V
0 1829 R
0 -31 V
699 300 M
0 31 V
0 1829 R
0 -31 V
728 300 M
0 31 V
0 1829 R
0 -31 V
0.050 UL
LT3
753 300 M
0 1860 V
1.000 UL
LTb
753 300 M
0 63 V
0 1797 R
0 -63 V
919 300 M
0 31 V
0 1829 R
0 -31 V
1017 300 M
0 31 V
0 1829 R
0 -31 V
1086 300 M
0 31 V
0 1829 R
0 -31 V
1140 300 M
0 31 V
0 1829 R
0 -31 V
1183 300 M
0 31 V
0 1829 R
0 -31 V
1220 300 M
0 31 V
0 1829 R
0 -31 V
1252 300 M
0 31 V
0 1829 R
0 -31 V
1281 300 M
0 31 V
0 1829 R
0 -31 V
0.050 UL
LT3
1306 300 M
0 1860 V
1.000 UL
LTb
1306 300 M
0 63 V
0 1797 R
0 -63 V
1472 300 M
0 31 V
0 1829 R
0 -31 V
1570 300 M
0 31 V
0 1829 R
0 -31 V
1639 300 M
0 31 V
0 1829 R
0 -31 V
1693 300 M
0 31 V
0 1829 R
0 -31 V
1736 300 M
0 31 V
0 1829 R
0 -31 V
1773 300 M
0 31 V
0 1829 R
0 -31 V
1805 300 M
0 31 V
0 1829 R
0 -31 V
1834 300 M
0 31 V
0 1829 R
0 -31 V
0.050 UL
LT3
1859 300 M
0 63 V
0 400 R
0 1397 V
1.000 UL
LTb
1859 300 M
0 63 V
0 1797 R
0 -63 V
2025 300 M
0 31 V
0 1829 R
0 -31 V
2123 300 M
0 31 V
0 1829 R
0 -31 V
2192 300 M
0 31 V
0 1829 R
0 -31 V
2246 300 M
0 31 V
0 1829 R
0 -31 V
2289 300 M
0 31 V
0 1829 R
0 -31 V
2326 300 M
0 31 V
0 1829 R
0 -31 V
2358 300 M
0 31 V
0 1829 R
0 -31 V
2387 300 M
0 31 V
0 1829 R
0 -31 V
0.050 UL
LT3
2412 300 M
0 1860 V
1.000 UL
LTb
2412 300 M
0 63 V
0 1797 R
0 -63 V
1.000 UL
LTb
200 300 M
2212 0 V
0 1860 V
-2212 0 V
200 300 L
1.000 UL
LTb
1475 363 M
0 400 V
887 0 V
0 -400 V
-887 0 V
0 400 R
887 0 V
0.000 UP
1.000 UL
LT0
2075 713 M
237 0 V
200 1850 M
2 0 V
4 0 V
10 0 V
19 0 V
31 124 V
45 -29 V
369 901 L
68 855 V
73 -161 V
76 276 V
76 8 V
75 63 V
72 -35 V
70 -56 V
66 -33 V
64 55 V
61 -21 V
59 20 V
56 -6 V
54 -13 V
51 -20 V
50 -14 V
47 36 V
46 7 V
44 -48 V
43 73 V
41 -50 V
40 8 V
38 4 V
38 -22 V
36 23 V
35 0 V
35 0 V
33 0 V
32 0 V
32 0 V
31 0 V
30 0 V
29 0 V
28 0 V
28 0 V
28 0 V
26 0 V
26 0 V
26 0 V
25 0 V
24 0 V
24 0 V
24 0 V
23 0 V
23 0 V
22 0 V
22 0 V
21 0 V
21 0 V
21 0 V
20 0 V
200 1850 Circle
200 1850 Circle
200 1850 Circle
202 1850 Circle
206 1850 Circle
216 1850 Circle
235 1850 Circle
266 1974 Circle
311 1945 Circle
369 901 Circle
437 1756 Circle
510 1595 Circle
586 1871 Circle
662 1879 Circle
737 1942 Circle
809 1907 Circle
879 1851 Circle
945 1818 Circle
1009 1873 Circle
1070 1852 Circle
1129 1872 Circle
1185 1866 Circle
1239 1853 Circle
1290 1833 Circle
1340 1819 Circle
1387 1855 Circle
1433 1862 Circle
1477 1814 Circle
1520 1887 Circle
1561 1837 Circle
1601 1845 Circle
1639 1849 Circle
1677 1827 Circle
1713 1850 Circle
1748 1850 Circle
1783 1850 Circle
1816 1850 Circle
1848 1850 Circle
1880 1850 Circle
1911 1850 Circle
1941 1850 Circle
1970 1850 Circle
1998 1850 Circle
2026 1850 Circle
2054 1850 Circle
2080 1850 Circle
2106 1850 Circle
2132 1850 Circle
2157 1850 Circle
2181 1850 Circle
2205 1850 Circle
2229 1850 Circle
2252 1850 Circle
2275 1850 Circle
2297 1850 Circle
2319 1850 Circle
2340 1850 Circle
2361 1850 Circle
2382 1850 Circle
2402 1850 Circle
2193 713 Circle
0.000 UP
1.000 UL
LT1
2075 613 M
237 0 V
200 1850 M
2 0 V
4 0 V
10 0 V
19 0 V
31 0 V
45 0 V
58 0 V
68 289 V
73 -276 V
76 -175 V
76 191 V
75 65 V
72 -37 V
70 -53 V
66 10 V
64 110 V
61 -26 V
59 17 V
56 38 V
54 19 V
51 33 V
50 -2 V
47 17 V
46 17 V
44 6 V
43 14 V
41 5 V
40 8 V
38 7 V
38 2 V
36 -279 V
35 0 V
35 0 V
33 0 V
32 0 V
32 0 V
31 0 V
30 0 V
29 0 V
28 0 V
28 0 V
28 0 V
26 0 V
26 0 V
26 0 V
25 0 V
24 0 V
24 0 V
24 0 V
23 0 V
23 0 V
22 0 V
22 0 V
21 0 V
21 0 V
21 0 V
20 0 V
200 1850 TriU
200 1850 TriU
200 1850 TriU
202 1850 TriU
206 1850 TriU
216 1850 TriU
235 1850 TriU
266 1850 TriU
311 1850 TriU
369 1850 TriU
437 2139 TriU
510 1863 TriU
586 1688 TriU
662 1879 TriU
737 1944 TriU
809 1907 TriU
879 1854 TriU
945 1864 TriU
1009 1974 TriU
1070 1948 TriU
1129 1965 TriU
1185 2003 TriU
1239 2022 TriU
1290 2055 TriU
1340 2053 TriU
1387 2070 TriU
1433 2087 TriU
1477 2093 TriU
1520 2107 TriU
1561 2112 TriU
1601 2120 TriU
1639 2127 TriU
1677 2129 TriU
1713 1850 TriU
1748 1850 TriU
1783 1850 TriU
1816 1850 TriU
1848 1850 TriU
1880 1850 TriU
1911 1850 TriU
1941 1850 TriU
1970 1850 TriU
1998 1850 TriU
2026 1850 TriU
2054 1850 TriU
2080 1850 TriU
2106 1850 TriU
2132 1850 TriU
2157 1850 TriU
2181 1850 TriU
2205 1850 TriU
2229 1850 TriU
2252 1850 TriU
2275 1850 TriU
2297 1850 TriU
2319 1850 TriU
2340 1850 TriU
2361 1850 TriU
2382 1850 TriU
2402 1850 TriU
2193 613 TriU
0.000 UP
1.000 UL
LT4
2075 513 M
237 0 V
200 1850 M
2 0 V
4 0 V
10 0 V
19 0 V
31 124 V
45 -29 V
369 901 L
68 339 V
73 -74 V
76 800 V
76 -74 V
75 -49 V
72 7 V
830 300 L
227 0 R
13 787 V
59 290 V
56 137 V
54 -53 V
51 93 V
50 -46 V
47 181 V
46 -13 V
44 -48 V
43 138 V
41 -45 V
40 18 V
38 25 V
38 -8 V
36 94 V
35 0 V
35 0 V
33 0 V
32 0 V
32 0 V
31 0 V
30 0 V
29 0 V
28 0 V
28 0 V
28 0 V
26 0 V
26 0 V
26 0 V
25 0 V
24 0 V
24 0 V
24 0 V
23 0 V
23 0 V
22 0 V
22 0 V
21 0 V
21 0 V
21 0 V
20 0 V
200 1850 Box
200 1850 Box
200 1850 Box
202 1850 Box
206 1850 Box
216 1850 Box
235 1850 Box
266 1974 Box
311 1945 Box
369 901 Box
437 1240 Box
510 1166 Box
586 1966 Box
662 1892 Box
737 1843 Box
809 1850 Box
1070 1087 Box
1129 1377 Box
1185 1514 Box
1239 1461 Box
1290 1554 Box
1340 1508 Box
1387 1689 Box
1433 1676 Box
1477 1628 Box
1520 1766 Box
1561 1721 Box
1601 1739 Box
1639 1764 Box
1677 1756 Box
1713 1850 Box
1748 1850 Box
1783 1850 Box
1816 1850 Box
1848 1850 Box
1880 1850 Box
1911 1850 Box
1941 1850 Box
1970 1850 Box
1998 1850 Box
2026 1850 Box
2054 1850 Box
2080 1850 Box
2106 1850 Box
2132 1850 Box
2157 1850 Box
2181 1850 Box
2205 1850 Box
2229 1850 Box
2252 1850 Box
2275 1850 Box
2297 1850 Box
2319 1850 Box
2340 1850 Box
2361 1850 Box
2382 1850 Box
2402 1850 Box
2193 513 Box
0.000 UP
1.000 UL
LT3
2075 413 M
237 0 V
200 1850 M
2 0 V
4 0 V
10 0 V
235 909 L
263 300 L
5 0 R
43 1849 V
58 -570 V
68 366 V
73 -156 V
76 180 V
76 -59 V
75 -128 V
72 191 V
846 300 L
162 0 R
1 38 V
61 594 V
59 318 V
56 176 V
54 144 V
51 62 V
50 90 V
47 31 V
46 -26 V
44 60 V
43 -47 V
41 62 V
40 -78 V
38 -44 V
38 -76 V
36 138 V
35 30 V
35 108 V
33 73 V
32 61 V
32 47 V
31 -58 V
30 28 V
29 6 V
28 47 V
28 -49 V
28 70 V
26 20 V
26 -20 V
26 -35 V
25 77 V
24 -19 V
24 -83 V
24 72 V
23 30 V
23 -21 V
22 15 V
22 12 V
21 -5 V
21 -21 V
21 27 V
20 -1 V
200 1850 Dia
200 1850 Dia
200 1850 Dia
202 1850 Dia
206 1850 Dia
216 1850 Dia
235 909 Dia
311 2149 Dia
369 1579 Dia
437 1945 Dia
510 1789 Dia
586 1969 Dia
662 1910 Dia
737 1782 Dia
809 1973 Dia
1009 338 Dia
1070 932 Dia
1129 1250 Dia
1185 1426 Dia
1239 1570 Dia
1290 1632 Dia
1340 1722 Dia
1387 1753 Dia
1433 1727 Dia
1477 1787 Dia
1520 1740 Dia
1561 1802 Dia
1601 1724 Dia
1639 1680 Dia
1677 1604 Dia
1713 1742 Dia
1748 1772 Dia
1783 1880 Dia
1816 1953 Dia
1848 2014 Dia
1880 2061 Dia
1911 2003 Dia
1941 2031 Dia
1970 2037 Dia
1998 2084 Dia
2026 2035 Dia
2054 2105 Dia
2080 2125 Dia
2106 2105 Dia
2132 2070 Dia
2157 2147 Dia
2181 2128 Dia
2205 2045 Dia
2229 2117 Dia
2252 2147 Dia
2275 2126 Dia
2297 2141 Dia
2319 2153 Dia
2340 2148 Dia
2361 2127 Dia
2382 2154 Dia
2402 2153 Dia
2193 413 Dia
stroke
grestore
end
showpage
}}%
\put(2025,413){\makebox(0,0)[r]{BP, $\tilde{t}=\tilde{t}_{\rm max}$}}%
\put(2025,513){\makebox(0,0)[r]{BP, $\tilde{t}=0$}}%
\put(2025,613){\makebox(0,0)[r]{EP, $\tilde{t}=0$}}%
\put(2025,713){\makebox(0,0)[r]{TP, $\tilde{t}=0$}}%
\put(1306,2310){\makebox(0,0){\textsf{Velocity anisotropy $\beta(\tilde{r})$}}}%
\put(1306,50){\makebox(0,0){\sffamily $\tilde{r}$}}%
\put(50,1230){%
\special{ps: gsave currentpoint currentpoint translate
270 rotate neg exch neg exch translate}%
\makebox(0,0)[b]{\shortstack{\sffamily $\beta$}}%
\special{ps: currentpoint grestore moveto}%
}%
\put(2412,200){\makebox(0,0){1000}}%
\put(1859,200){\makebox(0,0){100}}%
\put(1306,200){\makebox(0,0){10}}%
\put(753,200){\makebox(0,0){1}}%
\put(200,200){\makebox(0,0){0.1}}%
\put(150,2160){\makebox(0,0)[r]{1}}%
\put(150,1850){\makebox(0,0)[r]{0}}%
\put(150,1540){\makebox(0,0)[r]{-1}}%
\put(150,1230){\makebox(0,0)[r]{-2}}%
\put(150,920){\makebox(0,0)[r]{-3}}%
\put(150,610){\makebox(0,0)[r]{-4}}%
\put(150,300){\makebox(0,0)[r]{-5}}%
\end{picture}%
\endgroup
 

%% file: bhb_fig12.tex
\begingroup%
  \makeatletter%
  \newcommand{\GNUPLOTspecial}{%
    \@sanitize\catcode`\%=14\relax\special}%
  \setlength{\unitlength}{0.1bp}%
{\GNUPLOTspecial{!
/gnudict 256 dict def
gnudict begin
/Color false def
/Solid false def
/gnulinewidth 5.000 def
/userlinewidth gnulinewidth def
/vshift -33 def
/dl {10 mul} def
/hpt_ 31.5 def
/vpt_ 31.5 def
/hpt hpt_ def
/vpt vpt_ def
/M {moveto} bind def
/L {lineto} bind def
/R {rmoveto} bind def
/V {rlineto} bind def
/vpt2 vpt 2 mul def
/hpt2 hpt 2 mul def
/Lshow { currentpoint stroke M
  0 vshift R show } def
/Rshow { currentpoint stroke M
  dup stringwidth pop neg vshift R show } def
/Cshow { currentpoint stroke M
  dup stringwidth pop -2 div vshift R show } def
/UP { dup vpt_ mul /vpt exch def hpt_ mul /hpt exch def
  /hpt2 hpt 2 mul def /vpt2 vpt 2 mul def } def
/DL { Color {setrgbcolor Solid {pop []} if 0 setdash }
 {pop pop pop Solid {pop []} if 0 setdash} ifelse } def
/BL { stroke gnulinewidth 2 mul setlinewidth } def
/AL { stroke gnulinewidth 2 div setlinewidth } def
/UL { gnulinewidth mul /userlinewidth exch def } def
/PL { stroke userlinewidth setlinewidth } def
/LTb { BL [] 0 0 0 DL } def
/LTa { AL [1 dl 2 dl] 0 setdash 0 0 0 setrgbcolor } def
/LT0 { PL [] 1 0 0 DL } def
/LT1 { PL [4 dl 2 dl] 0 1 0 DL } def
/LT2 { PL [2 dl 3 dl] 0 0 1 DL } def
/LT3 { PL [1 dl 1.5 dl] 1 0 1 DL } def
/LT4 { PL [5 dl 2 dl 1 dl 2 dl] 0 1 1 DL } def
/LT5 { PL [4 dl 3 dl 1 dl 3 dl] 1 1 0 DL } def
/LT6 { PL [2 dl 2 dl 2 dl 4 dl] 0 0 0 DL } def
/LT7 { PL [2 dl 2 dl 2 dl 2 dl 2 dl 4 dl] 1 0.3 0 DL } def
/LT8 { PL [2 dl 2 dl 2 dl 2 dl 2 dl 2 dl 2 dl 4 dl] 0.5 0.5 0.5 DL } def
/Pnt { stroke [] 0 setdash
   gsave 1 setlinecap M 0 0 V stroke grestore } def
/Dia { stroke [] 0 setdash 2 copy vpt add M
  hpt neg vpt neg V hpt vpt neg V
  hpt vpt V hpt neg vpt V closepath stroke
  Pnt } def
/Pls { stroke [] 0 setdash vpt sub M 0 vpt2 V
  currentpoint stroke M
  hpt neg vpt neg R hpt2 0 V stroke
  } def
/Box { stroke [] 0 setdash 2 copy exch hpt sub exch vpt add M
  0 vpt2 neg V hpt2 0 V 0 vpt2 V
  hpt2 neg 0 V closepath stroke
  Pnt } def
/Crs { stroke [] 0 setdash exch hpt sub exch vpt add M
  hpt2 vpt2 neg V currentpoint stroke M
  hpt2 neg 0 R hpt2 vpt2 V stroke } def
/TriU { stroke [] 0 setdash 2 copy vpt 1.12 mul add M
  hpt neg vpt -1.62 mul V
  hpt 2 mul 0 V
  hpt neg vpt 1.62 mul V closepath stroke
  Pnt  } def
/Star { 2 copy Pls Crs } def
/BoxF { stroke [] 0 setdash exch hpt sub exch vpt add M
  0 vpt2 neg V  hpt2 0 V  0 vpt2 V
  hpt2 neg 0 V  closepath fill } def
/TriUF { stroke [] 0 setdash vpt 1.12 mul add M
  hpt neg vpt -1.62 mul V
  hpt 2 mul 0 V
  hpt neg vpt 1.62 mul V closepath fill } def
/TriD { stroke [] 0 setdash 2 copy vpt 1.12 mul sub M
  hpt neg vpt 1.62 mul V
  hpt 2 mul 0 V
  hpt neg vpt -1.62 mul V closepath stroke
  Pnt  } def
/TriDF { stroke [] 0 setdash vpt 1.12 mul sub M
  hpt neg vpt 1.62 mul V
  hpt 2 mul 0 V
  hpt neg vpt -1.62 mul V closepath fill} def
/DiaF { stroke [] 0 setdash vpt add M
  hpt neg vpt neg V hpt vpt neg V
  hpt vpt V hpt neg vpt V closepath fill } def
/Pent { stroke [] 0 setdash 2 copy gsave
  translate 0 hpt M 4 {72 rotate 0 hpt L} repeat
  closepath stroke grestore Pnt } def
/PentF { stroke [] 0 setdash gsave
  translate 0 hpt M 4 {72 rotate 0 hpt L} repeat
  closepath fill grestore } def
/Circle { stroke [] 0 setdash 2 copy
  hpt 0 360 arc stroke Pnt } def
/CircleF { stroke [] 0 setdash hpt 0 360 arc fill } def
/C0 { BL [] 0 setdash 2 copy moveto vpt 90 450  arc } bind def
/C1 { BL [] 0 setdash 2 copy        moveto
       2 copy  vpt 0 90 arc closepath fill
               vpt 0 360 arc closepath } bind def
/C2 { BL [] 0 setdash 2 copy moveto
       2 copy  vpt 90 180 arc closepath fill
               vpt 0 360 arc closepath } bind def
/C3 { BL [] 0 setdash 2 copy moveto
       2 copy  vpt 0 180 arc closepath fill
               vpt 0 360 arc closepath } bind def
/C4 { BL [] 0 setdash 2 copy moveto
       2 copy  vpt 180 270 arc closepath fill
               vpt 0 360 arc closepath } bind def
/C5 { BL [] 0 setdash 2 copy moveto
       2 copy  vpt 0 90 arc
       2 copy moveto
       2 copy  vpt 180 270 arc closepath fill
               vpt 0 360 arc } bind def
/C6 { BL [] 0 setdash 2 copy moveto
      2 copy  vpt 90 270 arc closepath fill
              vpt 0 360 arc closepath } bind def
/C7 { BL [] 0 setdash 2 copy moveto
      2 copy  vpt 0 270 arc closepath fill
              vpt 0 360 arc closepath } bind def
/C8 { BL [] 0 setdash 2 copy moveto
      2 copy vpt 270 360 arc closepath fill
              vpt 0 360 arc closepath } bind def
/C9 { BL [] 0 setdash 2 copy moveto
      2 copy  vpt 270 450 arc closepath fill
              vpt 0 360 arc closepath } bind def
/C10 { BL [] 0 setdash 2 copy 2 copy moveto vpt 270 360 arc closepath fill
       2 copy moveto
       2 copy vpt 90 180 arc closepath fill
               vpt 0 360 arc closepath } bind def
/C11 { BL [] 0 setdash 2 copy moveto
       2 copy  vpt 0 180 arc closepath fill
       2 copy moveto
       2 copy  vpt 270 360 arc closepath fill
               vpt 0 360 arc closepath } bind def
/C12 { BL [] 0 setdash 2 copy moveto
       2 copy  vpt 180 360 arc closepath fill
               vpt 0 360 arc closepath } bind def
/C13 { BL [] 0 setdash  2 copy moveto
       2 copy  vpt 0 90 arc closepath fill
       2 copy moveto
       2 copy  vpt 180 360 arc closepath fill
               vpt 0 360 arc closepath } bind def
/C14 { BL [] 0 setdash 2 copy moveto
       2 copy  vpt 90 360 arc closepath fill
               vpt 0 360 arc } bind def
/C15 { BL [] 0 setdash 2 copy vpt 0 360 arc closepath fill
               vpt 0 360 arc closepath } bind def
/Rec   { newpath 4 2 roll moveto 1 index 0 rlineto 0 exch rlineto
       neg 0 rlineto closepath } bind def
/Square { dup Rec } bind def
/Bsquare { vpt sub exch vpt sub exch vpt2 Square } bind def
/S0 { BL [] 0 setdash 2 copy moveto 0 vpt rlineto BL Bsquare } bind def
/S1 { BL [] 0 setdash 2 copy vpt Square fill Bsquare } bind def
/S2 { BL [] 0 setdash 2 copy exch vpt sub exch vpt Square fill Bsquare } bind def
/S3 { BL [] 0 setdash 2 copy exch vpt sub exch vpt2 vpt Rec fill Bsquare } bind def
/S4 { BL [] 0 setdash 2 copy exch vpt sub exch vpt sub vpt Square fill Bsquare } bind def
/S5 { BL [] 0 setdash 2 copy 2 copy vpt Square fill
       exch vpt sub exch vpt sub vpt Square fill Bsquare } bind def
/S6 { BL [] 0 setdash 2 copy exch vpt sub exch vpt sub vpt vpt2 Rec fill Bsquare } bind def
/S7 { BL [] 0 setdash 2 copy exch vpt sub exch vpt sub vpt vpt2 Rec fill
       2 copy vpt Square fill
       Bsquare } bind def
/S8 { BL [] 0 setdash 2 copy vpt sub vpt Square fill Bsquare } bind def
/S9 { BL [] 0 setdash 2 copy vpt sub vpt vpt2 Rec fill Bsquare } bind def
/S10 { BL [] 0 setdash 2 copy vpt sub vpt Square fill 2 copy exch vpt sub exch vpt Square fill
       Bsquare } bind def
/S11 { BL [] 0 setdash 2 copy vpt sub vpt Square fill 2 copy exch vpt sub exch vpt2 vpt Rec fill
       Bsquare } bind def
/S12 { BL [] 0 setdash 2 copy exch vpt sub exch vpt sub vpt2 vpt Rec fill Bsquare } bind def
/S13 { BL [] 0 setdash 2 copy exch vpt sub exch vpt sub vpt2 vpt Rec fill
       2 copy vpt Square fill Bsquare } bind def
/S14 { BL [] 0 setdash 2 copy exch vpt sub exch vpt sub vpt2 vpt Rec fill
       2 copy exch vpt sub exch vpt Square fill Bsquare } bind def
/S15 { BL [] 0 setdash 2 copy Bsquare fill Bsquare } bind def
/D0 { gsave translate 45 rotate 0 0 S0 stroke grestore } bind def
/D1 { gsave translate 45 rotate 0 0 S1 stroke grestore } bind def
/D2 { gsave translate 45 rotate 0 0 S2 stroke grestore } bind def
/D3 { gsave translate 45 rotate 0 0 S3 stroke grestore } bind def
/D4 { gsave translate 45 rotate 0 0 S4 stroke grestore } bind def
/D5 { gsave translate 45 rotate 0 0 S5 stroke grestore } bind def
/D6 { gsave translate 45 rotate 0 0 S6 stroke grestore } bind def
/D7 { gsave translate 45 rotate 0 0 S7 stroke grestore } bind def
/D8 { gsave translate 45 rotate 0 0 S8 stroke grestore } bind def
/D9 { gsave translate 45 rotate 0 0 S9 stroke grestore } bind def
/D10 { gsave translate 45 rotate 0 0 S10 stroke grestore } bind def
/D11 { gsave translate 45 rotate 0 0 S11 stroke grestore } bind def
/D12 { gsave translate 45 rotate 0 0 S12 stroke grestore } bind def
/D13 { gsave translate 45 rotate 0 0 S13 stroke grestore } bind def
/D14 { gsave translate 45 rotate 0 0 S14 stroke grestore } bind def
/D15 { gsave translate 45 rotate 0 0 S15 stroke grestore } bind def
/DiaE { stroke [] 0 setdash vpt add M
  hpt neg vpt neg V hpt vpt neg V
  hpt vpt V hpt neg vpt V closepath stroke } def
/BoxE { stroke [] 0 setdash exch hpt sub exch vpt add M
  0 vpt2 neg V hpt2 0 V 0 vpt2 V
  hpt2 neg 0 V closepath stroke } def
/TriUE { stroke [] 0 setdash vpt 1.12 mul add M
  hpt neg vpt -1.62 mul V
  hpt 2 mul 0 V
  hpt neg vpt 1.62 mul V closepath stroke } def
/TriDE { stroke [] 0 setdash vpt 1.12 mul sub M
  hpt neg vpt 1.62 mul V
  hpt 2 mul 0 V
  hpt neg vpt -1.62 mul V closepath stroke } def
/PentE { stroke [] 0 setdash gsave
  translate 0 hpt M 4 {72 rotate 0 hpt L} repeat
  closepath stroke grestore } def
/CircE { stroke [] 0 setdash 
  hpt 0 360 arc stroke } def
/Opaque { gsave closepath 1 setgray fill grestore 0 setgray closepath } def
/DiaW { stroke [] 0 setdash vpt add M
  hpt neg vpt neg V hpt vpt neg V
  hpt vpt V hpt neg vpt V Opaque stroke } def
/BoxW { stroke [] 0 setdash exch hpt sub exch vpt add M
  0 vpt2 neg V hpt2 0 V 0 vpt2 V
  hpt2 neg 0 V Opaque stroke } def
/TriUW { stroke [] 0 setdash vpt 1.12 mul add M
  hpt neg vpt -1.62 mul V
  hpt 2 mul 0 V
  hpt neg vpt 1.62 mul V Opaque stroke } def
/TriDW { stroke [] 0 setdash vpt 1.12 mul sub M
  hpt neg vpt 1.62 mul V
  hpt 2 mul 0 V
  hpt neg vpt -1.62 mul V Opaque stroke } def
/PentW { stroke [] 0 setdash gsave
  translate 0 hpt M 4 {72 rotate 0 hpt L} repeat
  Opaque stroke grestore } def
/CircW { stroke [] 0 setdash 
  hpt 0 360 arc Opaque stroke } def
/BoxFill { gsave Rec 1 setgray fill grestore } def
end
}}%
\begin{picture}(2412,2160)(0,0)%
{\GNUPLOTspecial{"
gnudict begin
gsave
0 0 translate
0.100 0.100 scale
0 setgray
newpath
1.000 UL
LTb
0.050 UL
LT3
400 300 M
2012 0 V
1.000 UL
LTb
400 300 M
63 0 V
1949 0 R
-63 0 V
400 393 M
31 0 V
1981 0 R
-31 0 V
400 517 M
31 0 V
1981 0 R
-31 0 V
400 580 M
31 0 V
1981 0 R
-31 0 V
0.050 UL
LT3
400 610 M
2012 0 V
1.000 UL
LTb
400 610 M
63 0 V
1949 0 R
-63 0 V
400 703 M
31 0 V
1981 0 R
-31 0 V
400 827 M
31 0 V
1981 0 R
-31 0 V
400 890 M
31 0 V
1981 0 R
-31 0 V
0.050 UL
LT3
400 920 M
2012 0 V
1.000 UL
LTb
400 920 M
63 0 V
1949 0 R
-63 0 V
400 1013 M
31 0 V
1981 0 R
-31 0 V
400 1137 M
31 0 V
1981 0 R
-31 0 V
400 1200 M
31 0 V
1981 0 R
-31 0 V
0.050 UL
LT3
400 1230 M
2012 0 V
1.000 UL
LTb
400 1230 M
63 0 V
1949 0 R
-63 0 V
400 1323 M
31 0 V
1981 0 R
-31 0 V
400 1447 M
31 0 V
1981 0 R
-31 0 V
400 1510 M
31 0 V
1981 0 R
-31 0 V
0.050 UL
LT3
400 1540 M
2012 0 V
1.000 UL
LTb
400 1540 M
63 0 V
1949 0 R
-63 0 V
400 1633 M
31 0 V
1981 0 R
-31 0 V
400 1757 M
31 0 V
1981 0 R
-31 0 V
400 1820 M
31 0 V
1981 0 R
-31 0 V
0.050 UL
LT3
400 1850 M
1375 0 V
587 0 R
50 0 V
1.000 UL
LTb
400 1850 M
63 0 V
1949 0 R
-63 0 V
400 1943 M
31 0 V
1981 0 R
-31 0 V
400 2067 M
31 0 V
1981 0 R
-31 0 V
400 2130 M
31 0 V
1981 0 R
-31 0 V
0.050 UL
LT3
400 2160 M
2012 0 V
1.000 UL
LTb
400 2160 M
63 0 V
1949 0 R
-63 0 V
0.050 UL
LT3
400 300 M
0 1860 V
1.000 UL
LTb
400 300 M
0 63 V
0 1797 R
0 -63 V
521 300 M
0 31 V
0 1829 R
0 -31 V
681 300 M
0 31 V
0 1829 R
0 -31 V
763 300 M
0 31 V
0 1829 R
0 -31 V
0.050 UL
LT3
802 300 M
0 1860 V
1.000 UL
LTb
802 300 M
0 63 V
0 1797 R
0 -63 V
924 300 M
0 31 V
0 1829 R
0 -31 V
1084 300 M
0 31 V
0 1829 R
0 -31 V
1166 300 M
0 31 V
0 1829 R
0 -31 V
0.050 UL
LT3
1205 300 M
0 1860 V
1.000 UL
LTb
1205 300 M
0 63 V
0 1797 R
0 -63 V
1326 300 M
0 31 V
0 1829 R
0 -31 V
1486 300 M
0 31 V
0 1829 R
0 -31 V
1568 300 M
0 31 V
0 1829 R
0 -31 V
0.050 UL
LT3
1607 300 M
0 1860 V
1.000 UL
LTb
1607 300 M
0 63 V
0 1797 R
0 -63 V
1728 300 M
0 31 V
0 1829 R
0 -31 V
1888 300 M
0 31 V
0 1829 R
0 -31 V
1971 300 M
0 31 V
0 1829 R
0 -31 V
0.050 UL
LT3
2010 300 M
0 1497 V
0 300 R
0 63 V
1.000 UL
LTb
2010 300 M
0 63 V
0 1797 R
0 -63 V
2131 300 M
0 31 V
0 1829 R
0 -31 V
2291 300 M
0 31 V
0 1829 R
0 -31 V
2373 300 M
0 31 V
0 1829 R
0 -31 V
0.050 UL
LT3
2412 300 M
0 1860 V
1.000 UL
LTb
2412 300 M
0 63 V
0 1797 R
0 -63 V
1.000 UL
LTb
400 300 M
2012 0 V
0 1860 V
-2012 0 V
400 300 L
1.000 UL
LTb
1775 1797 M
0 300 V
587 0 V
0 -300 V
-587 0 V
0 300 R
587 0 V
0.000 UP
1.000 UL
LT0
2075 2047 M
237 0 V
567 300 M
0 1484 V
73 -17 V
76 154 V
75 88 V
72 25 V
69 -28 V
64 -55 V
60 -42 V
57 -56 V
54 -73 V
50 -50 V
48 -94 V
46 -16 V
43 -72 V
42 -69 V
39 -68 V
38 -43 V
36 0 V
35 -60 V
34 -11 V
32 -45 V
31 9 V
30 -184 V
29 62 V
28 -124 V
28 31 V
26 -123 V
26 181 V
25 -132 V
24 -96 V
24 86 V
22 -24 V
23 -156 V
22 107 V
21 -16 V
21 56 V
20 -112 V
20 0 V
19 -48 V
19 -11 V
19 -41 V
18 27 V
18 32 V
17 -31 V
18 -99 V
16 19 V
17 -32 V
16 -2 V
16 113 V
15 -137 V
16 33 V
15 51 V
15 -147 V
567 1784 Circle
640 1767 Circle
716 1921 Circle
791 2009 Circle
863 2034 Circle
932 2006 Circle
996 1951 Circle
1056 1909 Circle
1113 1853 Circle
1167 1780 Circle
1217 1730 Circle
1265 1636 Circle
1311 1620 Circle
1354 1548 Circle
1396 1479 Circle
1435 1411 Circle
1473 1368 Circle
1509 1368 Circle
1544 1308 Circle
1578 1297 Circle
1610 1252 Circle
1641 1261 Circle
1671 1077 Circle
1700 1139 Circle
1728 1015 Circle
1756 1046 Circle
1782 923 Circle
1808 1104 Circle
1833 972 Circle
1857 876 Circle
1881 962 Circle
1903 938 Circle
1926 782 Circle
1948 889 Circle
1969 873 Circle
1990 929 Circle
2010 817 Circle
2030 817 Circle
2049 769 Circle
2068 758 Circle
2087 717 Circle
2105 744 Circle
2123 776 Circle
2140 745 Circle
2158 646 Circle
2174 665 Circle
2191 633 Circle
2207 631 Circle
2223 744 Circle
2238 607 Circle
2254 640 Circle
2269 691 Circle
2284 544 Circle
2193 2047 Circle
1.000 UL
LT1
2075 1947 M
237 0 V
495 386 M
1 -86 V
56 0 R
0 21 V
19 529 V
19 325 V
19 223 V
19 161 V
19 120 V
19 92 V
19 71 V
19 54 V
19 42 V
19 31 V
20 24 V
19 17 V
19 11 V
19 6 V
19 3 V
19 -1 V
19 -4 V
19 -6 V
19 -8 V
19 -10 V
19 -11 V
19 -13 V
19 -14 V
19 -15 V
19 -16 V
19 -16 V
19 -17 V
19 -18 V
19 -19 V
19 -18 V
19 -20 V
19 -19 V
19 -20 V
19 -20 V
19 -20 V
19 -21 V
19 -21 V
19 -21 V
19 -21 V
19 -21 V
19 -21 V
19 -21 V
19 -22 V
19 -21 V
19 -22 V
19 -21 V
19 -22 V
20 -21 V
19 -22 V
19 -22 V
19 -22 V
19 -22 V
19 -21 V
19 -22 V
19 -22 V
19 -22 V
19 -22 V
19 -22 V
19 -22 V
19 -22 V
19 -21 V
19 -22 V
19 -22 V
19 -22 V
19 -22 V
19 -22 V
19 -22 V
19 -22 V
19 -22 V
19 -22 V
19 -22 V
19 -22 V
19 -22 V
19 -22 V
19 -22 V
19 -22 V
19 -22 V
19 -22 V
19 -22 V
19 -22 V
19 -22 V
19 -22 V
19 -22 V
19 -22 V
20 -22 V
19 -22 V
19 -22 V
19 -22 V
19 -22 V
19 -22 V
19 -22 V
1.000 UL
LT3
2075 1847 M
237 0 V
779 2160 M
2 -1 V
19 -22 V
19 -22 V
19 -22 V
19 -22 V
19 -22 V
19 -22 V
19 -22 V
19 -22 V
19 -22 V
19 -22 V
19 -22 V
19 -22 V
19 -22 V
19 -22 V
19 -22 V
19 -22 V
19 -22 V
19 -22 V
19 -22 V
19 -22 V
19 -22 V
19 -22 V
19 -22 V
19 -22 V
19 -22 V
19 -22 V
19 -22 V
19 -22 V
19 -22 V
19 -22 V
19 -22 V
19 -22 V
19 -22 V
19 -22 V
19 -22 V
20 -22 V
19 -22 V
19 -22 V
19 -22 V
19 -22 V
19 -22 V
19 -22 V
19 -22 V
19 -22 V
19 -22 V
19 -22 V
19 -22 V
19 -22 V
19 -22 V
19 -22 V
19 -22 V
19 -22 V
19 -22 V
19 -22 V
19 -22 V
19 -22 V
19 -22 V
19 -22 V
19 -22 V
19 -22 V
19 -22 V
19 -22 V
19 -22 V
19 -22 V
19 -22 V
19 -22 V
19 -22 V
19 -22 V
19 -22 V
19 -22 V
19 -22 V
19 -22 V
20 -22 V
19 -21 V
19 -22 V
19 -22 V
19 -22 V
19 -22 V
19 -22 V
stroke
grestore
end
showpage
}}%
\put(2025,1847){\makebox(0,0)[r]{fit 2}}%
\put(2025,1947){\makebox(0,0)[r]{fit 1}}%
\put(2025,2047){\makebox(0,0)[r]{data}}%
\put(1406,2310){\makebox(0,0){\textsf{Angular momentum loss rate $\frac{d\tilde{L}_z}{d\tilde{t}},\,q=10$}}}%
\put(1406,50){\makebox(0,0){\sffamily $\tilde{t}$}}%
\put(50,1230){%
\special{ps: gsave currentpoint currentpoint translate
270 rotate neg exch neg exch translate}%
\makebox(0,0)[b]{\shortstack{\sffamily $\frac{d\tilde{L}_z}{d\tilde{t}}$}}%
\special{ps: currentpoint grestore moveto}%
}%
\put(2412,200){\makebox(0,0){1e+06}}%
\put(2010,200){\makebox(0,0){100000}}%
\put(1607,200){\makebox(0,0){10000}}%
\put(1205,200){\makebox(0,0){1000}}%
\put(802,200){\makebox(0,0){100}}%
\put(400,200){\makebox(0,0){10}}%
\put(350,2160){\makebox(0,0)[r]{0.001}}%
\put(350,1850){\makebox(0,0)[r]{0.0001}}%
\put(350,1540){\makebox(0,0)[r]{1e-05}}%
\put(350,1230){\makebox(0,0)[r]{1e-06}}%
\put(350,920){\makebox(0,0)[r]{1e-07}}%
\put(350,610){\makebox(0,0)[r]{1e-08}}%
\put(350,300){\makebox(0,0)[r]{1e-09}}%
\end{picture}%
\endgroup
 

%% file: bhb_fig13.tex
\begingroup%
  \makeatletter%
  \newcommand{\GNUPLOTspecial}{%
    \@sanitize\catcode`\%=14\relax\special}%
  \setlength{\unitlength}{0.1bp}%
{\GNUPLOTspecial{!
/gnudict 256 dict def
gnudict begin
/Color false def
/Solid false def
/gnulinewidth 5.000 def
/userlinewidth gnulinewidth def
/vshift -33 def
/dl {10 mul} def
/hpt_ 31.5 def
/vpt_ 31.5 def
/hpt hpt_ def
/vpt vpt_ def
/M {moveto} bind def
/L {lineto} bind def
/R {rmoveto} bind def
/V {rlineto} bind def
/vpt2 vpt 2 mul def
/hpt2 hpt 2 mul def
/Lshow { currentpoint stroke M
  0 vshift R show } def
/Rshow { currentpoint stroke M
  dup stringwidth pop neg vshift R show } def
/Cshow { currentpoint stroke M
  dup stringwidth pop -2 div vshift R show } def
/UP { dup vpt_ mul /vpt exch def hpt_ mul /hpt exch def
  /hpt2 hpt 2 mul def /vpt2 vpt 2 mul def } def
/DL { Color {setrgbcolor Solid {pop []} if 0 setdash }
 {pop pop pop Solid {pop []} if 0 setdash} ifelse } def
/BL { stroke gnulinewidth 2 mul setlinewidth } def
/AL { stroke gnulinewidth 2 div setlinewidth } def
/UL { gnulinewidth mul /userlinewidth exch def } def
/PL { stroke userlinewidth setlinewidth } def
/LTb { BL [] 0 0 0 DL } def
/LTa { AL [1 dl 2 dl] 0 setdash 0 0 0 setrgbcolor } def
/LT0 { PL [] 1 0 0 DL } def
/LT1 { PL [4 dl 2 dl] 0 1 0 DL } def
/LT2 { PL [2 dl 3 dl] 0 0 1 DL } def
/LT3 { PL [1 dl 1.5 dl] 1 0 1 DL } def
/LT4 { PL [5 dl 2 dl 1 dl 2 dl] 0 1 1 DL } def
/LT5 { PL [4 dl 3 dl 1 dl 3 dl] 1 1 0 DL } def
/LT6 { PL [2 dl 2 dl 2 dl 4 dl] 0 0 0 DL } def
/LT7 { PL [2 dl 2 dl 2 dl 2 dl 2 dl 4 dl] 1 0.3 0 DL } def
/LT8 { PL [2 dl 2 dl 2 dl 2 dl 2 dl 2 dl 2 dl 4 dl] 0.5 0.5 0.5 DL } def
/Pnt { stroke [] 0 setdash
   gsave 1 setlinecap M 0 0 V stroke grestore } def
/Dia { stroke [] 0 setdash 2 copy vpt add M
  hpt neg vpt neg V hpt vpt neg V
  hpt vpt V hpt neg vpt V closepath stroke
  Pnt } def
/Pls { stroke [] 0 setdash vpt sub M 0 vpt2 V
  currentpoint stroke M
  hpt neg vpt neg R hpt2 0 V stroke
  } def
/Box { stroke [] 0 setdash 2 copy exch hpt sub exch vpt add M
  0 vpt2 neg V hpt2 0 V 0 vpt2 V
  hpt2 neg 0 V closepath stroke
  Pnt } def
/Crs { stroke [] 0 setdash exch hpt sub exch vpt add M
  hpt2 vpt2 neg V currentpoint stroke M
  hpt2 neg 0 R hpt2 vpt2 V stroke } def
/TriU { stroke [] 0 setdash 2 copy vpt 1.12 mul add M
  hpt neg vpt -1.62 mul V
  hpt 2 mul 0 V
  hpt neg vpt 1.62 mul V closepath stroke
  Pnt  } def
/Star { 2 copy Pls Crs } def
/BoxF { stroke [] 0 setdash exch hpt sub exch vpt add M
  0 vpt2 neg V  hpt2 0 V  0 vpt2 V
  hpt2 neg 0 V  closepath fill } def
/TriUF { stroke [] 0 setdash vpt 1.12 mul add M
  hpt neg vpt -1.62 mul V
  hpt 2 mul 0 V
  hpt neg vpt 1.62 mul V closepath fill } def
/TriD { stroke [] 0 setdash 2 copy vpt 1.12 mul sub M
  hpt neg vpt 1.62 mul V
  hpt 2 mul 0 V
  hpt neg vpt -1.62 mul V closepath stroke
  Pnt  } def
/TriDF { stroke [] 0 setdash vpt 1.12 mul sub M
  hpt neg vpt 1.62 mul V
  hpt 2 mul 0 V
  hpt neg vpt -1.62 mul V closepath fill} def
/DiaF { stroke [] 0 setdash vpt add M
  hpt neg vpt neg V hpt vpt neg V
  hpt vpt V hpt neg vpt V closepath fill } def
/Pent { stroke [] 0 setdash 2 copy gsave
  translate 0 hpt M 4 {72 rotate 0 hpt L} repeat
  closepath stroke grestore Pnt } def
/PentF { stroke [] 0 setdash gsave
  translate 0 hpt M 4 {72 rotate 0 hpt L} repeat
  closepath fill grestore } def
/Circle { stroke [] 0 setdash 2 copy
  hpt 0 360 arc stroke Pnt } def
/CircleF { stroke [] 0 setdash hpt 0 360 arc fill } def
/C0 { BL [] 0 setdash 2 copy moveto vpt 90 450  arc } bind def
/C1 { BL [] 0 setdash 2 copy        moveto
       2 copy  vpt 0 90 arc closepath fill
               vpt 0 360 arc closepath } bind def
/C2 { BL [] 0 setdash 2 copy moveto
       2 copy  vpt 90 180 arc closepath fill
               vpt 0 360 arc closepath } bind def
/C3 { BL [] 0 setdash 2 copy moveto
       2 copy  vpt 0 180 arc closepath fill
               vpt 0 360 arc closepath } bind def
/C4 { BL [] 0 setdash 2 copy moveto
       2 copy  vpt 180 270 arc closepath fill
               vpt 0 360 arc closepath } bind def
/C5 { BL [] 0 setdash 2 copy moveto
       2 copy  vpt 0 90 arc
       2 copy moveto
       2 copy  vpt 180 270 arc closepath fill
               vpt 0 360 arc } bind def
/C6 { BL [] 0 setdash 2 copy moveto
      2 copy  vpt 90 270 arc closepath fill
              vpt 0 360 arc closepath } bind def
/C7 { BL [] 0 setdash 2 copy moveto
      2 copy  vpt 0 270 arc closepath fill
              vpt 0 360 arc closepath } bind def
/C8 { BL [] 0 setdash 2 copy moveto
      2 copy vpt 270 360 arc closepath fill
              vpt 0 360 arc closepath } bind def
/C9 { BL [] 0 setdash 2 copy moveto
      2 copy  vpt 270 450 arc closepath fill
              vpt 0 360 arc closepath } bind def
/C10 { BL [] 0 setdash 2 copy 2 copy moveto vpt 270 360 arc closepath fill
       2 copy moveto
       2 copy vpt 90 180 arc closepath fill
               vpt 0 360 arc closepath } bind def
/C11 { BL [] 0 setdash 2 copy moveto
       2 copy  vpt 0 180 arc closepath fill
       2 copy moveto
       2 copy  vpt 270 360 arc closepath fill
               vpt 0 360 arc closepath } bind def
/C12 { BL [] 0 setdash 2 copy moveto
       2 copy  vpt 180 360 arc closepath fill
               vpt 0 360 arc closepath } bind def
/C13 { BL [] 0 setdash  2 copy moveto
       2 copy  vpt 0 90 arc closepath fill
       2 copy moveto
       2 copy  vpt 180 360 arc closepath fill
               vpt 0 360 arc closepath } bind def
/C14 { BL [] 0 setdash 2 copy moveto
       2 copy  vpt 90 360 arc closepath fill
               vpt 0 360 arc } bind def
/C15 { BL [] 0 setdash 2 copy vpt 0 360 arc closepath fill
               vpt 0 360 arc closepath } bind def
/Rec   { newpath 4 2 roll moveto 1 index 0 rlineto 0 exch rlineto
       neg 0 rlineto closepath } bind def
/Square { dup Rec } bind def
/Bsquare { vpt sub exch vpt sub exch vpt2 Square } bind def
/S0 { BL [] 0 setdash 2 copy moveto 0 vpt rlineto BL Bsquare } bind def
/S1 { BL [] 0 setdash 2 copy vpt Square fill Bsquare } bind def
/S2 { BL [] 0 setdash 2 copy exch vpt sub exch vpt Square fill Bsquare } bind def
/S3 { BL [] 0 setdash 2 copy exch vpt sub exch vpt2 vpt Rec fill Bsquare } bind def
/S4 { BL [] 0 setdash 2 copy exch vpt sub exch vpt sub vpt Square fill Bsquare } bind def
/S5 { BL [] 0 setdash 2 copy 2 copy vpt Square fill
       exch vpt sub exch vpt sub vpt Square fill Bsquare } bind def
/S6 { BL [] 0 setdash 2 copy exch vpt sub exch vpt sub vpt vpt2 Rec fill Bsquare } bind def
/S7 { BL [] 0 setdash 2 copy exch vpt sub exch vpt sub vpt vpt2 Rec fill
       2 copy vpt Square fill
       Bsquare } bind def
/S8 { BL [] 0 setdash 2 copy vpt sub vpt Square fill Bsquare } bind def
/S9 { BL [] 0 setdash 2 copy vpt sub vpt vpt2 Rec fill Bsquare } bind def
/S10 { BL [] 0 setdash 2 copy vpt sub vpt Square fill 2 copy exch vpt sub exch vpt Square fill
       Bsquare } bind def
/S11 { BL [] 0 setdash 2 copy vpt sub vpt Square fill 2 copy exch vpt sub exch vpt2 vpt Rec fill
       Bsquare } bind def
/S12 { BL [] 0 setdash 2 copy exch vpt sub exch vpt sub vpt2 vpt Rec fill Bsquare } bind def
/S13 { BL [] 0 setdash 2 copy exch vpt sub exch vpt sub vpt2 vpt Rec fill
       2 copy vpt Square fill Bsquare } bind def
/S14 { BL [] 0 setdash 2 copy exch vpt sub exch vpt sub vpt2 vpt Rec fill
       2 copy exch vpt sub exch vpt Square fill Bsquare } bind def
/S15 { BL [] 0 setdash 2 copy Bsquare fill Bsquare } bind def
/D0 { gsave translate 45 rotate 0 0 S0 stroke grestore } bind def
/D1 { gsave translate 45 rotate 0 0 S1 stroke grestore } bind def
/D2 { gsave translate 45 rotate 0 0 S2 stroke grestore } bind def
/D3 { gsave translate 45 rotate 0 0 S3 stroke grestore } bind def
/D4 { gsave translate 45 rotate 0 0 S4 stroke grestore } bind def
/D5 { gsave translate 45 rotate 0 0 S5 stroke grestore } bind def
/D6 { gsave translate 45 rotate 0 0 S6 stroke grestore } bind def
/D7 { gsave translate 45 rotate 0 0 S7 stroke grestore } bind def
/D8 { gsave translate 45 rotate 0 0 S8 stroke grestore } bind def
/D9 { gsave translate 45 rotate 0 0 S9 stroke grestore } bind def
/D10 { gsave translate 45 rotate 0 0 S10 stroke grestore } bind def
/D11 { gsave translate 45 rotate 0 0 S11 stroke grestore } bind def
/D12 { gsave translate 45 rotate 0 0 S12 stroke grestore } bind def
/D13 { gsave translate 45 rotate 0 0 S13 stroke grestore } bind def
/D14 { gsave translate 45 rotate 0 0 S14 stroke grestore } bind def
/D15 { gsave translate 45 rotate 0 0 S15 stroke grestore } bind def
/DiaE { stroke [] 0 setdash vpt add M
  hpt neg vpt neg V hpt vpt neg V
  hpt vpt V hpt neg vpt V closepath stroke } def
/BoxE { stroke [] 0 setdash exch hpt sub exch vpt add M
  0 vpt2 neg V hpt2 0 V 0 vpt2 V
  hpt2 neg 0 V closepath stroke } def
/TriUE { stroke [] 0 setdash vpt 1.12 mul add M
  hpt neg vpt -1.62 mul V
  hpt 2 mul 0 V
  hpt neg vpt 1.62 mul V closepath stroke } def
/TriDE { stroke [] 0 setdash vpt 1.12 mul sub M
  hpt neg vpt 1.62 mul V
  hpt 2 mul 0 V
  hpt neg vpt -1.62 mul V closepath stroke } def
/PentE { stroke [] 0 setdash gsave
  translate 0 hpt M 4 {72 rotate 0 hpt L} repeat
  closepath stroke grestore } def
/CircE { stroke [] 0 setdash 
  hpt 0 360 arc stroke } def
/Opaque { gsave closepath 1 setgray fill grestore 0 setgray closepath } def
/DiaW { stroke [] 0 setdash vpt add M
  hpt neg vpt neg V hpt vpt neg V
  hpt vpt V hpt neg vpt V Opaque stroke } def
/BoxW { stroke [] 0 setdash exch hpt sub exch vpt add M
  0 vpt2 neg V hpt2 0 V 0 vpt2 V
  hpt2 neg 0 V Opaque stroke } def
/TriUW { stroke [] 0 setdash vpt 1.12 mul add M
  hpt neg vpt -1.62 mul V
  hpt 2 mul 0 V
  hpt neg vpt 1.62 mul V Opaque stroke } def
/TriDW { stroke [] 0 setdash vpt 1.12 mul sub M
  hpt neg vpt 1.62 mul V
  hpt 2 mul 0 V
  hpt neg vpt -1.62 mul V Opaque stroke } def
/PentW { stroke [] 0 setdash gsave
  translate 0 hpt M 4 {72 rotate 0 hpt L} repeat
  Opaque stroke grestore } def
/CircW { stroke [] 0 setdash 
  hpt 0 360 arc Opaque stroke } def
/BoxFill { gsave Rec 1 setgray fill grestore } def
end
}}%
\begin{picture}(2412,2160)(0,0)%
{\GNUPLOTspecial{"
gnudict begin
gsave
0 0 translate
0.100 0.100 scale
0 setgray
newpath
1.000 UL
LTb
0.050 UL
LT3
400 300 M
2012 0 V
1.000 UL
LTb
400 300 M
63 0 V
1949 0 R
-63 0 V
400 433 M
31 0 V
1981 0 R
-31 0 V
0.050 UL
LT3
400 566 M
2012 0 V
1.000 UL
LTb
400 566 M
63 0 V
1949 0 R
-63 0 V
400 699 M
31 0 V
1981 0 R
-31 0 V
0.050 UL
LT3
400 831 M
2012 0 V
1.000 UL
LTb
400 831 M
63 0 V
1949 0 R
-63 0 V
400 964 M
31 0 V
1981 0 R
-31 0 V
0.050 UL
LT3
400 1097 M
2012 0 V
1.000 UL
LTb
400 1097 M
63 0 V
1949 0 R
-63 0 V
400 1230 M
31 0 V
1981 0 R
-31 0 V
0.050 UL
LT3
400 1363 M
2012 0 V
1.000 UL
LTb
400 1363 M
63 0 V
1949 0 R
-63 0 V
400 1496 M
31 0 V
1981 0 R
-31 0 V
0.050 UL
LT3
400 1629 M
2012 0 V
1.000 UL
LTb
400 1629 M
63 0 V
1949 0 R
-63 0 V
400 1761 M
31 0 V
1981 0 R
-31 0 V
0.050 UL
LT3
400 1894 M
2012 0 V
1.000 UL
LTb
400 1894 M
63 0 V
1949 0 R
-63 0 V
400 2027 M
31 0 V
1981 0 R
-31 0 V
0.050 UL
LT3
400 2160 M
2012 0 V
1.000 UL
LTb
400 2160 M
63 0 V
1949 0 R
-63 0 V
0.050 UL
LT3
400 300 M
0 1860 V
1.000 UL
LTb
400 300 M
0 63 V
0 1797 R
0 -63 V
551 300 M
0 31 V
0 1829 R
0 -31 V
640 300 M
0 31 V
0 1829 R
0 -31 V
703 300 M
0 31 V
0 1829 R
0 -31 V
752 300 M
0 31 V
0 1829 R
0 -31 V
791 300 M
0 31 V
0 1829 R
0 -31 V
825 300 M
0 31 V
0 1829 R
0 -31 V
854 300 M
0 31 V
0 1829 R
0 -31 V
880 300 M
0 31 V
0 1829 R
0 -31 V
0.050 UL
LT3
903 300 M
0 63 V
0 200 R
0 1597 V
1.000 UL
LTb
903 300 M
0 63 V
0 1797 R
0 -63 V
1054 300 M
0 31 V
0 1829 R
0 -31 V
1143 300 M
0 31 V
0 1829 R
0 -31 V
1206 300 M
0 31 V
0 1829 R
0 -31 V
1255 300 M
0 31 V
0 1829 R
0 -31 V
1294 300 M
0 31 V
0 1829 R
0 -31 V
1328 300 M
0 31 V
0 1829 R
0 -31 V
1357 300 M
0 31 V
0 1829 R
0 -31 V
1383 300 M
0 31 V
0 1829 R
0 -31 V
0.050 UL
LT3
1406 300 M
0 63 V
0 200 R
0 1597 V
1.000 UL
LTb
1406 300 M
0 63 V
0 1797 R
0 -63 V
1557 300 M
0 31 V
0 1829 R
0 -31 V
1646 300 M
0 31 V
0 1829 R
0 -31 V
1709 300 M
0 31 V
0 1829 R
0 -31 V
1758 300 M
0 31 V
0 1829 R
0 -31 V
1797 300 M
0 31 V
0 1829 R
0 -31 V
1831 300 M
0 31 V
0 1829 R
0 -31 V
1860 300 M
0 31 V
0 1829 R
0 -31 V
1886 300 M
0 31 V
0 1829 R
0 -31 V
0.050 UL
LT3
1909 300 M
0 1860 V
1.000 UL
LTb
1909 300 M
0 63 V
0 1797 R
0 -63 V
2060 300 M
0 31 V
0 1829 R
0 -31 V
2149 300 M
0 31 V
0 1829 R
0 -31 V
2212 300 M
0 31 V
0 1829 R
0 -31 V
2261 300 M
0 31 V
0 1829 R
0 -31 V
2300 300 M
0 31 V
0 1829 R
0 -31 V
2334 300 M
0 31 V
0 1829 R
0 -31 V
2363 300 M
0 31 V
0 1829 R
0 -31 V
2389 300 M
0 31 V
0 1829 R
0 -31 V
0.050 UL
LT3
2412 300 M
0 1860 V
1.000 UL
LTb
2412 300 M
0 63 V
0 1797 R
0 -63 V
1.000 UL
LTb
400 300 M
2012 0 V
0 1860 V
-2012 0 V
400 300 L
1.000 UL
LTb
450 363 M
0 200 V
987 0 V
0 -200 V
-987 0 V
0 200 R
987 0 V
1.000 UL
LT0
1150 513 M
237 0 V
400 884 M
20 11 V
21 11 V
20 11 V
20 10 V
21 11 V
20 11 V
20 10 V
21 11 V
20 11 V
20 11 V
21 10 V
20 11 V
20 11 V
21 11 V
20 10 V
20 11 V
20 11 V
21 11 V
20 10 V
20 11 V
21 11 V
20 10 V
20 11 V
21 11 V
20 11 V
20 10 V
21 11 V
20 11 V
20 11 V
21 10 V
20 11 V
20 11 V
21 11 V
20 10 V
20 11 V
21 11 V
20 11 V
20 10 V
21 11 V
20 11 V
20 10 V
21 11 V
20 11 V
20 11 V
21 10 V
20 11 V
20 11 V
21 11 V
20 10 V
20 11 V
20 11 V
21 11 V
20 10 V
20 11 V
21 11 V
20 11 V
20 10 V
21 11 V
20 11 V
20 10 V
21 11 V
20 11 V
20 11 V
21 10 V
20 11 V
20 11 V
21 11 V
20 10 V
20 11 V
21 11 V
20 11 V
20 10 V
21 11 V
20 11 V
20 10 V
21 11 V
20 11 V
20 11 V
21 10 V
20 11 V
20 11 V
21 11 V
20 10 V
20 11 V
20 11 V
21 11 V
20 10 V
20 11 V
21 11 V
20 11 V
20 10 V
21 11 V
20 11 V
20 10 V
21 11 V
20 11 V
20 11 V
21 10 V
20 11 V
1.000 UL
LT2
1150 413 M
237 0 V
400 2004 M
20 -16 V
21 -16 V
20 -16 V
20 -16 V
21 -16 V
20 -16 V
20 -16 V
21 -17 V
20 -16 V
20 -16 V
21 -16 V
20 -16 V
20 -16 V
21 -16 V
20 -16 V
20 -16 V
20 -17 V
21 -16 V
20 -16 V
20 -16 V
21 -16 V
20 -16 V
20 -16 V
21 -16 V
20 -16 V
20 -16 V
21 -17 V
20 -16 V
20 -16 V
21 -16 V
20 -16 V
20 -16 V
21 -16 V
20 -16 V
20 -16 V
21 -17 V
20 -16 V
20 -16 V
21 -16 V
20 -16 V
20 -16 V
21 -16 V
20 -16 V
20 -16 V
21 -16 V
20 -17 V
20 -16 V
21 -16 V
20 -16 V
20 -16 V
20 -16 V
21 -16 V
20 -16 V
20 -16 V
21 -16 V
20 -17 V
20 -16 V
21 -16 V
20 -16 V
20 -16 V
21 -16 V
20 -16 V
20 -16 V
21 -16 V
20 -17 V
20 -16 V
21 -16 V
20 -16 V
20 -16 V
21 -16 V
20 -16 V
20 -16 V
21 -16 V
20 -16 V
20 -17 V
21 -16 V
20 -16 V
20 -16 V
21 -16 V
20 -16 V
20 -16 V
21 -16 V
20 -16 V
20 -16 V
20 -17 V
21 -16 V
20 -16 V
20 -16 V
21 -16 V
20 -16 V
20 -16 V
21 -16 V
20 -16 V
20 -17 V
21 -16 V
20 -16 V
20 -16 V
21 -16 V
20 -16 V
stroke
grestore
end
showpage
}}%
\put(1100,413){\makebox(0,0)[r]{grav. radiation}}%
\put(1100,513){\makebox(0,0)[r]{star ejection}}%
\put(1406,2310){\makebox(0,0){\textsf{Shrinking rate $\frac{d\tilde{a}}{d\tilde{t}}$}}}%
\put(1406,50){\makebox(0,0){\sffamily $\tilde{a}$}}%
\put(50,1230){%
\special{ps: gsave currentpoint currentpoint translate
270 rotate neg exch neg exch translate}%
\makebox(0,0)[b]{\shortstack{$\frac{d\tilde{a}}{d\tilde{t}}$}}%
\special{ps: currentpoint grestore moveto}%
}%
\put(2412,200){\makebox(0,0){1}}%
\put(1909,200){\makebox(0,0){0.1}}%
\put(1406,200){\makebox(0,0){0.01}}%
\put(903,200){\makebox(0,0){0.001}}%
\put(400,200){\makebox(0,0){0.0001}}%
\put(350,2160){\makebox(0,0)[r]{1}}%
\put(350,1894){\makebox(0,0)[r]{0.01}}%
\put(350,1629){\makebox(0,0)[r]{0.0001}}%
\put(350,1363){\makebox(0,0)[r]{1e-06}}%
\put(350,1097){\makebox(0,0)[r]{1e-08}}%
\put(350,831){\makebox(0,0)[r]{1e-10}}%
\put(350,566){\makebox(0,0)[r]{1e-12}}%
\put(350,300){\makebox(0,0)[r]{1e-14}}%
\end{picture}%
\endgroup
 

%% file: bhb_fig15.tex
\begingroup%
  \makeatletter%
  \newcommand{\GNUPLOTspecial}{%
    \@sanitize\catcode`\%=14\relax\special}%
  \setlength{\unitlength}{0.1bp}%
{\GNUPLOTspecial{!
/gnudict 256 dict def
gnudict begin
/Color false def
/Solid false def
/gnulinewidth 5.000 def
/userlinewidth gnulinewidth def
/vshift -33 def
/dl {10 mul} def
/hpt_ 31.5 def
/vpt_ 31.5 def
/hpt hpt_ def
/vpt vpt_ def
/M {moveto} bind def
/L {lineto} bind def
/R {rmoveto} bind def
/V {rlineto} bind def
/vpt2 vpt 2 mul def
/hpt2 hpt 2 mul def
/Lshow { currentpoint stroke M
  0 vshift R show } def
/Rshow { currentpoint stroke M
  dup stringwidth pop neg vshift R show } def
/Cshow { currentpoint stroke M
  dup stringwidth pop -2 div vshift R show } def
/UP { dup vpt_ mul /vpt exch def hpt_ mul /hpt exch def
  /hpt2 hpt 2 mul def /vpt2 vpt 2 mul def } def
/DL { Color {setrgbcolor Solid {pop []} if 0 setdash }
 {pop pop pop Solid {pop []} if 0 setdash} ifelse } def
/BL { stroke gnulinewidth 2 mul setlinewidth } def
/AL { stroke gnulinewidth 2 div setlinewidth } def
/UL { gnulinewidth mul /userlinewidth exch def } def
/PL { stroke userlinewidth setlinewidth } def
/LTb { BL [] 0 0 0 DL } def
/LTa { AL [1 dl 2 dl] 0 setdash 0 0 0 setrgbcolor } def
/LT0 { PL [] 1 0 0 DL } def
/LT1 { PL [4 dl 2 dl] 0 1 0 DL } def
/LT2 { PL [2 dl 3 dl] 0 0 1 DL } def
/LT3 { PL [1 dl 1.5 dl] 1 0 1 DL } def
/LT4 { PL [5 dl 2 dl 1 dl 2 dl] 0 1 1 DL } def
/LT5 { PL [4 dl 3 dl 1 dl 3 dl] 1 1 0 DL } def
/LT6 { PL [2 dl 2 dl 2 dl 4 dl] 0 0 0 DL } def
/LT7 { PL [2 dl 2 dl 2 dl 2 dl 2 dl 4 dl] 1 0.3 0 DL } def
/LT8 { PL [2 dl 2 dl 2 dl 2 dl 2 dl 2 dl 2 dl 4 dl] 0.5 0.5 0.5 DL } def
/Pnt { stroke [] 0 setdash
   gsave 1 setlinecap M 0 0 V stroke grestore } def
/Dia { stroke [] 0 setdash 2 copy vpt add M
  hpt neg vpt neg V hpt vpt neg V
  hpt vpt V hpt neg vpt V closepath stroke
  Pnt } def
/Pls { stroke [] 0 setdash vpt sub M 0 vpt2 V
  currentpoint stroke M
  hpt neg vpt neg R hpt2 0 V stroke
  } def
/Box { stroke [] 0 setdash 2 copy exch hpt sub exch vpt add M
  0 vpt2 neg V hpt2 0 V 0 vpt2 V
  hpt2 neg 0 V closepath stroke
  Pnt } def
/Crs { stroke [] 0 setdash exch hpt sub exch vpt add M
  hpt2 vpt2 neg V currentpoint stroke M
  hpt2 neg 0 R hpt2 vpt2 V stroke } def
/TriU { stroke [] 0 setdash 2 copy vpt 1.12 mul add M
  hpt neg vpt -1.62 mul V
  hpt 2 mul 0 V
  hpt neg vpt 1.62 mul V closepath stroke
  Pnt  } def
/Star { 2 copy Pls Crs } def
/BoxF { stroke [] 0 setdash exch hpt sub exch vpt add M
  0 vpt2 neg V  hpt2 0 V  0 vpt2 V
  hpt2 neg 0 V  closepath fill } def
/TriUF { stroke [] 0 setdash vpt 1.12 mul add M
  hpt neg vpt -1.62 mul V
  hpt 2 mul 0 V
  hpt neg vpt 1.62 mul V closepath fill } def
/TriD { stroke [] 0 setdash 2 copy vpt 1.12 mul sub M
  hpt neg vpt 1.62 mul V
  hpt 2 mul 0 V
  hpt neg vpt -1.62 mul V closepath stroke
  Pnt  } def
/TriDF { stroke [] 0 setdash vpt 1.12 mul sub M
  hpt neg vpt 1.62 mul V
  hpt 2 mul 0 V
  hpt neg vpt -1.62 mul V closepath fill} def
/DiaF { stroke [] 0 setdash vpt add M
  hpt neg vpt neg V hpt vpt neg V
  hpt vpt V hpt neg vpt V closepath fill } def
/Pent { stroke [] 0 setdash 2 copy gsave
  translate 0 hpt M 4 {72 rotate 0 hpt L} repeat
  closepath stroke grestore Pnt } def
/PentF { stroke [] 0 setdash gsave
  translate 0 hpt M 4 {72 rotate 0 hpt L} repeat
  closepath fill grestore } def
/Circle { stroke [] 0 setdash 2 copy
  hpt 0 360 arc stroke Pnt } def
/CircleF { stroke [] 0 setdash hpt 0 360 arc fill } def
/C0 { BL [] 0 setdash 2 copy moveto vpt 90 450  arc } bind def
/C1 { BL [] 0 setdash 2 copy        moveto
       2 copy  vpt 0 90 arc closepath fill
               vpt 0 360 arc closepath } bind def
/C2 { BL [] 0 setdash 2 copy moveto
       2 copy  vpt 90 180 arc closepath fill
               vpt 0 360 arc closepath } bind def
/C3 { BL [] 0 setdash 2 copy moveto
       2 copy  vpt 0 180 arc closepath fill
               vpt 0 360 arc closepath } bind def
/C4 { BL [] 0 setdash 2 copy moveto
       2 copy  vpt 180 270 arc closepath fill
               vpt 0 360 arc closepath } bind def
/C5 { BL [] 0 setdash 2 copy moveto
       2 copy  vpt 0 90 arc
       2 copy moveto
       2 copy  vpt 180 270 arc closepath fill
               vpt 0 360 arc } bind def
/C6 { BL [] 0 setdash 2 copy moveto
      2 copy  vpt 90 270 arc closepath fill
              vpt 0 360 arc closepath } bind def
/C7 { BL [] 0 setdash 2 copy moveto
      2 copy  vpt 0 270 arc closepath fill
              vpt 0 360 arc closepath } bind def
/C8 { BL [] 0 setdash 2 copy moveto
      2 copy vpt 270 360 arc closepath fill
              vpt 0 360 arc closepath } bind def
/C9 { BL [] 0 setdash 2 copy moveto
      2 copy  vpt 270 450 arc closepath fill
              vpt 0 360 arc closepath } bind def
/C10 { BL [] 0 setdash 2 copy 2 copy moveto vpt 270 360 arc closepath fill
       2 copy moveto
       2 copy vpt 90 180 arc closepath fill
               vpt 0 360 arc closepath } bind def
/C11 { BL [] 0 setdash 2 copy moveto
       2 copy  vpt 0 180 arc closepath fill
       2 copy moveto
       2 copy  vpt 270 360 arc closepath fill
               vpt 0 360 arc closepath } bind def
/C12 { BL [] 0 setdash 2 copy moveto
       2 copy  vpt 180 360 arc closepath fill
               vpt 0 360 arc closepath } bind def
/C13 { BL [] 0 setdash  2 copy moveto
       2 copy  vpt 0 90 arc closepath fill
       2 copy moveto
       2 copy  vpt 180 360 arc closepath fill
               vpt 0 360 arc closepath } bind def
/C14 { BL [] 0 setdash 2 copy moveto
       2 copy  vpt 90 360 arc closepath fill
               vpt 0 360 arc } bind def
/C15 { BL [] 0 setdash 2 copy vpt 0 360 arc closepath fill
               vpt 0 360 arc closepath } bind def
/Rec   { newpath 4 2 roll moveto 1 index 0 rlineto 0 exch rlineto
       neg 0 rlineto closepath } bind def
/Square { dup Rec } bind def
/Bsquare { vpt sub exch vpt sub exch vpt2 Square } bind def
/S0 { BL [] 0 setdash 2 copy moveto 0 vpt rlineto BL Bsquare } bind def
/S1 { BL [] 0 setdash 2 copy vpt Square fill Bsquare } bind def
/S2 { BL [] 0 setdash 2 copy exch vpt sub exch vpt Square fill Bsquare } bind def
/S3 { BL [] 0 setdash 2 copy exch vpt sub exch vpt2 vpt Rec fill Bsquare } bind def
/S4 { BL [] 0 setdash 2 copy exch vpt sub exch vpt sub vpt Square fill Bsquare } bind def
/S5 { BL [] 0 setdash 2 copy 2 copy vpt Square fill
       exch vpt sub exch vpt sub vpt Square fill Bsquare } bind def
/S6 { BL [] 0 setdash 2 copy exch vpt sub exch vpt sub vpt vpt2 Rec fill Bsquare } bind def
/S7 { BL [] 0 setdash 2 copy exch vpt sub exch vpt sub vpt vpt2 Rec fill
       2 copy vpt Square fill
       Bsquare } bind def
/S8 { BL [] 0 setdash 2 copy vpt sub vpt Square fill Bsquare } bind def
/S9 { BL [] 0 setdash 2 copy vpt sub vpt vpt2 Rec fill Bsquare } bind def
/S10 { BL [] 0 setdash 2 copy vpt sub vpt Square fill 2 copy exch vpt sub exch vpt Square fill
       Bsquare } bind def
/S11 { BL [] 0 setdash 2 copy vpt sub vpt Square fill 2 copy exch vpt sub exch vpt2 vpt Rec fill
       Bsquare } bind def
/S12 { BL [] 0 setdash 2 copy exch vpt sub exch vpt sub vpt2 vpt Rec fill Bsquare } bind def
/S13 { BL [] 0 setdash 2 copy exch vpt sub exch vpt sub vpt2 vpt Rec fill
       2 copy vpt Square fill Bsquare } bind def
/S14 { BL [] 0 setdash 2 copy exch vpt sub exch vpt sub vpt2 vpt Rec fill
       2 copy exch vpt sub exch vpt Square fill Bsquare } bind def
/S15 { BL [] 0 setdash 2 copy Bsquare fill Bsquare } bind def
/D0 { gsave translate 45 rotate 0 0 S0 stroke grestore } bind def
/D1 { gsave translate 45 rotate 0 0 S1 stroke grestore } bind def
/D2 { gsave translate 45 rotate 0 0 S2 stroke grestore } bind def
/D3 { gsave translate 45 rotate 0 0 S3 stroke grestore } bind def
/D4 { gsave translate 45 rotate 0 0 S4 stroke grestore } bind def
/D5 { gsave translate 45 rotate 0 0 S5 stroke grestore } bind def
/D6 { gsave translate 45 rotate 0 0 S6 stroke grestore } bind def
/D7 { gsave translate 45 rotate 0 0 S7 stroke grestore } bind def
/D8 { gsave translate 45 rotate 0 0 S8 stroke grestore } bind def
/D9 { gsave translate 45 rotate 0 0 S9 stroke grestore } bind def
/D10 { gsave translate 45 rotate 0 0 S10 stroke grestore } bind def
/D11 { gsave translate 45 rotate 0 0 S11 stroke grestore } bind def
/D12 { gsave translate 45 rotate 0 0 S12 stroke grestore } bind def
/D13 { gsave translate 45 rotate 0 0 S13 stroke grestore } bind def
/D14 { gsave translate 45 rotate 0 0 S14 stroke grestore } bind def
/D15 { gsave translate 45 rotate 0 0 S15 stroke grestore } bind def
/DiaE { stroke [] 0 setdash vpt add M
  hpt neg vpt neg V hpt vpt neg V
  hpt vpt V hpt neg vpt V closepath stroke } def
/BoxE { stroke [] 0 setdash exch hpt sub exch vpt add M
  0 vpt2 neg V hpt2 0 V 0 vpt2 V
  hpt2 neg 0 V closepath stroke } def
/TriUE { stroke [] 0 setdash vpt 1.12 mul add M
  hpt neg vpt -1.62 mul V
  hpt 2 mul 0 V
  hpt neg vpt 1.62 mul V closepath stroke } def
/TriDE { stroke [] 0 setdash vpt 1.12 mul sub M
  hpt neg vpt 1.62 mul V
  hpt 2 mul 0 V
  hpt neg vpt -1.62 mul V closepath stroke } def
/PentE { stroke [] 0 setdash gsave
  translate 0 hpt M 4 {72 rotate 0 hpt L} repeat
  closepath stroke grestore } def
/CircE { stroke [] 0 setdash 
  hpt 0 360 arc stroke } def
/Opaque { gsave closepath 1 setgray fill grestore 0 setgray closepath } def
/DiaW { stroke [] 0 setdash vpt add M
  hpt neg vpt neg V hpt vpt neg V
  hpt vpt V hpt neg vpt V Opaque stroke } def
/BoxW { stroke [] 0 setdash exch hpt sub exch vpt add M
  0 vpt2 neg V hpt2 0 V 0 vpt2 V
  hpt2 neg 0 V Opaque stroke } def
/TriUW { stroke [] 0 setdash vpt 1.12 mul add M
  hpt neg vpt -1.62 mul V
  hpt 2 mul 0 V
  hpt neg vpt 1.62 mul V Opaque stroke } def
/TriDW { stroke [] 0 setdash vpt 1.12 mul sub M
  hpt neg vpt 1.62 mul V
  hpt 2 mul 0 V
  hpt neg vpt -1.62 mul V Opaque stroke } def
/PentW { stroke [] 0 setdash gsave
  translate 0 hpt M 4 {72 rotate 0 hpt L} repeat
  Opaque stroke grestore } def
/CircW { stroke [] 0 setdash 
  hpt 0 360 arc Opaque stroke } def
/BoxFill { gsave Rec 1 setgray fill grestore } def
end
}}%
\begin{picture}(2412,2160)(0,0)%
{\GNUPLOTspecial{"
gnudict begin
gsave
0 0 translate
0.100 0.100 scale
0 setgray
newpath
1.000 UL
LTb
0.050 UL
LT3
350 300 M
2062 0 V
1.000 UL
LTb
350 300 M
63 0 V
1999 0 R
-63 0 V
0.050 UL
LT3
350 610 M
2062 0 V
1.000 UL
LTb
350 610 M
63 0 V
1999 0 R
-63 0 V
0.050 UL
LT3
350 920 M
2062 0 V
1.000 UL
LTb
350 920 M
63 0 V
1999 0 R
-63 0 V
0.050 UL
LT3
350 1230 M
2062 0 V
1.000 UL
LTb
350 1230 M
63 0 V
1999 0 R
-63 0 V
0.050 UL
LT3
350 1540 M
2062 0 V
1.000 UL
LTb
350 1540 M
63 0 V
1999 0 R
-63 0 V
0.050 UL
LT3
350 1850 M
2062 0 V
1.000 UL
LTb
350 1850 M
63 0 V
1999 0 R
-63 0 V
0.050 UL
LT3
350 2160 M
2062 0 V
1.000 UL
LTb
350 2160 M
63 0 V
1999 0 R
-63 0 V
0.050 UL
LT3
350 300 M
0 1860 V
1.000 UL
LTb
350 300 M
0 63 V
0 1797 R
0 -63 V
0.050 UL
LT3
556 300 M
0 1860 V
1.000 UL
LTb
556 300 M
0 63 V
0 1797 R
0 -63 V
0.050 UL
LT3
762 300 M
0 1860 V
1.000 UL
LTb
762 300 M
0 63 V
0 1797 R
0 -63 V
0.050 UL
LT3
969 300 M
0 1860 V
1.000 UL
LTb
969 300 M
0 63 V
0 1797 R
0 -63 V
0.050 UL
LT3
1175 300 M
0 1860 V
1.000 UL
LTb
1175 300 M
0 63 V
0 1797 R
0 -63 V
0.050 UL
LT3
1381 300 M
0 1860 V
1.000 UL
LTb
1381 300 M
0 63 V
0 1797 R
0 -63 V
0.050 UL
LT3
1587 300 M
0 1860 V
1.000 UL
LTb
1587 300 M
0 63 V
0 1797 R
0 -63 V
0.050 UL
LT3
1793 300 M
0 1597 V
0 200 R
0 63 V
1.000 UL
LTb
1793 300 M
0 63 V
0 1797 R
0 -63 V
0.050 UL
LT3
2000 300 M
0 1597 V
0 200 R
0 63 V
1.000 UL
LTb
2000 300 M
0 63 V
0 1797 R
0 -63 V
0.050 UL
LT3
2206 300 M
0 1597 V
0 200 R
0 63 V
1.000 UL
LTb
2206 300 M
0 63 V
0 1797 R
0 -63 V
0.050 UL
LT3
2412 300 M
0 1860 V
1.000 UL
LTb
2412 300 M
0 63 V
0 1797 R
0 -63 V
1.000 UL
LTb
350 300 M
2062 0 V
0 1860 V
-2062 0 V
350 300 L
1.000 UL
LTb
1775 1897 M
0 200 V
587 0 V
0 -200 V
-587 0 V
0 200 R
587 0 V
1.000 UL
LT2
2075 2047 M
237 0 V
384 976 M
405 872 L
20 -68 V
20 0 V
20 102 V
20 180 V
20 194 V
20 142 V
20 48 V
21 -27 V
20 -77 V
20 -102 V
20 -111 V
20 -110 V
20 -99 V
20 -76 V
21 -36 V
20 18 V
20 80 V
20 109 V
20 94 V
20 35 V
20 -25 V
20 -50 V
21 -38 V
20 1 V
20 29 V
20 41 V
20 39 V
20 38 V
20 49 V
21 73 V
20 89 V
20 64 V
20 -5 V
20 -105 V
20 -167 V
20 -162 V
20 -92 V
21 -5 V
20 45 V
20 51 V
20 25 V
20 -10 V
20 -48 V
20 -87 V
21 -100 V
20 -67 V
20 12 V
20 105 V
20 147 V
20 126 V
20 55 V
20 -1 V
21 -8 V
20 33 V
20 77 V
20 68 V
20 0 V
20 -107 V
20 -166 V
21 -152 V
20 -72 V
20 22 V
20 73 V
20 85 V
20 65 V
20 46 V
20 34 V
21 24 V
20 20 V
20 15 V
20 13 V
20 6 V
20 -9 V
20 -36 V
21 -68 V
20 -81 V
20 -64 V
20 -19 V
20 35 V
20 76 V
20 103 V
20 111 V
21 81 V
20 6 V
20 -104 V
20 -183 V
20 -180 V
20 -97 V
20 24 V
21 103 V
20 129 V
20 107 V
20 75 V
20 47 V
20 22 V
20 2 V
20 -11 V
21 -18 V
1.000 UL
LT0
2075 1947 M
237 0 V
384 527 M
21 76 V
20 63 V
20 39 V
20 3 V
20 -12 V
20 14 V
20 83 V
20 163 V
21 170 V
20 93 V
20 -54 V
20 -157 V
646 849 L
20 -61 V
20 74 V
21 145 V
20 144 V
20 84 V
20 15 V
20 -48 V
20 -101 V
827 988 L
20 -54 V
21 79 V
20 236 V
20 277 V
20 173 V
20 -57 V
20 -241 V
20 -274 V
21 -153 V
20 34 V
20 129 V
20 115 V
20 16 V
20 -38 V
20 0 V
20 131 V
21 248 V
20 236 V
20 88 V
20 -146 V
20 -273 V
20 -254 V
20 -96 V
21 63 V
20 123 V
20 84 V
20 -7 V
20 -38 V
20 2 V
20 104 V
20 161 V
21 130 V
20 11 V
20 -123 V
20 -168 V
20 -120 V
20 -4 V
20 66 V
21 56 V
20 -28 V
20 -114 V
20 -136 V
20 -95 V
20 -14 V
20 40 V
20 52 V
21 28 V
20 -8 V
20 -42 V
20 -74 V
20 -94 V
20 -83 V
20 -41 V
21 24 V
20 71 V
20 87 V
20 73 V
20 43 V
20 16 V
20 -10 V
20 -34 V
21 -52 V
20 -69 V
20 -81 V
20 -77 V
20 -49 V
20 3 V
20 53 V
21 58 V
20 11 V
20 -78 V
20 -140 V
20 -148 V
20 -102 V
20 -34 V
20 12 V
21 35 V
stroke
grestore
end
showpage
}}%
\put(2025,1947){\makebox(0,0)[r]{$\mli{\tilde{t}}{ej}$}}%
\put(2025,2047){\makebox(0,0)[r]{$\tilde{t}=0$}}%
\put(1381,2310){\makebox(0,0){\textsf{$\tilde{\rho}$ depending on the polar angle $\mli{\theta}{pos}$}}}%
\put(1381,50){\makebox(0,0){\sffamily $\cos\mli{\theta}{pos}$}}%
\put(50,1230){%
\special{ps: gsave currentpoint currentpoint translate
270 rotate neg exch neg exch translate}%
\makebox(0,0)[b]{\shortstack{\sffamily $\tilde{\rho}$}}%
\special{ps: currentpoint grestore moveto}%
}%
\put(2412,200){\makebox(0,0){1}}%
\put(2206,200){\makebox(0,0){0.8}}%
\put(2000,200){\makebox(0,0){0.6}}%
\put(1793,200){\makebox(0,0){0.4}}%
\put(1587,200){\makebox(0,0){0.2}}%
\put(1381,200){\makebox(0,0){0}}%
\put(1175,200){\makebox(0,0){-0.2}}%
\put(969,200){\makebox(0,0){-0.4}}%
\put(762,200){\makebox(0,0){-0.6}}%
\put(556,200){\makebox(0,0){-0.8}}%
\put(350,200){\makebox(0,0){-1}}%
\put(300,2160){\makebox(0,0)[r]{0.05}}%
\put(300,1850){\makebox(0,0)[r]{0.045}}%
\put(300,1540){\makebox(0,0)[r]{0.04}}%
\put(300,1230){\makebox(0,0)[r]{0.035}}%
\put(300,920){\makebox(0,0)[r]{0.03}}%
\put(300,610){\makebox(0,0)[r]{0.025}}%
\put(300,300){\makebox(0,0)[r]{0.02}}%
\end{picture}%
\endgroup
 

%% file: bhb_fig16.tex
\begingroup%
  \makeatletter%
  \newcommand{\GNUPLOTspecial}{%
    \@sanitize\catcode`\%=14\relax\special}%
  \setlength{\unitlength}{0.1bp}%
{\GNUPLOTspecial{!
/gnudict 256 dict def
gnudict begin
/Color false def
/Solid false def
/gnulinewidth 5.000 def
/userlinewidth gnulinewidth def
/vshift -33 def
/dl {10 mul} def
/hpt_ 31.5 def
/vpt_ 31.5 def
/hpt hpt_ def
/vpt vpt_ def
/M {moveto} bind def
/L {lineto} bind def
/R {rmoveto} bind def
/V {rlineto} bind def
/vpt2 vpt 2 mul def
/hpt2 hpt 2 mul def
/Lshow { currentpoint stroke M
  0 vshift R show } def
/Rshow { currentpoint stroke M
  dup stringwidth pop neg vshift R show } def
/Cshow { currentpoint stroke M
  dup stringwidth pop -2 div vshift R show } def
/UP { dup vpt_ mul /vpt exch def hpt_ mul /hpt exch def
  /hpt2 hpt 2 mul def /vpt2 vpt 2 mul def } def
/DL { Color {setrgbcolor Solid {pop []} if 0 setdash }
 {pop pop pop Solid {pop []} if 0 setdash} ifelse } def
/BL { stroke gnulinewidth 2 mul setlinewidth } def
/AL { stroke gnulinewidth 2 div setlinewidth } def
/UL { gnulinewidth mul /userlinewidth exch def } def
/PL { stroke userlinewidth setlinewidth } def
/LTb { BL [] 0 0 0 DL } def
/LTa { AL [1 dl 2 dl] 0 setdash 0 0 0 setrgbcolor } def
/LT0 { PL [] 1 0 0 DL } def
/LT1 { PL [4 dl 2 dl] 0 1 0 DL } def
/LT2 { PL [2 dl 3 dl] 0 0 1 DL } def
/LT3 { PL [1 dl 1.5 dl] 1 0 1 DL } def
/LT4 { PL [5 dl 2 dl 1 dl 2 dl] 0 1 1 DL } def
/LT5 { PL [4 dl 3 dl 1 dl 3 dl] 1 1 0 DL } def
/LT6 { PL [2 dl 2 dl 2 dl 4 dl] 0 0 0 DL } def
/LT7 { PL [2 dl 2 dl 2 dl 2 dl 2 dl 4 dl] 1 0.3 0 DL } def
/LT8 { PL [2 dl 2 dl 2 dl 2 dl 2 dl 2 dl 2 dl 4 dl] 0.5 0.5 0.5 DL } def
/Pnt { stroke [] 0 setdash
   gsave 1 setlinecap M 0 0 V stroke grestore } def
/Dia { stroke [] 0 setdash 2 copy vpt add M
  hpt neg vpt neg V hpt vpt neg V
  hpt vpt V hpt neg vpt V closepath stroke
  Pnt } def
/Pls { stroke [] 0 setdash vpt sub M 0 vpt2 V
  currentpoint stroke M
  hpt neg vpt neg R hpt2 0 V stroke
  } def
/Box { stroke [] 0 setdash 2 copy exch hpt sub exch vpt add M
  0 vpt2 neg V hpt2 0 V 0 vpt2 V
  hpt2 neg 0 V closepath stroke
  Pnt } def
/Crs { stroke [] 0 setdash exch hpt sub exch vpt add M
  hpt2 vpt2 neg V currentpoint stroke M
  hpt2 neg 0 R hpt2 vpt2 V stroke } def
/TriU { stroke [] 0 setdash 2 copy vpt 1.12 mul add M
  hpt neg vpt -1.62 mul V
  hpt 2 mul 0 V
  hpt neg vpt 1.62 mul V closepath stroke
  Pnt  } def
/Star { 2 copy Pls Crs } def
/BoxF { stroke [] 0 setdash exch hpt sub exch vpt add M
  0 vpt2 neg V  hpt2 0 V  0 vpt2 V
  hpt2 neg 0 V  closepath fill } def
/TriUF { stroke [] 0 setdash vpt 1.12 mul add M
  hpt neg vpt -1.62 mul V
  hpt 2 mul 0 V
  hpt neg vpt 1.62 mul V closepath fill } def
/TriD { stroke [] 0 setdash 2 copy vpt 1.12 mul sub M
  hpt neg vpt 1.62 mul V
  hpt 2 mul 0 V
  hpt neg vpt -1.62 mul V closepath stroke
  Pnt  } def
/TriDF { stroke [] 0 setdash vpt 1.12 mul sub M
  hpt neg vpt 1.62 mul V
  hpt 2 mul 0 V
  hpt neg vpt -1.62 mul V closepath fill} def
/DiaF { stroke [] 0 setdash vpt add M
  hpt neg vpt neg V hpt vpt neg V
  hpt vpt V hpt neg vpt V closepath fill } def
/Pent { stroke [] 0 setdash 2 copy gsave
  translate 0 hpt M 4 {72 rotate 0 hpt L} repeat
  closepath stroke grestore Pnt } def
/PentF { stroke [] 0 setdash gsave
  translate 0 hpt M 4 {72 rotate 0 hpt L} repeat
  closepath fill grestore } def
/Circle { stroke [] 0 setdash 2 copy
  hpt 0 360 arc stroke Pnt } def
/CircleF { stroke [] 0 setdash hpt 0 360 arc fill } def
/C0 { BL [] 0 setdash 2 copy moveto vpt 90 450  arc } bind def
/C1 { BL [] 0 setdash 2 copy        moveto
       2 copy  vpt 0 90 arc closepath fill
               vpt 0 360 arc closepath } bind def
/C2 { BL [] 0 setdash 2 copy moveto
       2 copy  vpt 90 180 arc closepath fill
               vpt 0 360 arc closepath } bind def
/C3 { BL [] 0 setdash 2 copy moveto
       2 copy  vpt 0 180 arc closepath fill
               vpt 0 360 arc closepath } bind def
/C4 { BL [] 0 setdash 2 copy moveto
       2 copy  vpt 180 270 arc closepath fill
               vpt 0 360 arc closepath } bind def
/C5 { BL [] 0 setdash 2 copy moveto
       2 copy  vpt 0 90 arc
       2 copy moveto
       2 copy  vpt 180 270 arc closepath fill
               vpt 0 360 arc } bind def
/C6 { BL [] 0 setdash 2 copy moveto
      2 copy  vpt 90 270 arc closepath fill
              vpt 0 360 arc closepath } bind def
/C7 { BL [] 0 setdash 2 copy moveto
      2 copy  vpt 0 270 arc closepath fill
              vpt 0 360 arc closepath } bind def
/C8 { BL [] 0 setdash 2 copy moveto
      2 copy vpt 270 360 arc closepath fill
              vpt 0 360 arc closepath } bind def
/C9 { BL [] 0 setdash 2 copy moveto
      2 copy  vpt 270 450 arc closepath fill
              vpt 0 360 arc closepath } bind def
/C10 { BL [] 0 setdash 2 copy 2 copy moveto vpt 270 360 arc closepath fill
       2 copy moveto
       2 copy vpt 90 180 arc closepath fill
               vpt 0 360 arc closepath } bind def
/C11 { BL [] 0 setdash 2 copy moveto
       2 copy  vpt 0 180 arc closepath fill
       2 copy moveto
       2 copy  vpt 270 360 arc closepath fill
               vpt 0 360 arc closepath } bind def
/C12 { BL [] 0 setdash 2 copy moveto
       2 copy  vpt 180 360 arc closepath fill
               vpt 0 360 arc closepath } bind def
/C13 { BL [] 0 setdash  2 copy moveto
       2 copy  vpt 0 90 arc closepath fill
       2 copy moveto
       2 copy  vpt 180 360 arc closepath fill
               vpt 0 360 arc closepath } bind def
/C14 { BL [] 0 setdash 2 copy moveto
       2 copy  vpt 90 360 arc closepath fill
               vpt 0 360 arc } bind def
/C15 { BL [] 0 setdash 2 copy vpt 0 360 arc closepath fill
               vpt 0 360 arc closepath } bind def
/Rec   { newpath 4 2 roll moveto 1 index 0 rlineto 0 exch rlineto
       neg 0 rlineto closepath } bind def
/Square { dup Rec } bind def
/Bsquare { vpt sub exch vpt sub exch vpt2 Square } bind def
/S0 { BL [] 0 setdash 2 copy moveto 0 vpt rlineto BL Bsquare } bind def
/S1 { BL [] 0 setdash 2 copy vpt Square fill Bsquare } bind def
/S2 { BL [] 0 setdash 2 copy exch vpt sub exch vpt Square fill Bsquare } bind def
/S3 { BL [] 0 setdash 2 copy exch vpt sub exch vpt2 vpt Rec fill Bsquare } bind def
/S4 { BL [] 0 setdash 2 copy exch vpt sub exch vpt sub vpt Square fill Bsquare } bind def
/S5 { BL [] 0 setdash 2 copy 2 copy vpt Square fill
       exch vpt sub exch vpt sub vpt Square fill Bsquare } bind def
/S6 { BL [] 0 setdash 2 copy exch vpt sub exch vpt sub vpt vpt2 Rec fill Bsquare } bind def
/S7 { BL [] 0 setdash 2 copy exch vpt sub exch vpt sub vpt vpt2 Rec fill
       2 copy vpt Square fill
       Bsquare } bind def
/S8 { BL [] 0 setdash 2 copy vpt sub vpt Square fill Bsquare } bind def
/S9 { BL [] 0 setdash 2 copy vpt sub vpt vpt2 Rec fill Bsquare } bind def
/S10 { BL [] 0 setdash 2 copy vpt sub vpt Square fill 2 copy exch vpt sub exch vpt Square fill
       Bsquare } bind def
/S11 { BL [] 0 setdash 2 copy vpt sub vpt Square fill 2 copy exch vpt sub exch vpt2 vpt Rec fill
       Bsquare } bind def
/S12 { BL [] 0 setdash 2 copy exch vpt sub exch vpt sub vpt2 vpt Rec fill Bsquare } bind def
/S13 { BL [] 0 setdash 2 copy exch vpt sub exch vpt sub vpt2 vpt Rec fill
       2 copy vpt Square fill Bsquare } bind def
/S14 { BL [] 0 setdash 2 copy exch vpt sub exch vpt sub vpt2 vpt Rec fill
       2 copy exch vpt sub exch vpt Square fill Bsquare } bind def
/S15 { BL [] 0 setdash 2 copy Bsquare fill Bsquare } bind def
/D0 { gsave translate 45 rotate 0 0 S0 stroke grestore } bind def
/D1 { gsave translate 45 rotate 0 0 S1 stroke grestore } bind def
/D2 { gsave translate 45 rotate 0 0 S2 stroke grestore } bind def
/D3 { gsave translate 45 rotate 0 0 S3 stroke grestore } bind def
/D4 { gsave translate 45 rotate 0 0 S4 stroke grestore } bind def
/D5 { gsave translate 45 rotate 0 0 S5 stroke grestore } bind def
/D6 { gsave translate 45 rotate 0 0 S6 stroke grestore } bind def
/D7 { gsave translate 45 rotate 0 0 S7 stroke grestore } bind def
/D8 { gsave translate 45 rotate 0 0 S8 stroke grestore } bind def
/D9 { gsave translate 45 rotate 0 0 S9 stroke grestore } bind def
/D10 { gsave translate 45 rotate 0 0 S10 stroke grestore } bind def
/D11 { gsave translate 45 rotate 0 0 S11 stroke grestore } bind def
/D12 { gsave translate 45 rotate 0 0 S12 stroke grestore } bind def
/D13 { gsave translate 45 rotate 0 0 S13 stroke grestore } bind def
/D14 { gsave translate 45 rotate 0 0 S14 stroke grestore } bind def
/D15 { gsave translate 45 rotate 0 0 S15 stroke grestore } bind def
/DiaE { stroke [] 0 setdash vpt add M
  hpt neg vpt neg V hpt vpt neg V
  hpt vpt V hpt neg vpt V closepath stroke } def
/BoxE { stroke [] 0 setdash exch hpt sub exch vpt add M
  0 vpt2 neg V hpt2 0 V 0 vpt2 V
  hpt2 neg 0 V closepath stroke } def
/TriUE { stroke [] 0 setdash vpt 1.12 mul add M
  hpt neg vpt -1.62 mul V
  hpt 2 mul 0 V
  hpt neg vpt 1.62 mul V closepath stroke } def
/TriDE { stroke [] 0 setdash vpt 1.12 mul sub M
  hpt neg vpt 1.62 mul V
  hpt 2 mul 0 V
  hpt neg vpt -1.62 mul V closepath stroke } def
/PentE { stroke [] 0 setdash gsave
  translate 0 hpt M 4 {72 rotate 0 hpt L} repeat
  closepath stroke grestore } def
/CircE { stroke [] 0 setdash 
  hpt 0 360 arc stroke } def
/Opaque { gsave closepath 1 setgray fill grestore 0 setgray closepath } def
/DiaW { stroke [] 0 setdash vpt add M
  hpt neg vpt neg V hpt vpt neg V
  hpt vpt V hpt neg vpt V Opaque stroke } def
/BoxW { stroke [] 0 setdash exch hpt sub exch vpt add M
  0 vpt2 neg V hpt2 0 V 0 vpt2 V
  hpt2 neg 0 V Opaque stroke } def
/TriUW { stroke [] 0 setdash vpt 1.12 mul add M
  hpt neg vpt -1.62 mul V
  hpt 2 mul 0 V
  hpt neg vpt 1.62 mul V Opaque stroke } def
/TriDW { stroke [] 0 setdash vpt 1.12 mul sub M
  hpt neg vpt 1.62 mul V
  hpt 2 mul 0 V
  hpt neg vpt -1.62 mul V Opaque stroke } def
/PentW { stroke [] 0 setdash gsave
  translate 0 hpt M 4 {72 rotate 0 hpt L} repeat
  Opaque stroke grestore } def
/CircW { stroke [] 0 setdash 
  hpt 0 360 arc Opaque stroke } def
/BoxFill { gsave Rec 1 setgray fill grestore } def
end
}}%
\begin{picture}(2412,2160)(0,0)%
{\GNUPLOTspecial{"
gnudict begin
gsave
0 0 translate
0.100 0.100 scale
0 setgray
newpath
1.000 UL
LTb
0.050 UL
LT3
300 300 M
2112 0 V
1.000 UL
LTb
300 300 M
63 0 V
2049 0 R
-63 0 V
0.050 UL
LT3
300 566 M
2112 0 V
1.000 UL
LTb
300 566 M
63 0 V
2049 0 R
-63 0 V
0.050 UL
LT3
300 831 M
2112 0 V
1.000 UL
LTb
300 831 M
63 0 V
2049 0 R
-63 0 V
0.050 UL
LT3
300 1097 M
2112 0 V
1.000 UL
LTb
300 1097 M
63 0 V
2049 0 R
-63 0 V
0.050 UL
LT3
300 1363 M
2112 0 V
1.000 UL
LTb
300 1363 M
63 0 V
2049 0 R
-63 0 V
0.050 UL
LT3
300 1629 M
2112 0 V
1.000 UL
LTb
300 1629 M
63 0 V
2049 0 R
-63 0 V
0.050 UL
LT3
300 1894 M
2112 0 V
1.000 UL
LTb
300 1894 M
63 0 V
2049 0 R
-63 0 V
0.050 UL
LT3
300 2160 M
2112 0 V
1.000 UL
LTb
300 2160 M
63 0 V
2049 0 R
-63 0 V
0.050 UL
LT3
300 300 M
0 1860 V
1.000 UL
LTb
300 300 M
0 63 V
0 1797 R
0 -63 V
0.050 UL
LT3
511 300 M
0 1597 V
0 200 R
0 63 V
1.000 UL
LTb
511 300 M
0 63 V
0 1797 R
0 -63 V
0.050 UL
LT3
722 300 M
0 1597 V
0 200 R
0 63 V
1.000 UL
LTb
722 300 M
0 63 V
0 1797 R
0 -63 V
0.050 UL
LT3
934 300 M
0 1597 V
0 200 R
0 63 V
1.000 UL
LTb
934 300 M
0 63 V
0 1797 R
0 -63 V
0.050 UL
LT3
1145 300 M
0 1860 V
1.000 UL
LTb
1145 300 M
0 63 V
0 1797 R
0 -63 V
0.050 UL
LT3
1356 300 M
0 1860 V
1.000 UL
LTb
1356 300 M
0 63 V
0 1797 R
0 -63 V
0.050 UL
LT3
1567 300 M
0 1860 V
1.000 UL
LTb
1567 300 M
0 63 V
0 1797 R
0 -63 V
0.050 UL
LT3
1778 300 M
0 1860 V
1.000 UL
LTb
1778 300 M
0 63 V
0 1797 R
0 -63 V
0.050 UL
LT3
1990 300 M
0 1860 V
1.000 UL
LTb
1990 300 M
0 63 V
0 1797 R
0 -63 V
0.050 UL
LT3
2201 300 M
0 1860 V
1.000 UL
LTb
2201 300 M
0 63 V
0 1797 R
0 -63 V
0.050 UL
LT3
2412 300 M
0 1860 V
1.000 UL
LTb
2412 300 M
0 63 V
0 1797 R
0 -63 V
1.000 UL
LTb
300 300 M
2112 0 V
0 1860 V
-2112 0 V
300 300 L
1.000 UL
LTb
350 1897 M
0 200 V
613 0 V
0 -200 V
-613 0 V
0 200 R
613 0 V
1.000 UL
LT2
650 2047 M
263 0 V
335 820 M
21 19 V
20 18 V
21 18 V
21 16 V
20 14 V
21 9 V
21 2 V
20 -7 V
21 -15 V
20 -23 V
21 -30 V
21 -29 V
20 -17 V
21 4 V
21 25 V
20 30 V
21 18 V
20 -7 V
21 -22 V
21 -19 V
20 -1 V
21 21 V
21 27 V
20 21 V
21 6 V
20 -2 V
21 1 V
21 12 V
20 17 V
21 8 V
20 -17 V
21 -43 V
21 -48 V
20 -30 V
21 7 V
21 33 V
20 39 V
21 22 V
20 0 V
21 -12 V
21 -12 V
20 -4 V
21 6 V
21 10 V
20 14 V
21 12 V
20 7 V
21 -3 V
21 -13 V
20 -12 V
21 5 V
21 31 V
20 47 V
21 46 V
20 26 V
21 1 V
21 -10 V
20 -6 V
21 8 V
21 16 V
20 11 V
21 -6 V
20 -24 V
21 -34 V
21 -37 V
20 -33 V
21 -23 V
21 -9 V
20 7 V
21 20 V
20 22 V
21 16 V
21 2 V
20 -10 V
21 -19 V
20 -28 V
21 -24 V
21 -9 V
20 18 V
21 42 V
21 44 V
20 25 V
21 -8 V
20 -27 V
21 -29 V
21 -13 V
20 4 V
21 11 V
21 10 V
20 5 V
21 7 V
20 17 V
21 32 V
21 39 V
20 26 V
21 -2 V
21 -34 V
20 -58 V
21 -68 V
1.000 UL
LT0
650 1947 M
263 0 V
335 523 M
21 -44 V
20 -33 V
21 -13 V
21 17 V
20 44 V
21 56 V
21 51 V
20 36 V
21 16 V
20 -10 V
21 -35 V
21 -47 V
20 -34 V
21 3 V
21 43 V
20 61 V
21 53 V
20 22 V
21 -2 V
21 -17 V
20 -19 V
21 -16 V
21 -17 V
20 -21 V
21 -22 V
20 -11 V
21 14 V
21 54 V
20 77 V
21 67 V
20 25 V
21 -29 V
21 -58 V
20 -56 V
21 -29 V
21 1 V
20 26 V
21 44 V
20 51 V
21 42 V
21 16 V
20 -19 V
21 -38 V
21 -33 V
20 -6 V
21 17 V
20 18 V
21 -4 V
21 -33 V
20 -40 V
21 -20 V
21 23 V
20 59 V
21 75 V
20 70 V
21 51 V
21 26 V
20 -5 V
21 -35 V
21 -35 V
20 2 V
21 74 V
20 130 V
21 124 V
21 55 V
20 -47 V
21 -101 V
21 -98 V
20 -39 V
21 16 V
20 38 V
21 24 V
21 -3 V
20 -17 V
21 -10 V
20 9 V
21 25 V
21 30 V
20 24 V
21 16 V
21 16 V
20 21 V
21 32 V
20 32 V
21 20 V
21 -3 V
20 -16 V
21 -8 V
21 20 V
20 56 V
21 72 V
20 64 V
21 37 V
21 21 V
20 32 V
21 65 V
21 108 V
20 136 V
21 151 V
stroke
grestore
end
showpage
}}%
\put(600,1947){\makebox(0,0)[r]{$\tilde{t}_{\rm ej}$}}%
\put(600,2047){\makebox(0,0)[r]{$\tilde{t}=0$}}%
\put(1356,2310){\makebox(0,0){\textsf{$\tilde{\rho}$ depending on the orientation of $\tilde{\VEC{L}}$}}}%
\put(1356,50){\makebox(0,0){\sffamily $\frac{\tilde{L}_z}{\tilde{L}}$}}%
\put(50,1230){%
\special{ps: gsave currentpoint currentpoint translate
270 rotate neg exch neg exch translate}%
\makebox(0,0)[b]{\shortstack{\sffamily $\tilde{\rho}$}}%
\special{ps: currentpoint grestore moveto}%
}%
\put(2412,200){\makebox(0,0){1}}%
\put(2201,200){\makebox(0,0){0.8}}%
\put(1990,200){\makebox(0,0){0.6}}%
\put(1778,200){\makebox(0,0){0.4}}%
\put(1567,200){\makebox(0,0){0.2}}%
\put(1356,200){\makebox(0,0){0}}%
\put(1145,200){\makebox(0,0){-0.2}}%
\put(934,200){\makebox(0,0){-0.4}}%
\put(722,200){\makebox(0,0){-0.6}}%
\put(511,200){\makebox(0,0){-0.8}}%
\put(300,200){\makebox(0,0){-1}}%
\put(250,2160){\makebox(0,0)[r]{0.08}}%
\put(250,1894){\makebox(0,0)[r]{0.07}}%
\put(250,1629){\makebox(0,0)[r]{0.06}}%
\put(250,1363){\makebox(0,0)[r]{0.05}}%
\put(250,1097){\makebox(0,0)[r]{0.04}}%
\put(250,831){\makebox(0,0)[r]{0.03}}%
\put(250,566){\makebox(0,0)[r]{0.02}}%
\put(250,300){\makebox(0,0)[r]{0.01}}%
\end{picture}%
\endgroup
 

%% file: bhb_fig18.tex
\begingroup%
  \makeatletter%
  \newcommand{\GNUPLOTspecial}{%
    \@sanitize\catcode`\%=14\relax\special}%
  \setlength{\unitlength}{0.1bp}%
{\GNUPLOTspecial{!
/gnudict 256 dict def
gnudict begin
/Color false def
/Solid false def
/gnulinewidth 5.000 def
/userlinewidth gnulinewidth def
/vshift -33 def
/dl {10 mul} def
/hpt_ 31.5 def
/vpt_ 31.5 def
/hpt hpt_ def
/vpt vpt_ def
/M {moveto} bind def
/L {lineto} bind def
/R {rmoveto} bind def
/V {rlineto} bind def
/vpt2 vpt 2 mul def
/hpt2 hpt 2 mul def
/Lshow { currentpoint stroke M
  0 vshift R show } def
/Rshow { currentpoint stroke M
  dup stringwidth pop neg vshift R show } def
/Cshow { currentpoint stroke M
  dup stringwidth pop -2 div vshift R show } def
/UP { dup vpt_ mul /vpt exch def hpt_ mul /hpt exch def
  /hpt2 hpt 2 mul def /vpt2 vpt 2 mul def } def
/DL { Color {setrgbcolor Solid {pop []} if 0 setdash }
 {pop pop pop Solid {pop []} if 0 setdash} ifelse } def
/BL { stroke gnulinewidth 2 mul setlinewidth } def
/AL { stroke gnulinewidth 2 div setlinewidth } def
/UL { gnulinewidth mul /userlinewidth exch def } def
/PL { stroke userlinewidth setlinewidth } def
/LTb { BL [] 0 0 0 DL } def
/LTa { AL [1 dl 2 dl] 0 setdash 0 0 0 setrgbcolor } def
/LT0 { PL [] 1 0 0 DL } def
/LT1 { PL [4 dl 2 dl] 0 1 0 DL } def
/LT2 { PL [2 dl 3 dl] 0 0 1 DL } def
/LT3 { PL [1 dl 1.5 dl] 1 0 1 DL } def
/LT4 { PL [5 dl 2 dl 1 dl 2 dl] 0 1 1 DL } def
/LT5 { PL [4 dl 3 dl 1 dl 3 dl] 1 1 0 DL } def
/LT6 { PL [2 dl 2 dl 2 dl 4 dl] 0 0 0 DL } def
/LT7 { PL [2 dl 2 dl 2 dl 2 dl 2 dl 4 dl] 1 0.3 0 DL } def
/LT8 { PL [2 dl 2 dl 2 dl 2 dl 2 dl 2 dl 2 dl 4 dl] 0.5 0.5 0.5 DL } def
/Pnt { stroke [] 0 setdash
   gsave 1 setlinecap M 0 0 V stroke grestore } def
/Dia { stroke [] 0 setdash 2 copy vpt add M
  hpt neg vpt neg V hpt vpt neg V
  hpt vpt V hpt neg vpt V closepath stroke
  Pnt } def
/Pls { stroke [] 0 setdash vpt sub M 0 vpt2 V
  currentpoint stroke M
  hpt neg vpt neg R hpt2 0 V stroke
  } def
/Box { stroke [] 0 setdash 2 copy exch hpt sub exch vpt add M
  0 vpt2 neg V hpt2 0 V 0 vpt2 V
  hpt2 neg 0 V closepath stroke
  Pnt } def
/Crs { stroke [] 0 setdash exch hpt sub exch vpt add M
  hpt2 vpt2 neg V currentpoint stroke M
  hpt2 neg 0 R hpt2 vpt2 V stroke } def
/TriU { stroke [] 0 setdash 2 copy vpt 1.12 mul add M
  hpt neg vpt -1.62 mul V
  hpt 2 mul 0 V
  hpt neg vpt 1.62 mul V closepath stroke
  Pnt  } def
/Star { 2 copy Pls Crs } def
/BoxF { stroke [] 0 setdash exch hpt sub exch vpt add M
  0 vpt2 neg V  hpt2 0 V  0 vpt2 V
  hpt2 neg 0 V  closepath fill } def
/TriUF { stroke [] 0 setdash vpt 1.12 mul add M
  hpt neg vpt -1.62 mul V
  hpt 2 mul 0 V
  hpt neg vpt 1.62 mul V closepath fill } def
/TriD { stroke [] 0 setdash 2 copy vpt 1.12 mul sub M
  hpt neg vpt 1.62 mul V
  hpt 2 mul 0 V
  hpt neg vpt -1.62 mul V closepath stroke
  Pnt  } def
/TriDF { stroke [] 0 setdash vpt 1.12 mul sub M
  hpt neg vpt 1.62 mul V
  hpt 2 mul 0 V
  hpt neg vpt -1.62 mul V closepath fill} def
/DiaF { stroke [] 0 setdash vpt add M
  hpt neg vpt neg V hpt vpt neg V
  hpt vpt V hpt neg vpt V closepath fill } def
/Pent { stroke [] 0 setdash 2 copy gsave
  translate 0 hpt M 4 {72 rotate 0 hpt L} repeat
  closepath stroke grestore Pnt } def
/PentF { stroke [] 0 setdash gsave
  translate 0 hpt M 4 {72 rotate 0 hpt L} repeat
  closepath fill grestore } def
/Circle { stroke [] 0 setdash 2 copy
  hpt 0 360 arc stroke Pnt } def
/CircleF { stroke [] 0 setdash hpt 0 360 arc fill } def
/C0 { BL [] 0 setdash 2 copy moveto vpt 90 450  arc } bind def
/C1 { BL [] 0 setdash 2 copy        moveto
       2 copy  vpt 0 90 arc closepath fill
               vpt 0 360 arc closepath } bind def
/C2 { BL [] 0 setdash 2 copy moveto
       2 copy  vpt 90 180 arc closepath fill
               vpt 0 360 arc closepath } bind def
/C3 { BL [] 0 setdash 2 copy moveto
       2 copy  vpt 0 180 arc closepath fill
               vpt 0 360 arc closepath } bind def
/C4 { BL [] 0 setdash 2 copy moveto
       2 copy  vpt 180 270 arc closepath fill
               vpt 0 360 arc closepath } bind def
/C5 { BL [] 0 setdash 2 copy moveto
       2 copy  vpt 0 90 arc
       2 copy moveto
       2 copy  vpt 180 270 arc closepath fill
               vpt 0 360 arc } bind def
/C6 { BL [] 0 setdash 2 copy moveto
      2 copy  vpt 90 270 arc closepath fill
              vpt 0 360 arc closepath } bind def
/C7 { BL [] 0 setdash 2 copy moveto
      2 copy  vpt 0 270 arc closepath fill
              vpt 0 360 arc closepath } bind def
/C8 { BL [] 0 setdash 2 copy moveto
      2 copy vpt 270 360 arc closepath fill
              vpt 0 360 arc closepath } bind def
/C9 { BL [] 0 setdash 2 copy moveto
      2 copy  vpt 270 450 arc closepath fill
              vpt 0 360 arc closepath } bind def
/C10 { BL [] 0 setdash 2 copy 2 copy moveto vpt 270 360 arc closepath fill
       2 copy moveto
       2 copy vpt 90 180 arc closepath fill
               vpt 0 360 arc closepath } bind def
/C11 { BL [] 0 setdash 2 copy moveto
       2 copy  vpt 0 180 arc closepath fill
       2 copy moveto
       2 copy  vpt 270 360 arc closepath fill
               vpt 0 360 arc closepath } bind def
/C12 { BL [] 0 setdash 2 copy moveto
       2 copy  vpt 180 360 arc closepath fill
               vpt 0 360 arc closepath } bind def
/C13 { BL [] 0 setdash  2 copy moveto
       2 copy  vpt 0 90 arc closepath fill
       2 copy moveto
       2 copy  vpt 180 360 arc closepath fill
               vpt 0 360 arc closepath } bind def
/C14 { BL [] 0 setdash 2 copy moveto
       2 copy  vpt 90 360 arc closepath fill
               vpt 0 360 arc } bind def
/C15 { BL [] 0 setdash 2 copy vpt 0 360 arc closepath fill
               vpt 0 360 arc closepath } bind def
/Rec   { newpath 4 2 roll moveto 1 index 0 rlineto 0 exch rlineto
       neg 0 rlineto closepath } bind def
/Square { dup Rec } bind def
/Bsquare { vpt sub exch vpt sub exch vpt2 Square } bind def
/S0 { BL [] 0 setdash 2 copy moveto 0 vpt rlineto BL Bsquare } bind def
/S1 { BL [] 0 setdash 2 copy vpt Square fill Bsquare } bind def
/S2 { BL [] 0 setdash 2 copy exch vpt sub exch vpt Square fill Bsquare } bind def
/S3 { BL [] 0 setdash 2 copy exch vpt sub exch vpt2 vpt Rec fill Bsquare } bind def
/S4 { BL [] 0 setdash 2 copy exch vpt sub exch vpt sub vpt Square fill Bsquare } bind def
/S5 { BL [] 0 setdash 2 copy 2 copy vpt Square fill
       exch vpt sub exch vpt sub vpt Square fill Bsquare } bind def
/S6 { BL [] 0 setdash 2 copy exch vpt sub exch vpt sub vpt vpt2 Rec fill Bsquare } bind def
/S7 { BL [] 0 setdash 2 copy exch vpt sub exch vpt sub vpt vpt2 Rec fill
       2 copy vpt Square fill
       Bsquare } bind def
/S8 { BL [] 0 setdash 2 copy vpt sub vpt Square fill Bsquare } bind def
/S9 { BL [] 0 setdash 2 copy vpt sub vpt vpt2 Rec fill Bsquare } bind def
/S10 { BL [] 0 setdash 2 copy vpt sub vpt Square fill 2 copy exch vpt sub exch vpt Square fill
       Bsquare } bind def
/S11 { BL [] 0 setdash 2 copy vpt sub vpt Square fill 2 copy exch vpt sub exch vpt2 vpt Rec fill
       Bsquare } bind def
/S12 { BL [] 0 setdash 2 copy exch vpt sub exch vpt sub vpt2 vpt Rec fill Bsquare } bind def
/S13 { BL [] 0 setdash 2 copy exch vpt sub exch vpt sub vpt2 vpt Rec fill
       2 copy vpt Square fill Bsquare } bind def
/S14 { BL [] 0 setdash 2 copy exch vpt sub exch vpt sub vpt2 vpt Rec fill
       2 copy exch vpt sub exch vpt Square fill Bsquare } bind def
/S15 { BL [] 0 setdash 2 copy Bsquare fill Bsquare } bind def
/D0 { gsave translate 45 rotate 0 0 S0 stroke grestore } bind def
/D1 { gsave translate 45 rotate 0 0 S1 stroke grestore } bind def
/D2 { gsave translate 45 rotate 0 0 S2 stroke grestore } bind def
/D3 { gsave translate 45 rotate 0 0 S3 stroke grestore } bind def
/D4 { gsave translate 45 rotate 0 0 S4 stroke grestore } bind def
/D5 { gsave translate 45 rotate 0 0 S5 stroke grestore } bind def
/D6 { gsave translate 45 rotate 0 0 S6 stroke grestore } bind def
/D7 { gsave translate 45 rotate 0 0 S7 stroke grestore } bind def
/D8 { gsave translate 45 rotate 0 0 S8 stroke grestore } bind def
/D9 { gsave translate 45 rotate 0 0 S9 stroke grestore } bind def
/D10 { gsave translate 45 rotate 0 0 S10 stroke grestore } bind def
/D11 { gsave translate 45 rotate 0 0 S11 stroke grestore } bind def
/D12 { gsave translate 45 rotate 0 0 S12 stroke grestore } bind def
/D13 { gsave translate 45 rotate 0 0 S13 stroke grestore } bind def
/D14 { gsave translate 45 rotate 0 0 S14 stroke grestore } bind def
/D15 { gsave translate 45 rotate 0 0 S15 stroke grestore } bind def
/DiaE { stroke [] 0 setdash vpt add M
  hpt neg vpt neg V hpt vpt neg V
  hpt vpt V hpt neg vpt V closepath stroke } def
/BoxE { stroke [] 0 setdash exch hpt sub exch vpt add M
  0 vpt2 neg V hpt2 0 V 0 vpt2 V
  hpt2 neg 0 V closepath stroke } def
/TriUE { stroke [] 0 setdash vpt 1.12 mul add M
  hpt neg vpt -1.62 mul V
  hpt 2 mul 0 V
  hpt neg vpt 1.62 mul V closepath stroke } def
/TriDE { stroke [] 0 setdash vpt 1.12 mul sub M
  hpt neg vpt 1.62 mul V
  hpt 2 mul 0 V
  hpt neg vpt -1.62 mul V closepath stroke } def
/PentE { stroke [] 0 setdash gsave
  translate 0 hpt M 4 {72 rotate 0 hpt L} repeat
  closepath stroke grestore } def
/CircE { stroke [] 0 setdash 
  hpt 0 360 arc stroke } def
/Opaque { gsave closepath 1 setgray fill grestore 0 setgray closepath } def
/DiaW { stroke [] 0 setdash vpt add M
  hpt neg vpt neg V hpt vpt neg V
  hpt vpt V hpt neg vpt V Opaque stroke } def
/BoxW { stroke [] 0 setdash exch hpt sub exch vpt add M
  0 vpt2 neg V hpt2 0 V 0 vpt2 V
  hpt2 neg 0 V Opaque stroke } def
/TriUW { stroke [] 0 setdash vpt 1.12 mul add M
  hpt neg vpt -1.62 mul V
  hpt 2 mul 0 V
  hpt neg vpt 1.62 mul V Opaque stroke } def
/TriDW { stroke [] 0 setdash vpt 1.12 mul sub M
  hpt neg vpt 1.62 mul V
  hpt 2 mul 0 V
  hpt neg vpt -1.62 mul V Opaque stroke } def
/PentW { stroke [] 0 setdash gsave
  translate 0 hpt M 4 {72 rotate 0 hpt L} repeat
  Opaque stroke grestore } def
/CircW { stroke [] 0 setdash 
  hpt 0 360 arc Opaque stroke } def
/BoxFill { gsave Rec 1 setgray fill grestore } def
end
}}%
\begin{picture}(2412,2160)(0,0)%
{\GNUPLOTspecial{"
gnudict begin
gsave
0 0 translate
0.100 0.100 scale
0 setgray
newpath
1.000 UL
LTb
0.050 UL
LT3
400 300 M
2012 0 V
1.000 UL
LTb
400 300 M
63 0 V
1949 0 R
-63 0 V
400 469 M
31 0 V
1981 0 R
-31 0 V
0.050 UL
LT3
400 638 M
2012 0 V
1.000 UL
LTb
400 638 M
63 0 V
1949 0 R
-63 0 V
400 807 M
31 0 V
1981 0 R
-31 0 V
0.050 UL
LT3
400 976 M
2012 0 V
1.000 UL
LTb
400 976 M
63 0 V
1949 0 R
-63 0 V
400 1145 M
31 0 V
1981 0 R
-31 0 V
0.050 UL
LT3
400 1315 M
2012 0 V
1.000 UL
LTb
400 1315 M
63 0 V
1949 0 R
-63 0 V
400 1484 M
31 0 V
1981 0 R
-31 0 V
0.050 UL
LT3
400 1653 M
2012 0 V
1.000 UL
LTb
400 1653 M
63 0 V
1949 0 R
-63 0 V
400 1822 M
31 0 V
1981 0 R
-31 0 V
0.050 UL
LT3
400 1991 M
1075 0 V
887 0 R
50 0 V
1.000 UL
LTb
400 1991 M
63 0 V
1949 0 R
-63 0 V
400 2160 M
31 0 V
1981 0 R
-31 0 V
0.050 UL
LT3
400 300 M
0 1860 V
1.000 UL
LTb
400 300 M
0 63 V
0 1797 R
0 -63 V
551 300 M
0 31 V
0 1829 R
0 -31 V
640 300 M
0 31 V
0 1829 R
0 -31 V
703 300 M
0 31 V
0 1829 R
0 -31 V
752 300 M
0 31 V
0 1829 R
0 -31 V
791 300 M
0 31 V
0 1829 R
0 -31 V
825 300 M
0 31 V
0 1829 R
0 -31 V
854 300 M
0 31 V
0 1829 R
0 -31 V
880 300 M
0 31 V
0 1829 R
0 -31 V
0.050 UL
LT3
903 300 M
0 1860 V
1.000 UL
LTb
903 300 M
0 63 V
0 1797 R
0 -63 V
1054 300 M
0 31 V
0 1829 R
0 -31 V
1143 300 M
0 31 V
0 1829 R
0 -31 V
1206 300 M
0 31 V
0 1829 R
0 -31 V
1255 300 M
0 31 V
0 1829 R
0 -31 V
1294 300 M
0 31 V
0 1829 R
0 -31 V
1328 300 M
0 31 V
0 1829 R
0 -31 V
1357 300 M
0 31 V
0 1829 R
0 -31 V
1383 300 M
0 31 V
0 1829 R
0 -31 V
0.050 UL
LT3
1406 300 M
0 1860 V
1.000 UL
LTb
1406 300 M
0 63 V
0 1797 R
0 -63 V
1557 300 M
0 31 V
0 1829 R
0 -31 V
1646 300 M
0 31 V
0 1829 R
0 -31 V
1709 300 M
0 31 V
0 1829 R
0 -31 V
1758 300 M
0 31 V
0 1829 R
0 -31 V
1797 300 M
0 31 V
0 1829 R
0 -31 V
1831 300 M
0 31 V
0 1829 R
0 -31 V
1860 300 M
0 31 V
0 1829 R
0 -31 V
1886 300 M
0 31 V
0 1829 R
0 -31 V
0.050 UL
LT3
1909 300 M
0 1397 V
0 400 R
0 63 V
1.000 UL
LTb
1909 300 M
0 63 V
0 1797 R
0 -63 V
2060 300 M
0 31 V
0 1829 R
0 -31 V
2149 300 M
0 31 V
0 1829 R
0 -31 V
2212 300 M
0 31 V
0 1829 R
0 -31 V
2261 300 M
0 31 V
0 1829 R
0 -31 V
2300 300 M
0 31 V
0 1829 R
0 -31 V
2334 300 M
0 31 V
0 1829 R
0 -31 V
2363 300 M
0 31 V
0 1829 R
0 -31 V
2389 300 M
0 31 V
0 1829 R
0 -31 V
0.050 UL
LT3
2412 300 M
0 1860 V
1.000 UL
LTb
2412 300 M
0 63 V
0 1797 R
0 -63 V
1.000 UL
LTb
400 300 M
2012 0 V
0 1860 V
-2012 0 V
400 300 L
1.000 UL
LTb
1475 1697 M
0 400 V
887 0 V
0 -400 V
-887 0 V
0 400 R
887 0 V
0.000 UP
1.000 UL
LT0
2075 2047 M
237 0 V
460 300 M
0 1766 V
41 67 V
53 -199 V
61 137 V
67 -86 V
69 -22 V
69 -55 V
68 -40 V
66 -31 V
63 -50 V
61 -44 V
58 -48 V
56 -37 V
53 -33 V
51 -34 V
49 -30 V
46 -36 V
45 -30 V
44 -23 V
41 -32 V
41 -27 V
38 -29 V
38 -24 V
36 -22 V
35 -23 V
34 -28 V
0 -987 V
460 2066 Circle
501 2133 Circle
554 1934 Circle
615 2071 Circle
682 1985 Circle
751 1963 Circle
820 1908 Circle
888 1868 Circle
954 1837 Circle
1017 1787 Circle
1078 1743 Circle
1136 1695 Circle
1192 1658 Circle
1245 1625 Circle
1296 1591 Circle
1345 1561 Circle
1391 1525 Circle
1436 1495 Circle
1480 1472 Circle
1521 1440 Circle
1562 1413 Circle
1600 1384 Circle
1638 1360 Circle
1674 1338 Circle
1709 1315 Circle
1743 1287 Circle
2193 2047 Circle
0.000 UP
1.000 UL
LT1
2075 1947 M
237 0 V
682 300 M
0 1572 V
69 -9 V
69 -80 V
820 300 L
197 0 R
0 1160 V
61 60 V
58 63 V
56 -9 V
53 -24 V
51 -12 V
49 -28 V
46 -22 V
45 -27 V
44 -26 V
41 -37 V
41 -19 V
38 -30 V
38 -30 V
36 -30 V
35 -37 V
34 -33 V
33 -83 V
32 -65 V
31 -72 V
31 -36 V
29 -42 V
29 -45 V
28 -34 V
27 -50 V
27 -53 V
26 -47 V
25 6 V
25 -59 V
24 -7 V
24 -95 V
23 23 V
23 -79 V
22 13 V
22 -71 V
21 21 V
21 -31 V
21 1 V
20 -19 V
20 -40 V
20 -48 V
19 33 V
11 -40 V
682 1872 TriU
751 1863 TriU
820 1783 TriU
1017 1460 TriU
1078 1520 TriU
1136 1583 TriU
1192 1574 TriU
1245 1550 TriU
1296 1538 TriU
1345 1510 TriU
1391 1488 TriU
1436 1461 TriU
1480 1435 TriU
1521 1398 TriU
1562 1379 TriU
1600 1349 TriU
1638 1319 TriU
1674 1289 TriU
1709 1252 TriU
1743 1219 TriU
1776 1136 TriU
1808 1071 TriU
1839 999 TriU
1870 963 TriU
1899 921 TriU
1928 876 TriU
1956 842 TriU
1983 792 TriU
2010 739 TriU
2036 692 TriU
2061 698 TriU
2086 639 TriU
2110 632 TriU
2134 537 TriU
2157 560 TriU
2180 481 TriU
2202 494 TriU
2224 423 TriU
2245 444 TriU
2266 413 TriU
2287 414 TriU
2307 395 TriU
2327 355 TriU
2347 307 TriU
2366 340 TriU
2193 1947 TriU
0.000 UP
1.000 UL
LT4
2075 1847 M
237 0 V
432 300 M
0 1827 V
28 20 V
41 -94 V
53 42 V
61 -56 V
67 -65 V
69 -65 V
69 -126 V
68 -147 V
66 -115 V
63 -61 V
61 -7 V
58 141 V
56 -58 V
53 25 V
51 -38 V
49 -7 V
46 -26 V
45 -23 V
44 -41 V
41 -21 V
41 -23 V
38 -33 V
38 -32 V
36 -31 V
35 -37 V
34 -36 V
33 -71 V
32 -60 V
31 -84 V
31 -18 V
29 -58 V
29 -34 V
28 -40 V
27 -51 V
27 -44 V
26 -35 V
25 -7 V
25 -37 V
24 -52 V
24 -64 V
23 14 V
23 -67 V
22 30 V
22 -163 V
21 81 V
21 10 V
21 -58 V
20 7 V
20 -57 V
20 -31 V
19 6 V
19 -29 V
1 -1 V
432 2127 Box
460 2147 Box
501 2053 Box
554 2095 Box
615 2039 Box
682 1974 Box
751 1909 Box
820 1783 Box
888 1636 Box
954 1521 Box
1017 1460 Box
1078 1453 Box
1136 1594 Box
1192 1536 Box
1245 1561 Box
1296 1523 Box
1345 1516 Box
1391 1490 Box
1436 1467 Box
1480 1426 Box
1521 1405 Box
1562 1382 Box
1600 1349 Box
1638 1317 Box
1674 1286 Box
1709 1249 Box
1743 1213 Box
1776 1142 Box
1808 1082 Box
1839 998 Box
1870 980 Box
1899 922 Box
1928 888 Box
1956 848 Box
1983 797 Box
2010 753 Box
2036 718 Box
2061 711 Box
2086 674 Box
2110 622 Box
2134 558 Box
2157 572 Box
2180 505 Box
2202 535 Box
2224 372 Box
2245 453 Box
2266 463 Box
2287 405 Box
2307 412 Box
2327 355 Box
2347 324 Box
2366 330 Box
2385 301 Box
2193 1847 Box
0.000 UP
1.000 UL
LT3
2075 1747 M
237 0 V
460 300 M
0 1817 V
41 3 V
53 -55 V
61 10 V
67 -75 V
69 -52 V
69 -76 V
68 -42 V
66 -115 V
63 -28 V
61 -35 V
58 -16 V
56 -34 V
53 -29 V
51 -29 V
49 -21 V
46 -25 V
45 -21 V
44 -25 V
41 -30 V
41 -32 V
38 -31 V
38 -31 V
36 -40 V
35 -32 V
34 -33 V
33 -80 V
32 -55 V
31 -87 V
31 -34 V
29 -46 V
29 -43 V
28 -30 V
27 -39 V
27 -50 V
26 -41 V
25 -24 V
25 -40 V
24 -32 V
24 -17 V
23 -18 V
23 -72 V
22 -12 V
22 -21 V
21 -12 V
21 -57 V
21 1 V
20 -53 V
20 -19 V
20 -35 V
19 12 V
11 -19 V
460 2117 Dia
501 2120 Dia
554 2065 Dia
615 2075 Dia
682 2000 Dia
751 1948 Dia
820 1872 Dia
888 1830 Dia
954 1715 Dia
1017 1687 Dia
1078 1652 Dia
1136 1636 Dia
1192 1602 Dia
1245 1573 Dia
1296 1544 Dia
1345 1523 Dia
1391 1498 Dia
1436 1477 Dia
1480 1452 Dia
1521 1422 Dia
1562 1390 Dia
1600 1359 Dia
1638 1328 Dia
1674 1288 Dia
1709 1256 Dia
1743 1223 Dia
1776 1143 Dia
1808 1088 Dia
1839 1001 Dia
1870 967 Dia
1899 921 Dia
1928 878 Dia
1956 848 Dia
1983 809 Dia
2010 759 Dia
2036 718 Dia
2061 694 Dia
2086 654 Dia
2110 622 Dia
2134 605 Dia
2157 587 Dia
2180 515 Dia
2202 503 Dia
2224 482 Dia
2245 470 Dia
2266 413 Dia
2287 414 Dia
2307 361 Dia
2327 342 Dia
2347 307 Dia
2366 319 Dia
2193 1747 Dia
stroke
grestore
end
showpage
}}%
\put(2025,1747){\makebox(0,0)[r]{$q=100$}}%
\put(2025,1847){\makebox(0,0)[r]{$q=10$}}%
\put(2025,1947){\makebox(0,0)[r]{$q=1$}}%
\put(2025,2047){\makebox(0,0)[r]{TP at $\tilde{t}=0$}}%
\put(1406,2310){\makebox(0,0){\textsf{Density $\tilde{\rho}(\tilde{r})$ at the end $\tilde{t}_{\rm max}$}}}%
\put(1406,50){\makebox(0,0){\sffamily $\tilde{r}$}}%
\put(100,1230){%
\special{ps: gsave currentpoint currentpoint translate
270 rotate neg exch neg exch translate}%
\makebox(0,0)[b]{\shortstack{\sffamily $\tilde{\rho}$}}%
\special{ps: currentpoint grestore moveto}%
}%
\put(2412,200){\makebox(0,0){1000}}%
\put(1909,200){\makebox(0,0){100}}%
\put(1406,200){\makebox(0,0){10}}%
\put(903,200){\makebox(0,0){1}}%
\put(400,200){\makebox(0,0){0.1}}%
\put(350,1991){\makebox(0,0)[r]{0.01}}%
\put(350,1653){\makebox(0,0)[r]{0.0001}}%
\put(350,1315){\makebox(0,0)[r]{1e-06}}%
\put(350,976){\makebox(0,0)[r]{1e-08}}%
\put(350,638){\makebox(0,0)[r]{1e-10}}%
\put(350,300){\makebox(0,0)[r]{1e-12}}%
\end{picture}%
\endgroup
 

%% file: bhb_fig23.tex
\begingroup%
  \makeatletter%
  \newcommand{\GNUPLOTspecial}{%
    \@sanitize\catcode`\%=14\relax\special}%
  \setlength{\unitlength}{0.1bp}%
{\GNUPLOTspecial{!
/gnudict 256 dict def
gnudict begin
/Color false def
/Solid false def
/gnulinewidth 5.000 def
/userlinewidth gnulinewidth def
/vshift -33 def
/dl {10 mul} def
/hpt_ 31.5 def
/vpt_ 31.5 def
/hpt hpt_ def
/vpt vpt_ def
/M {moveto} bind def
/L {lineto} bind def
/R {rmoveto} bind def
/V {rlineto} bind def
/vpt2 vpt 2 mul def
/hpt2 hpt 2 mul def
/Lshow { currentpoint stroke M
  0 vshift R show } def
/Rshow { currentpoint stroke M
  dup stringwidth pop neg vshift R show } def
/Cshow { currentpoint stroke M
  dup stringwidth pop -2 div vshift R show } def
/UP { dup vpt_ mul /vpt exch def hpt_ mul /hpt exch def
  /hpt2 hpt 2 mul def /vpt2 vpt 2 mul def } def
/DL { Color {setrgbcolor Solid {pop []} if 0 setdash }
 {pop pop pop Solid {pop []} if 0 setdash} ifelse } def
/BL { stroke gnulinewidth 2 mul setlinewidth } def
/AL { stroke gnulinewidth 2 div setlinewidth } def
/UL { gnulinewidth mul /userlinewidth exch def } def
/PL { stroke userlinewidth setlinewidth } def
/LTb { BL [] 0 0 0 DL } def
/LTa { AL [1 dl 2 dl] 0 setdash 0 0 0 setrgbcolor } def
/LT0 { PL [] 1 0 0 DL } def
/LT1 { PL [4 dl 2 dl] 0 1 0 DL } def
/LT2 { PL [2 dl 3 dl] 0 0 1 DL } def
/LT3 { PL [1 dl 1.5 dl] 1 0 1 DL } def
/LT4 { PL [5 dl 2 dl 1 dl 2 dl] 0 1 1 DL } def
/LT5 { PL [4 dl 3 dl 1 dl 3 dl] 1 1 0 DL } def
/LT6 { PL [2 dl 2 dl 2 dl 4 dl] 0 0 0 DL } def
/LT7 { PL [2 dl 2 dl 2 dl 2 dl 2 dl 4 dl] 1 0.3 0 DL } def
/LT8 { PL [2 dl 2 dl 2 dl 2 dl 2 dl 2 dl 2 dl 4 dl] 0.5 0.5 0.5 DL } def
/Pnt { stroke [] 0 setdash
   gsave 1 setlinecap M 0 0 V stroke grestore } def
/Dia { stroke [] 0 setdash 2 copy vpt add M
  hpt neg vpt neg V hpt vpt neg V
  hpt vpt V hpt neg vpt V closepath stroke
  Pnt } def
/Pls { stroke [] 0 setdash vpt sub M 0 vpt2 V
  currentpoint stroke M
  hpt neg vpt neg R hpt2 0 V stroke
  } def
/Box { stroke [] 0 setdash 2 copy exch hpt sub exch vpt add M
  0 vpt2 neg V hpt2 0 V 0 vpt2 V
  hpt2 neg 0 V closepath stroke
  Pnt } def
/Crs { stroke [] 0 setdash exch hpt sub exch vpt add M
  hpt2 vpt2 neg V currentpoint stroke M
  hpt2 neg 0 R hpt2 vpt2 V stroke } def
/TriU { stroke [] 0 setdash 2 copy vpt 1.12 mul add M
  hpt neg vpt -1.62 mul V
  hpt 2 mul 0 V
  hpt neg vpt 1.62 mul V closepath stroke
  Pnt  } def
/Star { 2 copy Pls Crs } def
/BoxF { stroke [] 0 setdash exch hpt sub exch vpt add M
  0 vpt2 neg V  hpt2 0 V  0 vpt2 V
  hpt2 neg 0 V  closepath fill } def
/TriUF { stroke [] 0 setdash vpt 1.12 mul add M
  hpt neg vpt -1.62 mul V
  hpt 2 mul 0 V
  hpt neg vpt 1.62 mul V closepath fill } def
/TriD { stroke [] 0 setdash 2 copy vpt 1.12 mul sub M
  hpt neg vpt 1.62 mul V
  hpt 2 mul 0 V
  hpt neg vpt -1.62 mul V closepath stroke
  Pnt  } def
/TriDF { stroke [] 0 setdash vpt 1.12 mul sub M
  hpt neg vpt 1.62 mul V
  hpt 2 mul 0 V
  hpt neg vpt -1.62 mul V closepath fill} def
/DiaF { stroke [] 0 setdash vpt add M
  hpt neg vpt neg V hpt vpt neg V
  hpt vpt V hpt neg vpt V closepath fill } def
/Pent { stroke [] 0 setdash 2 copy gsave
  translate 0 hpt M 4 {72 rotate 0 hpt L} repeat
  closepath stroke grestore Pnt } def
/PentF { stroke [] 0 setdash gsave
  translate 0 hpt M 4 {72 rotate 0 hpt L} repeat
  closepath fill grestore } def
/Circle { stroke [] 0 setdash 2 copy
  hpt 0 360 arc stroke Pnt } def
/CircleF { stroke [] 0 setdash hpt 0 360 arc fill } def
/C0 { BL [] 0 setdash 2 copy moveto vpt 90 450  arc } bind def
/C1 { BL [] 0 setdash 2 copy        moveto
       2 copy  vpt 0 90 arc closepath fill
               vpt 0 360 arc closepath } bind def
/C2 { BL [] 0 setdash 2 copy moveto
       2 copy  vpt 90 180 arc closepath fill
               vpt 0 360 arc closepath } bind def
/C3 { BL [] 0 setdash 2 copy moveto
       2 copy  vpt 0 180 arc closepath fill
               vpt 0 360 arc closepath } bind def
/C4 { BL [] 0 setdash 2 copy moveto
       2 copy  vpt 180 270 arc closepath fill
               vpt 0 360 arc closepath } bind def
/C5 { BL [] 0 setdash 2 copy moveto
       2 copy  vpt 0 90 arc
       2 copy moveto
       2 copy  vpt 180 270 arc closepath fill
               vpt 0 360 arc } bind def
/C6 { BL [] 0 setdash 2 copy moveto
      2 copy  vpt 90 270 arc closepath fill
              vpt 0 360 arc closepath } bind def
/C7 { BL [] 0 setdash 2 copy moveto
      2 copy  vpt 0 270 arc closepath fill
              vpt 0 360 arc closepath } bind def
/C8 { BL [] 0 setdash 2 copy moveto
      2 copy vpt 270 360 arc closepath fill
              vpt 0 360 arc closepath } bind def
/C9 { BL [] 0 setdash 2 copy moveto
      2 copy  vpt 270 450 arc closepath fill
              vpt 0 360 arc closepath } bind def
/C10 { BL [] 0 setdash 2 copy 2 copy moveto vpt 270 360 arc closepath fill
       2 copy moveto
       2 copy vpt 90 180 arc closepath fill
               vpt 0 360 arc closepath } bind def
/C11 { BL [] 0 setdash 2 copy moveto
       2 copy  vpt 0 180 arc closepath fill
       2 copy moveto
       2 copy  vpt 270 360 arc closepath fill
               vpt 0 360 arc closepath } bind def
/C12 { BL [] 0 setdash 2 copy moveto
       2 copy  vpt 180 360 arc closepath fill
               vpt 0 360 arc closepath } bind def
/C13 { BL [] 0 setdash  2 copy moveto
       2 copy  vpt 0 90 arc closepath fill
       2 copy moveto
       2 copy  vpt 180 360 arc closepath fill
               vpt 0 360 arc closepath } bind def
/C14 { BL [] 0 setdash 2 copy moveto
       2 copy  vpt 90 360 arc closepath fill
               vpt 0 360 arc } bind def
/C15 { BL [] 0 setdash 2 copy vpt 0 360 arc closepath fill
               vpt 0 360 arc closepath } bind def
/Rec   { newpath 4 2 roll moveto 1 index 0 rlineto 0 exch rlineto
       neg 0 rlineto closepath } bind def
/Square { dup Rec } bind def
/Bsquare { vpt sub exch vpt sub exch vpt2 Square } bind def
/S0 { BL [] 0 setdash 2 copy moveto 0 vpt rlineto BL Bsquare } bind def
/S1 { BL [] 0 setdash 2 copy vpt Square fill Bsquare } bind def
/S2 { BL [] 0 setdash 2 copy exch vpt sub exch vpt Square fill Bsquare } bind def
/S3 { BL [] 0 setdash 2 copy exch vpt sub exch vpt2 vpt Rec fill Bsquare } bind def
/S4 { BL [] 0 setdash 2 copy exch vpt sub exch vpt sub vpt Square fill Bsquare } bind def
/S5 { BL [] 0 setdash 2 copy 2 copy vpt Square fill
       exch vpt sub exch vpt sub vpt Square fill Bsquare } bind def
/S6 { BL [] 0 setdash 2 copy exch vpt sub exch vpt sub vpt vpt2 Rec fill Bsquare } bind def
/S7 { BL [] 0 setdash 2 copy exch vpt sub exch vpt sub vpt vpt2 Rec fill
       2 copy vpt Square fill
       Bsquare } bind def
/S8 { BL [] 0 setdash 2 copy vpt sub vpt Square fill Bsquare } bind def
/S9 { BL [] 0 setdash 2 copy vpt sub vpt vpt2 Rec fill Bsquare } bind def
/S10 { BL [] 0 setdash 2 copy vpt sub vpt Square fill 2 copy exch vpt sub exch vpt Square fill
       Bsquare } bind def
/S11 { BL [] 0 setdash 2 copy vpt sub vpt Square fill 2 copy exch vpt sub exch vpt2 vpt Rec fill
       Bsquare } bind def
/S12 { BL [] 0 setdash 2 copy exch vpt sub exch vpt sub vpt2 vpt Rec fill Bsquare } bind def
/S13 { BL [] 0 setdash 2 copy exch vpt sub exch vpt sub vpt2 vpt Rec fill
       2 copy vpt Square fill Bsquare } bind def
/S14 { BL [] 0 setdash 2 copy exch vpt sub exch vpt sub vpt2 vpt Rec fill
       2 copy exch vpt sub exch vpt Square fill Bsquare } bind def
/S15 { BL [] 0 setdash 2 copy Bsquare fill Bsquare } bind def
/D0 { gsave translate 45 rotate 0 0 S0 stroke grestore } bind def
/D1 { gsave translate 45 rotate 0 0 S1 stroke grestore } bind def
/D2 { gsave translate 45 rotate 0 0 S2 stroke grestore } bind def
/D3 { gsave translate 45 rotate 0 0 S3 stroke grestore } bind def
/D4 { gsave translate 45 rotate 0 0 S4 stroke grestore } bind def
/D5 { gsave translate 45 rotate 0 0 S5 stroke grestore } bind def
/D6 { gsave translate 45 rotate 0 0 S6 stroke grestore } bind def
/D7 { gsave translate 45 rotate 0 0 S7 stroke grestore } bind def
/D8 { gsave translate 45 rotate 0 0 S8 stroke grestore } bind def
/D9 { gsave translate 45 rotate 0 0 S9 stroke grestore } bind def
/D10 { gsave translate 45 rotate 0 0 S10 stroke grestore } bind def
/D11 { gsave translate 45 rotate 0 0 S11 stroke grestore } bind def
/D12 { gsave translate 45 rotate 0 0 S12 stroke grestore } bind def
/D13 { gsave translate 45 rotate 0 0 S13 stroke grestore } bind def
/D14 { gsave translate 45 rotate 0 0 S14 stroke grestore } bind def
/D15 { gsave translate 45 rotate 0 0 S15 stroke grestore } bind def
/DiaE { stroke [] 0 setdash vpt add M
  hpt neg vpt neg V hpt vpt neg V
  hpt vpt V hpt neg vpt V closepath stroke } def
/BoxE { stroke [] 0 setdash exch hpt sub exch vpt add M
  0 vpt2 neg V hpt2 0 V 0 vpt2 V
  hpt2 neg 0 V closepath stroke } def
/TriUE { stroke [] 0 setdash vpt 1.12 mul add M
  hpt neg vpt -1.62 mul V
  hpt 2 mul 0 V
  hpt neg vpt 1.62 mul V closepath stroke } def
/TriDE { stroke [] 0 setdash vpt 1.12 mul sub M
  hpt neg vpt 1.62 mul V
  hpt 2 mul 0 V
  hpt neg vpt -1.62 mul V closepath stroke } def
/PentE { stroke [] 0 setdash gsave
  translate 0 hpt M 4 {72 rotate 0 hpt L} repeat
  closepath stroke grestore } def
/CircE { stroke [] 0 setdash 
  hpt 0 360 arc stroke } def
/Opaque { gsave closepath 1 setgray fill grestore 0 setgray closepath } def
/DiaW { stroke [] 0 setdash vpt add M
  hpt neg vpt neg V hpt vpt neg V
  hpt vpt V hpt neg vpt V Opaque stroke } def
/BoxW { stroke [] 0 setdash exch hpt sub exch vpt add M
  0 vpt2 neg V hpt2 0 V 0 vpt2 V
  hpt2 neg 0 V Opaque stroke } def
/TriUW { stroke [] 0 setdash vpt 1.12 mul add M
  hpt neg vpt -1.62 mul V
  hpt 2 mul 0 V
  hpt neg vpt 1.62 mul V Opaque stroke } def
/TriDW { stroke [] 0 setdash vpt 1.12 mul sub M
  hpt neg vpt 1.62 mul V
  hpt 2 mul 0 V
  hpt neg vpt -1.62 mul V Opaque stroke } def
/PentW { stroke [] 0 setdash gsave
  translate 0 hpt M 4 {72 rotate 0 hpt L} repeat
  Opaque stroke grestore } def
/CircW { stroke [] 0 setdash 
  hpt 0 360 arc Opaque stroke } def
/BoxFill { gsave Rec 1 setgray fill grestore } def
end
}}%
\begin{picture}(2412,2160)(0,0)%
{\GNUPLOTspecial{"
gnudict begin
gsave
0 0 translate
0.100 0.100 scale
0 setgray
newpath
1.000 UL
LTb
0.050 UL
LT3
400 300 M
2012 0 V
1.000 UL
LTb
400 300 M
63 0 V
1949 0 R
-63 0 V
400 380 M
31 0 V
1981 0 R
-31 0 V
400 486 M
31 0 V
1981 0 R
-31 0 V
400 540 M
31 0 V
1981 0 R
-31 0 V
0.050 UL
LT3
400 566 M
2012 0 V
1.000 UL
LTb
400 566 M
63 0 V
1949 0 R
-63 0 V
400 646 M
31 0 V
1981 0 R
-31 0 V
400 751 M
31 0 V
1981 0 R
-31 0 V
400 806 M
31 0 V
1981 0 R
-31 0 V
0.050 UL
LT3
400 831 M
2012 0 V
1.000 UL
LTb
400 831 M
63 0 V
1949 0 R
-63 0 V
400 911 M
31 0 V
1981 0 R
-31 0 V
400 1017 M
31 0 V
1981 0 R
-31 0 V
400 1071 M
31 0 V
1981 0 R
-31 0 V
0.050 UL
LT3
400 1097 M
2012 0 V
1.000 UL
LTb
400 1097 M
63 0 V
1949 0 R
-63 0 V
400 1177 M
31 0 V
1981 0 R
-31 0 V
400 1283 M
31 0 V
1981 0 R
-31 0 V
400 1337 M
31 0 V
1981 0 R
-31 0 V
0.050 UL
LT3
400 1363 M
2012 0 V
1.000 UL
LTb
400 1363 M
63 0 V
1949 0 R
-63 0 V
400 1443 M
31 0 V
1981 0 R
-31 0 V
400 1549 M
31 0 V
1981 0 R
-31 0 V
400 1603 M
31 0 V
1981 0 R
-31 0 V
0.050 UL
LT3
400 1629 M
2012 0 V
1.000 UL
LTb
400 1629 M
63 0 V
1949 0 R
-63 0 V
400 1709 M
31 0 V
1981 0 R
-31 0 V
400 1814 M
31 0 V
1981 0 R
-31 0 V
400 1869 M
31 0 V
1981 0 R
-31 0 V
0.050 UL
LT3
400 1894 M
1275 0 V
687 0 R
50 0 V
1.000 UL
LTb
400 1894 M
63 0 V
1949 0 R
-63 0 V
400 1974 M
31 0 V
1981 0 R
-31 0 V
400 2080 M
31 0 V
1981 0 R
-31 0 V
400 2134 M
31 0 V
1981 0 R
-31 0 V
0.050 UL
LT3
400 2160 M
2012 0 V
1.000 UL
LTb
400 2160 M
63 0 V
1949 0 R
-63 0 V
0.050 UL
LT3
400 300 M
0 1860 V
1.000 UL
LTb
400 300 M
0 63 V
0 1797 R
0 -63 V
521 300 M
0 31 V
0 1829 R
0 -31 V
681 300 M
0 31 V
0 1829 R
0 -31 V
763 300 M
0 31 V
0 1829 R
0 -31 V
0.050 UL
LT3
802 300 M
0 1860 V
1.000 UL
LTb
802 300 M
0 63 V
0 1797 R
0 -63 V
924 300 M
0 31 V
0 1829 R
0 -31 V
1084 300 M
0 31 V
0 1829 R
0 -31 V
1166 300 M
0 31 V
0 1829 R
0 -31 V
0.050 UL
LT3
1205 300 M
0 1860 V
1.000 UL
LTb
1205 300 M
0 63 V
0 1797 R
0 -63 V
1326 300 M
0 31 V
0 1829 R
0 -31 V
1486 300 M
0 31 V
0 1829 R
0 -31 V
1568 300 M
0 31 V
0 1829 R
0 -31 V
0.050 UL
LT3
1607 300 M
0 1860 V
1.000 UL
LTb
1607 300 M
0 63 V
0 1797 R
0 -63 V
1728 300 M
0 31 V
0 1829 R
0 -31 V
1888 300 M
0 31 V
0 1829 R
0 -31 V
1971 300 M
0 31 V
0 1829 R
0 -31 V
0.050 UL
LT3
2010 300 M
0 1497 V
0 300 R
0 63 V
1.000 UL
LTb
2010 300 M
0 63 V
0 1797 R
0 -63 V
2131 300 M
0 31 V
0 1829 R
0 -31 V
2291 300 M
0 31 V
0 1829 R
0 -31 V
2373 300 M
0 31 V
0 1829 R
0 -31 V
0.050 UL
LT3
2412 300 M
0 1860 V
1.000 UL
LTb
2412 300 M
0 63 V
0 1797 R
0 -63 V
1.000 UL
LTb
400 300 M
2012 0 V
0 1860 V
-2012 0 V
400 300 L
1.000 UL
LTb
1675 1797 M
0 300 V
687 0 V
0 -300 V
-687 0 V
0 300 R
687 0 V
0.000 UP
1.000 UL
LT0
2075 2047 M
237 0 V
640 300 M
0 1514 V
76 114 V
75 11 V
72 -39 V
69 -64 V
64 -74 V
60 -71 V
57 -91 V
54 -49 V
50 -103 V
48 -40 V
46 -56 V
43 -94 V
42 -10 V
39 -87 V
38 39 V
36 -53 V
35 -72 V
34 -9 V
32 -7 V
31 -24 V
30 -128 V
29 57 V
28 -66 V
28 22 V
26 -23 V
26 20 V
25 -147 V
24 -8 V
24 -5 V
22 105 V
23 -158 V
22 25 V
21 -45 V
21 33 V
20 14 V
20 -109 V
19 -2 V
19 -55 V
19 81 V
18 -103 V
18 106 V
17 -108 V
18 -16 V
16 -47 V
17 155 V
16 -175 V
16 -2 V
15 -61 V
16 -14 V
15 168 V
15 -81 V
640 1814 Circle
716 1928 Circle
791 1939 Circle
863 1900 Circle
932 1836 Circle
996 1762 Circle
1056 1691 Circle
1113 1600 Circle
1167 1551 Circle
1217 1448 Circle
1265 1408 Circle
1311 1352 Circle
1354 1258 Circle
1396 1248 Circle
1435 1161 Circle
1473 1200 Circle
1509 1147 Circle
1544 1075 Circle
1578 1066 Circle
1610 1059 Circle
1641 1035 Circle
1671 907 Circle
1700 964 Circle
1728 898 Circle
1756 920 Circle
1782 897 Circle
1808 917 Circle
1833 770 Circle
1857 762 Circle
1881 757 Circle
1903 862 Circle
1926 704 Circle
1948 729 Circle
1969 684 Circle
1990 717 Circle
2010 731 Circle
2030 622 Circle
2049 620 Circle
2068 565 Circle
2087 646 Circle
2105 543 Circle
2123 649 Circle
2140 541 Circle
2158 525 Circle
2174 478 Circle
2191 633 Circle
2207 458 Circle
2223 456 Circle
2238 395 Circle
2254 381 Circle
2269 549 Circle
2284 468 Circle
2193 2047 Circle
0.000 UP
1.000 UL
LT1
2075 1947 M
237 0 V
567 300 M
0 1272 V
73 -14 V
76 132 V
75 75 V
72 22 V
69 -25 V
64 -47 V
60 -36 V
57 -48 V
54 -63 V
50 -42 V
48 -81 V
46 -14 V
43 -61 V
42 -60 V
39 -57 V
38 -38 V
36 0 V
35 -51 V
34 -10 V
32 -38 V
31 8 V
30 -158 V
29 53 V
28 -106 V
28 26 V
26 -105 V
26 155 V
25 -113 V
24 -82 V
24 73 V
22 -20 V
23 -134 V
22 92 V
21 -14 V
21 48 V
20 -96 V
20 0 V
19 -41 V
19 -9 V
19 -36 V
18 24 V
18 27 V
17 -27 V
18 -84 V
16 16 V
17 -28 V
16 -1 V
16 97 V
15 -118 V
16 29 V
15 43 V
15 -126 V
567 1572 TriU
640 1558 TriU
716 1690 TriU
791 1765 TriU
863 1787 TriU
932 1762 TriU
996 1715 TriU
1056 1679 TriU
1113 1631 TriU
1167 1568 TriU
1217 1526 TriU
1265 1445 TriU
1311 1431 TriU
1354 1370 TriU
1396 1310 TriU
1435 1253 TriU
1473 1215 TriU
1509 1215 TriU
1544 1164 TriU
1578 1154 TriU
1610 1116 TriU
1641 1124 TriU
1671 966 TriU
1700 1019 TriU
1728 913 TriU
1756 939 TriU
1782 834 TriU
1808 989 TriU
1833 876 TriU
1857 794 TriU
1881 867 TriU
1903 847 TriU
1926 713 TriU
1948 805 TriU
1969 791 TriU
1990 839 TriU
2010 743 TriU
2030 743 TriU
2049 702 TriU
2068 693 TriU
2087 657 TriU
2105 681 TriU
2123 708 TriU
2140 681 TriU
2158 597 TriU
2174 613 TriU
2191 585 TriU
2207 584 TriU
2223 681 TriU
2238 563 TriU
2254 592 TriU
2269 635 TriU
2284 509 TriU
2193 1947 TriU
0.000 UP
1.000 UL
LT3
2075 1847 M
237 0 V
716 300 M
0 1199 V
716 300 L
147 0 R
0 1015 V
69 74 V
64 -82 V
60 5 V
57 12 V
54 -50 V
50 -51 V
48 37 V
46 57 V
43 -14 V
42 71 V
39 -33 V
38 -33 V
36 -18 V
35 -10 V
34 -36 V
32 -2 V
31 -32 V
30 -1 V
29 -39 V
28 15 V
28 -38 V
26 -26 V
26 -25 V
25 28 V
24 -41 V
24 -42 V
22 5 V
23 -76 V
22 4 V
21 -3 V
21 -29 V
20 -97 V
20 53 V
19 -11 V
19 -61 V
19 -8 V
18 -39 V
18 -37 V
17 9 V
18 62 V
16 -67 V
17 78 V
16 -126 V
16 -54 V
15 36 V
16 8 V
15 -106 V
15 90 V
716 1499 Dia
863 1315 Dia
932 1389 Dia
996 1307 Dia
1056 1312 Dia
1113 1324 Dia
1167 1274 Dia
1217 1223 Dia
1265 1260 Dia
1311 1317 Dia
1354 1303 Dia
1396 1374 Dia
1435 1341 Dia
1473 1308 Dia
1509 1290 Dia
1544 1280 Dia
1578 1244 Dia
1610 1242 Dia
1641 1210 Dia
1671 1209 Dia
1700 1170 Dia
1728 1185 Dia
1756 1147 Dia
1782 1121 Dia
1808 1096 Dia
1833 1124 Dia
1857 1083 Dia
1881 1041 Dia
1903 1046 Dia
1926 970 Dia
1948 974 Dia
1969 971 Dia
1990 942 Dia
2010 845 Dia
2030 898 Dia
2049 887 Dia
2068 826 Dia
2087 818 Dia
2105 779 Dia
2123 742 Dia
2140 751 Dia
2158 813 Dia
2174 746 Dia
2191 824 Dia
2207 698 Dia
2223 644 Dia
2238 680 Dia
2254 688 Dia
2269 582 Dia
2284 672 Dia
2193 1847 Dia
stroke
grestore
end
showpage
}}%
\put(2025,1847){\makebox(0,0)[r]{$q=100$}}%
\put(2025,1947){\makebox(0,0)[r]{$q=10$}}%
\put(2025,2047){\makebox(0,0)[r]{$q=1$}}%
\put(1406,2310){\makebox(0,0){\textsf{Angular momentum loss rate $\frac{d\tilde{L}_z}{d\tilde{t}}$}}}%
\put(1406,50){\makebox(0,0){\sffamily $\tilde{t}$}}%
\put(50,1230){%
\special{ps: gsave currentpoint currentpoint translate
270 rotate neg exch neg exch translate}%
\makebox(0,0)[b]{\shortstack{\sffamily $\frac{d\tilde{L}_z}{d\tilde{t}}$}}%
\special{ps: currentpoint grestore moveto}%
}%
\put(2412,200){\makebox(0,0){1e+06}}%
\put(2010,200){\makebox(0,0){100000}}%
\put(1607,200){\makebox(0,0){10000}}%
\put(1205,200){\makebox(0,0){1000}}%
\put(802,200){\makebox(0,0){100}}%
\put(400,200){\makebox(0,0){10}}%
\put(350,2160){\makebox(0,0)[r]{0.01}}%
\put(350,1894){\makebox(0,0)[r]{0.001}}%
\put(350,1629){\makebox(0,0)[r]{0.0001}}%
\put(350,1363){\makebox(0,0)[r]{1e-05}}%
\put(350,1097){\makebox(0,0)[r]{1e-06}}%
\put(350,831){\makebox(0,0)[r]{1e-07}}%
\put(350,566){\makebox(0,0)[r]{1e-08}}%
\put(350,300){\makebox(0,0)[r]{1e-09}}%
\end{picture}%
\endgroup
 